\documentstyle[aps,prb,multicol,graphicx]{revtex}

\newcommand{\keq}{\! = \!}

\title{Topological phase-fluctuations, amplitude fluctuations, and
  criticality \\ in extreme type-II superconductors} 

\author{A. K.  Nguyen$^1$ and A. Sudb{\o}$^{1,2}$} 

\address{$^1$ Norwegian University of Science and Technology, N-7034
  Trondheim, Norway \\ $^2$ California Institute of Technology,
  Pasadena, CA91125, USA }

\begin{document}
  \maketitle

\begin{abstract}
  We study the effect of critical fluctuations on the $(B,T)$ phase
  diagram in extreme type-II superconductors in zero and finite
  magnetic field.  In zero magnetic field the critical fluctuations
  are transverse phase-fluctuations of the complex scalar
  Ginzurg-Landau order parameter, which when excited thermally will
  induce line-defects in the form of closed vortex loops into the
  system. The distribution function $D(p)$ of vortex loops of
  perimeter $p$ changes from an exponential function $D(p) \sim
  p^{-\alpha} ~ \exp(-\varepsilon(T) p/k_BT)$ to a power law
  distribution $D(p) \sim p^{-\alpha}$ at the zero-field critical
  temperature $T=T_c$. We find that the long-wavelength vortex-line
  tension vanishes as $\varepsilon(T) \sim |T-T_c|^{\gamma}; \gamma
  \approx 1.45$, as $T \to T_c$.  At $T=T_c$, an extreme type-II
  superconductor suffers an unbinding of large vortex loops of order
  the system size. When this happens, the connectivity of the
  thermally excited vortex-tangle of the system changes abruptly.
  {\it When amplitude fluctuations are included, it is shown that they
    are far from being critical at the superconducting transition
    temperature $T_c$. The vortex-loop unbinding can therefore not be
    reparametrized in terms of critical amplitude fluctuations of the
    original local Ginzburg-Landau order parameter.} The loss of
  phase-stiffness in the Ginzburg-Landau order parameter, the anomaly
  in specific heat, the loss of vortex-line tension, and the change in
  the connectivity of the vortex-tangle are all found at the same
  temperature, the critical temperature of the superconductor.  At
  zero magnetic field, unbinding of vortex-loops of order the system
  size can be phrased in terms of a global $U(1)$-symmetry breaking
  involving a local complex disorder field which is dual to the order
  parameter of the usual Ginzburg-Landau theory. There is one
  parameter in the theory that controls the width of the critical
  region, and for the parameters we have used, we show that a
  vortex-loop unbinding gives a correct picture of the zero-field
  transition even in the presence of amplitude fluctuations.  {\it A
    key result is the extraction of the anomalous dimension of the
    dual field directly from the statistics of the vortex-loop
    excitations of the Ginzburg-Landau theory in the phase-only
    approximation.} In finite magnetic fields, the first order
  vortex-line lattice (VLL) melting transition is accompanied by a
  loss of longitudinal superfluid stiffness; this is true also for the
  isotropic case. A scaling analysis of the vortex lattice melting
  line is carried out, yielding {\it two different scaling regimes}
  for the vortex lattice melting line, namely a high-field scaling
  regime and a distinct low-field $3DXY$ scaling regime. We also find
  indications of an abrupt change in the connectivity of the
  vortex-tangle in the vortex liquid along a line $T_L(B)$, which at
  low enough fields appears to coincide with the VLL melting
  transition line within the resolution of our numerical calculations.
  We study the temperature at which this phenomenon takes place as a
  function of system size and shape. Our results show that this
  temperature decreases and appears to saturate with increasing system
  size, and is insensitive to aspect ratios of the systems on which
  the simulations are performed on, for large enough systems.  A
  necessary, but not sufficient, condition for the vortex-line tension
  to vanish is a change in the connectivity of the vortex tangle in an
  extreme type-II superconductor.

Pacs-numbers: 74.20.De, 74.25.Dw, 74.25.Ha,74.60.Ec
\end{abstract}

\begin{multicols}{2}

\section{Introduction}

Ten years after Abrikosov's classic prediction of a lattice of
quantized vortices, the Abrikosov vortex-line lattice (VLL),
\cite{Abrikosov:ZETP57} as the ground state of type-II superconductors
when the magnetic field is tuned beyond a lower critical value
\cite{Footnote:VLL}, Gerd Eilenberger suggested that the VLL could
melt close to the critical temperature of the system
\cite{Eilenberger:PR67}. The magnetic field versus temperature
$(B,T)$-phase diagram of extreme type-II superconductors has for some
time been under intense investigation both theoretically and
experimentally, following suggestions that the VLL could undergo a
melting transition in regime of the $(B,T)$-phase diagram of the
high-temperature superconductor that could be experimentally resolved
\cite{Gammel:L87,Nelson:L88}. This was soon confirmed by a more
thorough theoretical analysis \cite{Houghton:B89} where it was shown
that the VLL of the high-temperature superconductors was particularly
susceptible to thermal fluctuations due to the large anisotropy of
these compounds. The anisotropy only affects the melting line of the
VLL when the pronounced nonlocal elastic properties of the VLL in
strong type-II superconductors, first discussed for the isotropic case
in the pioneering works of Brandt
\cite{Brandt:JLTP77a,Brandt:JLTP77b}, are taken into account
\cite{Houghton:B89}.

While it now appears well established both experimentally
\cite{Safar:L92} and theoretically for three dimensional vortex
systems \cite{Brezin:B85,Hetzel:L92} that the VLL in the clean limit
of type-II superconductors melts in a first order phase-transition,
much less consensus has been reached on how to describe the state
which the VLL melts into, even in the clean limit. Only very recently
has it been established, through numerical simulations
\cite{Hu:L97,Nguyen:B98b} that the vortex-liquid is always {\it
  incoherent}, i.e phase-coherence is destroyed in all direction,
including the direction of the induction, as soon as the VLL melts.
Inside the vortex-liquid {\it regime}, there is therefore no
transition from a disentangled to an entangled vortex-liqiud. For such
a transition to occur inside the vortex-liquid, the longitudinal
superfluid density would have to be non-zero above the melting
temperature. This however does not happen in the clean limit
\cite{Hu:L97,Nguyen:B98b}, even in the isotropic case
\cite{Chin:cm98}. Recently, questions have also been raised whether
the vortex-line picture of the molten phase of the Abrikosov VLL is
viable at all at low fields $B < 1T$
\cite{Tesanovic:B95,Nguyen:L96,Nguyen:B98a,Nguyen:B98b,Tesanovic:cm98}
$\!\!^,$\cite{Chin:cm98}.

In terms of fundamental physics, extreme type-II superconductors are
interesting due to their large fluctuation effects not commonly seen
in condensed matter systems. This is ultimately due to the fact that
they are strong coupling superconductors arising out of doped
Mott-Hubbard insulators. The latter fact gives rise to the effect that
the phase-stiffness of the superconducting order parameter is low, due
to a low value of the superfluid density $\rho_s$
\begin{eqnarray*}
\rho_s  \sim \frac{\partial^2 f}{\partial (\Delta \theta)^2},
\end{eqnarray*}
where $\Delta \theta$ is a twist in the superconducting order
parameter over the size of the system, and $f$ is the free energy
density. This particular and important aspect of doped
Mott-Hubbard insulators has been quite strongly 
emphasized already for some time 
\cite{Emery:N95,Emery:JPCS98,Emery:cm99}, 
see also Ref. \onlinecite{Carlson:cm99}. The strong 
coupling effect gives rise to a large $T_c$, so that the Ginzburg-Landau 
parameter $\kappa \sim T_c/\sqrt{\rho_s}$ is large. This also softens 
the vortex matter in these systems, particularly when coupled with 
their strong layeredness \cite{Houghton:B89}.

There is also a close connection between thermodynamic phase
transitions in these systems, and phase-transitions in superfluids
\cite{Griffin:Bo93}, liquid crystals \cite{Gennes:Bo93}, crystals
\cite{Kleinert:Bo89}, and cosmology
\cite{Farakos:NPB94,Kajantie:NPB98,Hindmarsh:RPP95,Zurek:PR96,Vilenkin:Bo94}.
The close connection between these apparently different physical
problems, is due to the similarity of the topological objects that
appear in these problems. Particularly in the context of superfluid
$He^4$, the proposition that an unbinding of topological
phase-fluctuations in the form of vortex-loops is the microscopic
mechanism for the superfluid-normal state transition, has been
extensively studied in the past
\cite{Dasgupta:L81,Williams:L87,Williams:PhysB:90,Shenoy:B89,Shenoy:L94}
$^,$\cite{Williams:L98}, and early attempts at formulating a
field-theory of this in the context of charged superfluids in zero
magnetic field has also appeared in the literature
\cite{Halperin:LesHouches80}.  Effective gauge-field theories with an
internal $U(1)$-symmetry all have in common that they support {\it
  line-defect} in the form of vortex-loop excitations as stable
topological objects. Understanding the role of such excitations on the
$(B,T)$ phase-diagram of type-II superconductors is an important
problem in physics, and presumably will shed light on the related
problems mentioned above as well.

In conventional low-temperature superconductors, the temperature where
Cooper pairs start to form, $T_{MF}$, is practically identical to the
true superconducting transition temperature $T_c$. The commonly
applied Ginzburg-criterion provides a useful estimate for the width of
the critical regions in systems with weak fluctuation effects, showing
that the width of the critical region is of order $|t| \sim
(T_c-T)/T_c \sim 10^{-6}-10^{-4}$.  A mean field description of the
S-N phase transition is appropriate `for all practical purposes". In
high-$T_c$ superconductors, this may no longer be the case.  There
appears to be mounting experimental evidence that the width of the
critical region is as large as a few Kelvin in YBCO
\cite{Salamon:B93,Junod:C97,Roulin:C97,Roulin:Thesis,Roulin:L98},
which would encompass the melting line of the flux-line lattice up to
a field of order $1T$ \cite{Houghton:B89}.

In zero field the superconducting-normal phase transition is
exclusively caused by a vortex loop unbinding
\cite{Nguyen:B98a,Nguyen:cm98,Chin:cm98,Antunes:L98a,Antunes:L98b}.
Below the critical temperature $T_c$ vortex loops are confined to a
typical perimeter $L_0$, and cause only local disturbances in the
macroscopic superconducting state. Recently, this has been
demonstrated clearly, by correlating an abrupt change in vortex tangle
connectivity, a loss of vortex-line tension, loss of superfluid
stiffness and specific heat anomaly precisely at the critical
temperature of the superconductor, even for the isotropic case
\cite{Nguyen:cm98,Chin:cm98}. At $T_c$, thermally induced vortex loops
loose their effective line tension and therefore unbind.

In this scenario, at low fields, thermally induced vortex loops could
conceivably interact strongly with field induced flux lines. This
interaction is ignored in models using the line-only approximation,
where the thermally induced vortex loops degrees of freedom are
neglected, effectively being considered as irrelevant relativistic
corrections in a corresponding $2D$ quantum boson system.
\cite{Nelson:L88,Blatter:RMP94,Nordborg:Thesis}. This model has met
with considerable success in describing parts of the flux-line lattice
melting curve at intermediate to elevated fields, where its position
the phase-diagram as well as its dependence on anisotropy was
explained using the Lindemann-criterion
\cite{Houghton:B89,Footnote:Houghton,Blatter:RMP94}. The melting line
for fields of more than a few Tesla is little affected by the
vortex-loop unbinding, as pointed out recently
\cite{Nguyen:cm98,Nonomura:cm98}.

However, the question arises whether this is a tenable conclusion for
low fields as well. The fact that the zero-field transition can be
characterized precisely by a loss of line-tension of thermally induced
flux lines, implies that there is a sharp change in the distribution
function $D(p)$ for vortex loops of a given perimeter $p$. It changes
from an exponential form $D(p) \sim p^{-5/2} ~ \exp(-\varepsilon(T)
p/k_B T)$ to a power law $D(p) \sim p^{-5/2}$ at $T_c$
\cite{Chin:cm98}. This implies the existence of a diverging length
scale $L_0(T) = k_B T/\varepsilon(T)$ \cite{Nguyen:cm98,Chin:cm98}.
Given this fact, it raises the question of whether the critical
fluctuations can affect the melting line in a sizeable
field-temperature regime, rendering the vortex lines tension-less.
The vortex line tension is analagous to the mass of the bosons in a
$2D$ non-relativistic boson-analogy of the vortex system.  If the
vortex-line tension were to vanish, it would mean that the boson-mass
wouild vanish in the corresponding analogy. There is no
non-relativistic limit of any mass-less theory. The conclusion would
be that any $2D$ boson-model which is non-relativistic, is
inapplicable in the part of the phase-diagram where the vortex-line
tension vanishes.  We reemphasize that at elevated fields, where the
first order flux-line lattice melting line splits off from the new
transition line proposed here, the Lindemann-criterion of flux-line
lattice melting \cite{Houghton:B89} is expected to correctly locate
the position of the melting line \cite{Nguyen:cm98}.

The outline of this paper is as follows. In Section II, we introduce
the Ginzburg-Landau model studied in this paper, and various
approximations and reformulations of it, as well as their
inter-relations. In Section III we present the ideas underlying the
simulations that are presented in this paper, and introduce and
discuss the quantities we study. In Section IV, we present results of
the simulations in zero magnetic field. {\it In particular, we present
  results which demonstrate that the zero-field transition in an
  extreme type-II superconductor is driven by a proliferation of
  unbound vortex loops, which therefore constitute the critical
  fluctuations of this system}. In Section V, finite-field results are
given. Summary and conclusions are presented in Section VI, and in
this section we also list point by point the new results obtained in
this paper.

\section{Models}

In this section we define the models considered in this paper: 1) the
continuum Ginzburg-Landau model, 2) the lattice Ginzburg-Landau model
in a frozen gauge approximation, and 3) the uniformly frustrated 3D XY
model. We also discuss the approximations involved and the validity of
the models.

\subsection{Ginzburg-Landau model}

Our starting point is the continuum Ginzburg-Landau (GL) model
\cite{Ginzburg:ZETF50}. In quantum field theory, the GL model is also
referred as the scalar QED or the $U(1)$+Higgs model or the Abelian
Higgs model. The effective Hamiltonian for the GL model in an
anisotropic system is given by \cite{Takanaka:PSS75}
\begin{eqnarray}
  H_{GL} &=& \int d^3r \biggl[\alpha(T)~|\psi|^2+\frac{g}{2} |\psi|^4 
             \nonumber \\ 
         &+& \sum_{\mu = x,y,z} \frac{\hbar^2}{2m_\mu} \left |
             \left ( \nabla_\mu - i \frac{2\pi}{\Phi_0} A_\mu \right ) \psi
             \right |^2 \nonumber \\
         &+& \frac{1}{2\mu_0} ( {\mathbf \nabla} \times {\mathbf
    A})^2 \biggr].
\label{GL:Hamiltonian}
\end{eqnarray}
Here, $\psi({\mathbf r}) = |\psi({\mathbf r})| e^{i\theta({\mathbf
    r})}$ is a complex order field representing the superconducting
condensate. In superconductors, the amplitude $|\psi({\mathbf r})|^2$
should, be interpreted as the local Cooper-pair density.  Furthermore,
$m_\mu$ is the effective mass for {\em one} Cooper pair when moving
along the $\mu$-direction, $\Phi_0=h/2e$ is the flux quantum, and
$\mu_0$ is the vacuum permeability. In Eq.  \ref{GL:Hamiltonian}, the
gauge field ${\mathbf A}$ is related to the local magnetic induction,
$b({\mathbf r}) = {\mathbf \nabla} \times {\mathbf A}({\mathbf r})$.
Finally, the GL parameter $g$ is assumed to be temperature
independent, while $\alpha = \alpha(T)$ changes sign at a mean field
critical temperature $T_{MF}(B)$, where Cooper pairs start to form.
$B$ is the spatial average of the magnetic induction. The critical
temperature $T_c$ where phase-coherence develops, is always smaller
than $T_{MF}$; the existence a finite Copper-pair density does not
imply that the system is in a superconducting state.

Later on, we shall recast the Ginzburg-Landau theory in a quite
different form that also exhibts a $U(1)$-symmetry, but where the
field conjugate to the relevant phase is the number operator for the
topological excitations destroying the order of the Ginzburg-Landau
theory itself. Although this may seem like an unnecessary
complication, it has the advantage of facilitating a detailed
discussion of the vortex-liquid phase of the GL-theory in terms of the
ordering of some local field, namely the complex scalar field
$\phi({\bf r})$ to be introduced and discussed in Section IIF. This is
not possible using the Ginzburg-Landay function, $\psi({\bf r})$,
since $<\psi({\bf r})>$ is always zero in the vortex liquid phase
\cite{Hu:L97,Nguyen:B98b}. In the zero-field low-temperature ordered
phase, the system spontaneously chooses a preferred phase angle
$\Theta$, and explicitly breaks the $U(1)$ symmetry. The vortex-sector
of the GL-theory also exhibits a $U(1)$-symmetry breaking, but where
$U(1)$-symmetry is broken in the high-temperature phase, and restored
in the low-temperature phase.

Eq. \ref{GL:Hamiltonian} has two intrinsic length scales, the
mean-field coherence length
\begin{eqnarray}
   \xi_\mu^2(T) = \frac{\hbar^2}{2m_\mu |\alpha(T)|},
\end{eqnarray}
and the magnetic penetration depth
\begin{eqnarray}
   \lambda_\mu^2 = \frac{m_\mu \beta}{4 \mu_0 e^2 |\alpha(T)|}.
\label{Pene.Depht}
\end{eqnarray}
$\xi_\mu$ is the characteristic length of the variation of
$|\psi({\mathbf r})|$ along the $\mu$-direction, and $\lambda_\mu$ is
the characteristic length of the variation of the current flowing
along the $\mu$-direction.

%

In order to carry out Monte Carlo simulations of the GL model, the
model is discretized by replacing the covariant derivative in the
continuum GL Hamiltonian, Eq.  \ref{GL:Hamiltonian}, with a covariant
lattice derivative,
\begin{eqnarray}
  D_\mu \psi & = & ( \nabla_\mu - i \frac{2\pi}{\Phi_0} A_\mu )\psi
  \nonumber \\ 
  \rightarrow {\makebox[0pt][l]{\bf $\! -$}D}_\mu \psi &
  = & \frac{1}{a_\mu} \left ( \psi({\mathbf r}+ \hat{\mu}) e^{-i
      \frac{2\pi}{\Phi_0} a_\mu A_\mu({\mathbf r})} - \psi({\mathbf
      r}) \right ).
\end{eqnarray}
The resulting model is a version of the Lawrence-Doniach model
\cite{Lawrence:PICLTP71} with all three directions discretized. The
effective Hamiltonian for the lattice GL model is given by,

\begin{eqnarray}
  H_{LGL} &=& a_xa_ya_z \sum_{\mathbf r} \biggl[ \alpha |\psi|^2 +
              \frac{g}{2} |\psi|^4 \nonumber \\ 
          &+& \sum_{\mu = x,y,z} \frac{\hbar^2}{2m_\mu a_\mu^2} \left 
              |\psi({\mathbf r}+\hat{\mu}) 
              e^{-i \frac{2\pi}{\Phi_0} a_\mu A_\mu({\mathbf r})} 
              -\psi({\mathbf r}) \right |^2 \nonumber \\ 
          &+& \sum_{\mu=x,y,z} \frac{1}{2\mu_0 a_\mu^2} 
              ({\mathbf \Delta} \times {\mathbf A})_\mu^2 \biggr]
\label{LGL:Hamiltonian}.
\end{eqnarray}

\hspace{-0.5cm} Here, $a_\mu$ and $\hat{\mu}$ is a lattice constant
and a unit vector along the $\mu$-axis, respectively.  Furthermore,
the lattice derivative is defined as
\begin{eqnarray*}
  \Delta_\mu \psi({\mathbf r}) = \psi({\mathbf r} + \hat{\mu}) -
  \psi({\mathbf r}).
\end{eqnarray*}
Taking the continuum limit $a_\mu \rightarrow 0$, the effective
Hamiltonian for the lattice GL model (Eq.  \ref{LGL:Hamiltonian})
reduces correctly to the GL effective Hamiltonian in the continuum
(Eq.  \ref{GL:Hamiltonian}). As defined in Eq.  \ref{LGL:Hamiltonian},
the lattice GL model does not contain vortices.  To reintroduce the
vortices in the model, we must compactify the gauge-theory by
requiring that the gauge invariant phase differences satisfy
\cite{Polyakov:PL75},
\begin{eqnarray}
  [\theta({\mathbf x}+\hat{\mu}) - \theta({\mathbf x})
  - \frac{2\pi}{\Phi_0} a_\mu A_\mu({\mathbf x})]
  \in [-\pi, \pi>.
\end{eqnarray}
Whenever this constraint is used to bring the gauge invariant phase
differences back to its primary interval, we automaticall introduce a
unit closed vortex loop, and the net vorticity of the system is
guaranteed to be conserved at every stage of the Monte-Carlo
simulation. From the renormalization group point of view the continuum
GL model and the lattice GL model belong to the same universality
class \cite{Kleinert:Bo89}. We therefore expect the lattice GL model
and the continuum GL model to give, qualitatively, the same results.

\subsection{Lattice Ginzburg-Landau model in a frozen gauge approximation}

In extreme type-II superconductors, the zero temperature mean-field
penetration depth is much greater than the zero temperature coherence
length, $\lambda_\mu(T \keq 0) \gg \xi_\mu(T \keq 0)$. Thus,
fluctuations of the gauge field represented by the last term in Eq.
\ref{GL:Hamiltonian}, around the extremal field configuration are
strongly suppressed and can therefore be neglected. The effective
Hamiltonian for the frozen gauge (FG) model can be written as

\begin{eqnarray}
  H_{FG} &=& \frac{|\alpha(0)|^2}{g} a_xa_ya_z \sum_{\mathbf r}
  \biggl[ \frac{\alpha(T)}{\alpha(0)} |\psi'|^2 + \frac{1}{2}
  |\psi'|^4 \nonumber \\ &+& \!\!\!\! \sum_{\mu = x,y,z} \!\!
  \frac{\xi_\mu^2}{a_\mu^2} |\psi'({\mathbf
    r}+\hat{\mu})||\psi'({\mathbf r})| \left [2 \!-\!2\cos(\Delta_\mu
    \theta - \! {\mathcal A}_\mu) \right ] \biggr].
\label{FG:Hamiltonian}
\end{eqnarray}

\hspace{-0.5cm} Here, we have defined a dimensionless order
field and vector potential
\begin{eqnarray*}
  \psi' = \frac{\psi}{\sqrt{\frac{|\alpha(0)|}{g}}} 
  ~~~ \rightarrow ~~~ |\psi'| \sim [0,1],
\end{eqnarray*}
\begin{eqnarray*}
  {\mathcal A}_\mu = \frac{2\pi}{\Phi_0} a_\mu A_\mu.
\end{eqnarray*}
The natural energy scale along the $\mu$-direction is,
\begin{eqnarray*}
  J_\mu = 2 \frac{|\alpha(0)|^2}{g} a_xa_ya_z
  \frac{\xi_\mu^2}{a_\mu^2}.
\end{eqnarray*}
Assuming a uniaxial anisotropy along the $z$-axis, the natural energy
scale for the FG model is
\begin{eqnarray}
  J_0 = J_x = \frac{2|\alpha(0)|^2}{g} \xi_{ab}^2 a_z
  = \frac{\Phi_0^2 d} {4\pi^2\mu_0 \lambda_{ab}^2}.
\label{Energy.Scale}
\end{eqnarray}
Here, we have put our coordinates ($x,y,z$)-axis parallel to the
crystals ($a,b,c$)-axis.  Furthermore, $\xi_x \keq \xi_y \keq
\xi_{ab}$ and $\xi_z \keq \xi_c$ is the coherence length in the
$CuO$-planes and along the crystal's $c$-axis, respectively.
Furthermore, $\lambda_x \keq \lambda_y \keq \lambda_{ab}$ and
$\lambda_z \keq \lambda_c$ is the penetration depth in the
$CuO$-planes and along the crystals $c$-axis, respectively. In
Eq.\ref{Energy.Scale}, $d$ is the distance between two $CuO$
superconducting planes in adjacent unit cells.  The energy scale $J_0$
is roughly the energy scale of exciting a unit vortex loop
\cite{Emery:N95,Li:L91,Nguyen:B98b}.

The ratio between the energy scales $J_x$ and $J_z$ serves as an
anisotropy parameter,
\begin{eqnarray}
  \Gamma = \sqrt{\frac{J_x}{J_z}} = \frac{\xi_{ab}a_z}{\xi_ca_x}
  = \frac{\lambda_c a_z}{\lambda_{ab}a_x}.
\label{Anisotropy:Parameter}
\end{eqnarray}

In this model, the lattice constant $a_\mu$ should be defined as
\begin{eqnarray*}
  a_\mu = \rm{max}(d_\mu,C_0\xi_\mu).
\end{eqnarray*}
Here, $d_\mu$ is an intrinsic length along the $\mu$-direction in the
underlying superconductor to be modeled.  Examples of such intrinsic
length are the distance between $CuO$-planes in adjacent unit cells,
the (a,b)-dimension of the unit cell. To be consistent, the constant
$C_0$ should be larger than $\sim 4$. This requirement $a_\mu/\xi_\mu
> 4$ ensures that the amplitude of the order field does not overlap
\cite{Tinkham:Bo96}. Such overlap will give rise to a domain wall term
($\nabla |\psi|$), which is absent in the lattice GL model.

Within the frozen gauge approximation, the gauge field serves only as
a constraint in fixing the value of the uniform induction. In terms of
magnetic induction this approximation is valid when $B \gg B_{c1}(T)$,
where the field distribution from individual flux lines overlap
strongly, giving uniform induction. Note that $B_{c1}(T)$ also
vanishes when the temperature approaches $T_c$. In zero field, this
approximation is valid for all temperatures except an inaccessible
temperature region around $T_c$ \cite{Friesen:C98}.

In our simulations on the FG model, we allow for both phase- and
amplitude fluctuations of the superconducting order parameter
$\psi({\bf r})=|\psi({\bf r})| \exp[i \theta({\bf r})]$. Details of
the Monte-Carlo procedure for this case will be given below.

\subsection{Uniformly frustrated 3D XY model}

The uniformly frustrated 3D XY model was first used as a
phenomenological model for extreme type-II superconductors by Li et
al. \cite {Li:L91} and Hetzel et al. \cite{Hetzel:L92}. To obtain the
uniformly frustrated 3D XY (3DXY) model from the FG model, we freeze
the amplitude of the complex order field in Eq.  \ref{FG:Hamiltonian},
$|\psi'| = 1$. This is the London approximation. The resulting
effective Hamiltonian for the 3DXY model is given by
\begin{eqnarray}
  H_{XY} = - \frac{2|\alpha(0)|^2}{g} a_xa_ya_z \sum_{{\mathbf
      r},\mu} \frac{\xi_\mu^2}{a_\mu^2} \cos \left ( \Delta_\mu \theta
    - {\mathcal A}_\mu \right ).
\label{XY:Hamiltonian}
\end{eqnarray}
The lattice constants in the 3DXY model should be defined as
\begin{eqnarray*}
  a_\mu = \rm{max}(d_\mu,\xi_\mu).
\end{eqnarray*}
Assuming uniaxial anisotropy, the energy scales and the anisotropy
parameter of the 3DXY model are the same as for the FG model, Eqs.
\ref{Energy.Scale}, \ref{Anisotropy:Parameter}.  Note that both the FG
model and the 3DXY model contain precisely the same topological
objects, i.e. vortex loops and vortex lines, as for the GL model. The
local gauge symmetry in the GL model is however reduced to a global
$U(1)$-symmetry in the FG model and the 3DXY model.

\subsection{Villain-approximation and vortex representation}
To further corroborate interpretations of the results from our
Monte-Carlo simulations using the uniformly frustrated $3DXY$-model,
to be detailed in the next section, it is useful to provide an
alternative, but entirely equivalent formulation of the GL-theory.
This formulation replaces a description in terms of the GL-function
$\psi$ by vortex-degrees of freedom, where the interaction between
vortex-segments is mediated by a gauge-field, which we denote by ${\bf
  h}$. This gauge-field is {\it not} the electromagnetic vector
potential ${\bf A}$, but will couple to it. The resulting structure of
the theory makes it possible, {\it in three dimensions, and three
  dimensions only}, to reformulate the vortex-content of the GL-theory
as a theory of a complex matter field $\phi$ coupled to the
gauge-fields ${\bf h}$ and ${\bf A}$. Although this may seem as an
unnecessary detour, the great advantage of this approach, is that
certain {\it vortex-correlators}, notably our quantity $O_L$ to be
defined below, can be directly related to a $U(1)$-symmetry of the
$\phi$-theory.

To proceed with this, we introduce the well-known
Villain-approximation to the 3DXY model.  The Villain approximation
consists of replacing the cosine potential in the uniformly frustrated
$3DXY$-model by a Gaussian $2 \pi$-periodic potential. In this way the
longitudinal spin-wave excitations of the $\theta$-field decouple from
transverse vortex-excitations of the theory. {\it This decoupling does
  not alter the critical behavior of the system}. The partition
function for this theory reads, after a rescaling of the vector
potential and charge
\begin{eqnarray*}
  Z_{V} &=& \prod_{{\bf r}} ~ \int ~D \theta({\bf r}) \int ~ D A({\bf
    r}) \sum_{{\bf n}({\bf r}) = - \infty}^{\infty} ~~ e^S \\
   S    &=& - \sum_{{\bf r}} [\frac{\beta}{2}(\nabla \theta -2
  e {\bf A} - 2 \pi {\bf n})^2
       + \frac{1}{2} (\nabla \times {\bf A})^2 ],
\end{eqnarray*}
where ${\bf n}({\bf r})$ is an integer-valued field. The kinetic term
is linearized by a Hubbard-Stratonovich decoupling, introducing an
auxiliary velocity-field ${\bf v}({\bf r})$ and using the identity
\begin{eqnarray*}
  \prod_{{\bf r}}e^{-\beta u^2({\bf r}) /2} \! \sim \! \prod_{{\bf r}}
  \int \! \! D {\bf v} \exp \bigl(- \! \sum_{{\bf r}} [ \frac{{\bf
      v}({\bf r})^2}{2 \beta} - i {\bf v}({\bf r}) \cdot {\bf u}({\bf
    r})] \bigr).
\end{eqnarray*}
This is now inserted back into the original partition function, using
${\bf u} = \nabla \theta - 2 e {\bf A} - 2 \pi {\bf n}({\bf r})$.  The
sum over the integers ${\bf n({\bf r})}$ may be carried out, yielding
the constraint that ${\bf v({\bf r})}$ is integer valued, say ${\bf
  v({\bf r})} = {\bf l({\bf r})}$. The next step is to integrate out
the $\theta({\bf r})$-variable, which yields the constraint $\nabla
\cdot {\bf v({\bf r})}=0$, which is solved by introducing an integer
valued field ${\bf h}({\bf r})$ such that ${\bf l}({\bf r}) = \nabla
\times {\bf h}({\bf r})$. In order to be able to treat ${\bf h}$ as a
continuous variable, we introduce a new integer-valued field ${\bf m}$
and apply the Poisson-summation formula
\begin{eqnarray*}
  \sum_{{\bf m}=-\infty}^{{\bf m}=\infty} e^{2 \pi i {\bf m} \cdot
    {\bf h} } = \sum_{{\bf l} = -\infty}^{\infty} \delta({\bf l} -
  {\bf h}),
\end{eqnarray*}
Note that this procedure does not involve any approximations.
Finally, we  write the partition function for the GL-theory in
phase-only and Villain-approximations as
\begin{eqnarray}
  Z &=& \prod_{{\bf r}} \int D {\bf h} ~ D {\bf A} \sum_{{\bf
      m}=-\infty}^{\infty} ~ e^{S_{\rm{eff}} [{\bf m},{\bf h},{\bf
      A}]},
\label{Shma} \\
  S_{\rm{eff}} &=& -\sum_{{\bf r}} \biggl[ 2 \pi i {\bf m} \cdot
  {\bf h} + \frac{1}{2 \beta} (\nabla \! \times \! {\bf h})^2 + 2 i e
  (\nabla \! \times \! {\bf h}) \cdot {\bf A} \nonumber \\
  &+& \frac{1}{2} (\nabla \! \times \! {\bf A})^2 \biggr], \nonumber
\end{eqnarray}
where the following constraints apply in the functional integral:
$\nabla \cdot {\bf A} = \nabla \cdot {\bf h} = \nabla \cdot {\bf m} =
0$.  The effective action, Eq. \ref{Shma}, is invariant under
\begin{eqnarray}
{\bf h}  & \to & {\bf h} + \nabla \omega_{\rm h} \nonumber \\
{\bf A}  & \to & {\bf A} + \nabla \omega_{\rm A}. \nonumber
\end{eqnarray}
The field ${\bf h}$ is readily
interpreted as a fictitious gauge-field that mediates an interaction
between vortex-segments ${\bf m}$. This is easily seen by integrating
out the ${\bf h}$-field.

\subsection{Dual (disorder field) representation}
Whenever a field theory sustains topological defects, it is often
useful to formulate a {\it new} field-theory of the topologial
excitations of the original theory {\it per se}, and this forms a dual
description of the original theory.  We will do this for the present
problem also, following Ref.  \onlinecite{Kleinert:Bo89}.  This means
that the vortex-content Eq. \ref{Shma} of the Ginzburg-Landau theory
Eq. \ref{LGL:Hamiltonian} in the phase-only Villain-approximation is
cast into the form of a local field theory involving a complex scalar
mass-field describing local vortex-fluctuations, coupled to a dual
gauge-field that mediates an interaction between the vortex-segments.
The resulting theory will exhibit explicitly a $U(1)$-symmetry, and as
always in such cases, the question to be asked is under what
circumstances, if any, the symmetry will be spontaneously broken
\cite{Footnote:U1symmetry}.

The purpose of this reformulation is to provide a point of contact
between on the one hand a quantity to be introduced in Section IIIA
and studied in Sections IVB and VB, probing vortex-tangle connectivity
and denoted $O_L$, and on the other hand thermodynamics.  The key
point is that in zero magnetic field, the two-point correlation
function of the complex scalar mass-field $\phi({\bf r})$ of the dual
theory, is precisely the probability of finding a {\it connected}
vortex path between the two points of the correlation function,
regardless of by which path the two points are connected
\cite{Kleinert:Bo89}.  Long-range order in $G({\bf x},{\bf y})$
implies a broken $U(1)$-symmetry. {\it Equivalently, long-range vortex
  connectivity in zero magnetic field implies a broken
  $U(1)$-symmetry}, which is ``hidden" at the level of Eq. \ref{Shma},
but is brought out when Eq. \ref{Shma} is reformulated to the dual
form, to be described below.

In three dimensions, {\it and three dimensions only}, a vortex-loop
system interacting with a long-range Biot-Savart interaction and
steric repulsion, may {\it in the continuum limit} be written as a
gauge theory of a local complex matter field $\phi$ coupled to ${\bf
  h}$ \cite{Kleinert:Bo89,Kiometzis:PoP95,Tesanovic:cm98}. We may
extend the results of this work including fluctuations in ${\bf A}$ in
a finite magnetic field, in which case the vortex-content of the
Villain-approximation to the GL-theory corresponds precisely to an
action of the following form
\begin{eqnarray}
  Z & = & \prod_{{\bf r}} \int D \phi({\bf r}) ~ D \phi^*({\bf r}) ~ D
  {\bf h}({\bf r}) D{\bf A}({\bf r}) e^{\tilde
    S_{\rm{eff}}[\phi,\phi^*,{\bf h},{\bf A}]} \nonumber \\ \tilde
  S_{\rm{eff}} & = & - \sum_{{\bf r}} \biggl[ \alpha' |\phi|^2 +
  \frac{g'}{2} |\phi|^4 + \frac{1}{2} |(\frac{\nabla}{i} - e' {\bf h})
  \phi|^2 \nonumber \\ & + & \frac{1}{2 \beta} (\nabla \times {\bf
    h})^2 + 2 i ~ e ~ (\nabla \times {\bf h}) \cdot {\bf A} + \frac{1}{2}
  (\nabla \times {\bf A})^2 \biggr],
\label{Sphiha}
\end{eqnarray}
where the coefficients $(\alpha',e',g')$ appearing in the theory are
given in terms of the parameters entering Eq. \ref{Shma}
\cite{Kiometzis:PoP95}. For our discussion, their precise values are
of no importance. The interpretation of the $\phi$-field is that it is
a local field describing local fluctuations in the topological
excitations of the GL-theory, namely line-defects in the form of
vortex-lines. An analogue of this dual description of the present
$U(1)$-symmetric theory is the dual description of the Ising-model,
where a local field is introduced to describe the local fluctuations
in the topological defects of that model, which are domain walls
separating different spin-ordered regions of the magnet. The effective
action, Eq. \ref{Sphiha}, is invariant under the set of
transformations
\begin{eqnarray}
  \phi & \to & \phi \exp(i \omega_{\rm h}) \nonumber \\ {\bf h} & \to
  & {\bf h} + \frac{1}{e'} ~ \nabla \omega_{\rm h} \nonumber \\ {\bf
    A} & \to & {\bf A} + \nabla \omega_{\rm A}.
\label{symmetry}
\end{eqnarray}
By rewriting the theory in Eq. \ref{Shma} in this way one observes
that it explicitly exhibits a $U(1)$-symmetry. Note that this relies
entirely on the possibility of reformulating the interacting loop-gas,
including Biot-Savart interactions, in terms of a complex matter field
$\phi$ coupled to a gauge-field ${\bf h}$.

Consider now for the moment the case of zero magnetic field.
The main advantage of the above formulation is that the probability of
finding a connected path of vortex segments, starting at ${\bf x}$ and
ending at ${\bf y}$, $G({\bf x},{\bf y})$, is given by the two-point
correlation
function of the $\phi$-field \cite{Kleinert:Bo89}
\begin{eqnarray}
G({\bf x},{\bf y}) = < \phi^*({\bf x}) ~ \phi({\bf y})>.
\end{eqnarray}
A vortex-loop unbinding will lead to a finite $G({\bf x},{\bf y})$
when $|{\bf x}-{\bf y}| \to \infty$, because infinitely large loops
will connect opposite sides of the vortex system. On the other hand,
if $\lim_{|{\bf x} - {\bf y}| \to \infty} G({\bf x},{\bf y}) \neq 0$,
this implies that $<\phi^*(x)> \neq 0$, corresponding to a broken
$U(1)$-symmetry.  {\it Therefore, the dual field $\phi({\bf r})$ is an
  order parameter of a vortex-loop unbinding transition}. The broken
$U(1)$-symmetry is associated with the loss of number conservation of
connected vortex-paths threading the system in any direction
(including direction perpendicular to an applied magnetic field, if
that is present). This limit of the two-point correlation function is
closely related to the quantity $O_L$ we introduce in Section IIIA,
{\it which probes the connectivity of the vortex tangle in an extreme
  type-II superconductor}. The above connection makes it at the very
least plausible that an abrupt change in this connectivity, as probed
by the change in $O_L$, is associated with breaking a $U(1)$-symmetry
of the {\it vortex}-sector of the GL-theory, equivalently an onset
$<\phi^*>$ or $<\phi>$.  Since this only happens above a critical
temperature, we may view $\phi$ as a {\it disorder}-field, in contrast
to the order-parameter field $\psi$ of the original GL-theory. We will
make explicit use of this connection in Section IV.

In zero magnetic field, we will show that the loss of superfluid
density, specific heat anomaly, change in vortex-loop distribution,
loss of {\it long-wavelength} vortex-line tension, and abrupt change
in vortex tangle connectivity abruptly occurs  precisely at the same
temperature both for the $3DXY$-model, also when amplitude
fluctuations are included.

At finite magnetic fields, the situation is complicated by the fact
that the vortex system is always connected across the system in at
least one direction, namely the field direction, at all temperatures.
One may however still extract information of the type encoded in
$<\phi^*({\bf r}) \phi({\bf r})>$ at zero field by performing a
singular gauge-transformation of the type used in Ref.
\onlinecite{Tesanovic:B95,Tesanovic:cm98}, which roughly speaking
amounts to subtracting out the field-induced vortices and studying the
remaining loop-gas, which has a field-theory description very much
like the zero-field version of Eq.  \ref{Sphiha}. The obvious
advantage of this is that one removes the asymmetry of the system
imposed by the magnetic field.  A twopoint correlator of this theory
then probes the connectivity of {\it non-field induced vortex paths
  across the system}, which in turns probes the possibility of having
a broken $U(1)$-symmetry and hence an onset of the order parameter
$<\phi({\bf r})> \neq 0$.

We will perform a numerical analogue of this in our simulations,
namely we will probe the connectivity of the vortex tangle of the
superconductor in directions perpendicular to the magnetic field.
Ideally, what one should do is to generate phase-configurations (and
vortex-configurations) of the extreme type-II superconductor, subtract
out from each configuration a number of vortex paths that connects the
system along the field direction precisely corresponding to the number
of field induced vortices in the system, which is a fixed number in a
canonical ensemble usually studied for this problem. Out of the
remaining vortex tangle one may then try to find vortex paths
connecting opposite sides of the system. Numerically this procedure is
entirely prohibitive and we therefore opt for the algorithm of
calculating $O_L$, to be described in detail in Section IIIA.

We stress that the procedure of computing $O_L$ described in Section
IIIA unquestionably probes the connectivity of a vortex-tangle across
the system, {\it not} associated with magnetic field, precisely as in
the zero-field case. The objective is to probe the breaking of a
$U(1)$-symmetry associated with the proliferation of unbound
vortex-loops in the system, as pointed out in Ref.
\onlinecite{Nguyen:cm98}. This will be shown to be precisely borne out
in zero magnetic field. In finite magnetic field we also obtain a weak
specific heat anomaly at the temperature where $O_L$ changes abruptly,
as the system size is increased.

\section{Definitions, simulation procedure, and model parameters}

In this section, we define the physical quantities considered,
describe our Monte Carlo procedure, and present the values of the
model parameters used in the simulations.

\subsection{Definitions}

\subsubsection{Specific heat $C$}

To calculate the  specific heat per site $C$, we use the
fluctuation formula,
\begin{eqnarray}
  \frac{C}{k_B} = \frac{1}{\mathcal V} \frac{<H^2> - <H>^2}{(k_BT)^2}.
\label{Spes.Heat.Fluc}
\end  {eqnarray}
Here, the dimensionless volume ${\mathcal V} = L_xL_yL_z$, and $L_\mu$
is the dimensionless linear dimension of the system along the
$\mu$-direction.  $L_\mu$ is measured in units of the lattice constant
$a_\mu$.  As a check of consistency, we also calculate the specific
heat per site using the numerical derivation of the internal energy
$U$ \cite{Dodgson:L98,Nguyen:B98b,Nordborg:B98},
\begin{eqnarray}
  C_U = \frac{1}{\mathcal V} \frac{\partial U}
  {\partial T}.
\label{Spes.Heat.Diff}
\end  {eqnarray}
Note that for the FG model, where the effective Hamiltonian depends
explicitly on the temperature, there strictly speaking is an
additional term in the expression for the internal energy
\cite{Nguyen:B98b},
\begin{eqnarray*}
  U = - \frac{\partial \ln Z}{\partial \beta} = <H> - T \left \langle
    \frac{\partial H}{\partial T} \right \rangle.
\end  {eqnarray*}
For the 3DXY model, if the simulations are properly done, $C \cong
C_U$. For the FG model, where the effective Hamiltonian explicitly
depends on the temperature, $C \not = C_U$. Note however that
$<\partial H/\partial T>$ arises out of introducing a temperature
dependence of the coefficients of the GL-theory
\cite{Hu:B97,Dodgson:L98,Nguyen:B98b}. The temperature dependence of
these coefficients always has a temperature dependence set by the {\it
  mean-field} critical temperature.  Thus, close to the true $T_c$
these corrections, arising from integrating out the fermions of the
underlying microscopic description of the superconductor
\cite{Hu:B97}, are always negligible compared to the contribution
coming from the true critical fluctuations of the order parameter,
i.e. the transverse phase-fluctuations.  In terms of Eq.
\ref{Energy.Scale} the term $-T <\partial H/\partial T>$ orginates
from the temperature dependence of the {\it amplitude} of the
order-parameter. Were this to actually vanish at $T = T_c$,
substantial corrections to the specific heat and entropy would result.
As we will see later, at $T=T_c$, the ensemble average of the {\it
  amplitude} of the order parameter is far from renormalized to zero
by vortex-loop fluctuations. Hence, at the critical point the
correction term $-T ~ <\partial H/\partial T>$ in the internal energy,
with its resulting corrections to entropy \cite{Dodgson:L98}, is
negligible \cite{Nguyen:B98b}. There is now ample experimental
evidence that critical fluctuations are indeed important over a
sizeable temperature window in the high-$T_c$ cuprates of order
several Kelvin below $T_c$
\cite{Salamon:B93,Junod:C97,Roulin:JLTP96,Roulin:Sc96,Roulin:C97} a
window which is consistent with a coherence length of order $10$ \AA, 
about two orders of magnitude shorter than in conventional 
superconductors.

\subsubsection{Local Cooper-pair density $<|\psi'|^2 \! >$}

As a probe for the local Cooper-pair density, we calculate
\begin{eqnarray}
  <|\psi'|^2 \! > \equiv \frac{1}{\mathcal V}
  \sum_{\mathbf r}  < |\psi'({\mathbf r})|^2 \! >.
\label{CoPaDens}
\end{eqnarray}
We see in Eq. \ref{CoPaDens} that $<|\psi'|^2 \! >$ involves both
thermal and space average.  Recall that $\psi' \equiv \psi /
\sqrt{|\alpha(0)| / g}$.  At the mean field level, we expect
$<|\psi'|^2 \! >$ to develop an expectation value below the mean field
critical temperature $T_{MF}(B)$.

\subsubsection{Superfluid condensate  density $|< \! \psi' \! >|^2$}

As a probe for the local condensate density (density of Cooper pairs
{\em participating in the superconducting condensate}), we calculate
\begin{eqnarray}
  |< \! \psi' \! >|^2 \equiv \frac{1}{\mathcal V}
  \sum_{\mathbf r} |< \! \psi'({\mathbf r}) \! >|^2.
\label{SuFluidDens}
\end{eqnarray}
Note the difference between $<|\psi'|^2>$ and $|< \! \psi' \! >|^2$.
The former describes local Cooper-pair density, while the latter
describes what is commonly known as the condensate density in
$^4He$-physics.  In zero field, we expect $|< \! \psi' \! >|^2$ to
develop an expectation value below the critical temperature $T_c$.

\subsubsection{Distribution of the order field phase angle}

To probe the distribution of the phase angle in $\psi'({\mathbf r}) =
|\psi'({\mathbf r})| e^{i\theta({\mathbf r})}$, we define the
distribution function
\begin{eqnarray}
  D_\theta(\theta') = \frac{1}{\mathcal V}
                     < \sum_{\mathbf r}
                     \delta_{\theta({\bf r}), \theta'} >.
\end{eqnarray}
Here, $\delta_{i,j}$ is the Kronecker delta function. In the
simulations, we have chosen to work with a discrete set of phase
angles, $\theta',\theta({\mathbf r}) = 2\pi n/N_\theta$.  Here, $n \in
[0,N_\theta]$ is an integer, and $N_\theta$ is the number of allowed
discrete phase angles. In our experience, the simulation results do
not depend on $N_\theta$, provided $N_\theta \stackrel{>}{\sim} 16$.
In zero field, when the phase is disordered, we expect
$D_\theta(\theta)$ to be uniformly distributed, $D(\theta) =
1/N_\theta$.  In the ordered phase, we expect $D_\theta(\theta)$ to
show a peak around a preferred phase angle.

\subsubsection{Helicity modulus $\Upsilon_\mu$}

To probe the global superconducting phase coherence across the system,
we consider the helicity modulus $\Upsilon_\mu$, defined as the second
derivate of the free energy with respect to an infinitesimal phase
twist in the $\mu$-direction \cite{Fisher:A73,Li:B93,Nguyen:B98a}. 
Finite $\Upsilon_\mu$ means that the system can carry
a supercurrent along the $\mu$-direction. Within the 3DXY-model, the
helicity modulus along the $\mu$-direction becomes,
\begin{eqnarray*}
  \frac{\Upsilon_\mu}{J_\mu} 
  &=& \frac{1}{\mathcal V} \left \langle \sum_{\mathbf r} 
      \cos[\Delta_\mu \theta - {\mathcal A}_\mu] \right \rangle \\ 
  &-& \frac{J_\mu}{k_B T {\mathcal V}} \left \langle \left [\sum_{\mathbf r} 
      \sin[\Delta_\mu \theta - {\mathcal A}_\mu] \right ]^2 \right \rangle.
\end{eqnarray*}
For the FG case,
\begin{eqnarray*}
  \frac{\Upsilon_\mu}{J_\mu} &=& \frac{1}{\mathcal V} \left \langle
    \sum_{\mathbf r} |\psi'({\mathbf r})||\psi'({\mathbf
      r}+\hat{\mu})| \cos[\Delta_\mu \theta - {\mathcal
      A}_\mu] \right \rangle \\ 
      &-& \frac{J_\mu}{k_B T {\mathcal V}} \times \\
      & & \left \langle \left [ \sum_{\mathbf r}
      |\psi'({\mathbf r})||\psi'({\mathbf r}+\hat{\mu})|
      \sin[\Delta_\mu \theta - {\mathcal A}_\mu]
    \right ]^2 \right \rangle.
\end{eqnarray*}
Note the difference between $|<\!\psi'\!>|^2$ and $\Upsilon_\mu$, they
are not {\it identical}. The former quantity probes the  superfluid 
condensate density, which is a locally defined quantity, while the latter 
quantity probes a global phase coherence along a given direction $\mu$. 
Since $<\psi'>$ is the order parameter of the Ginzburg-Landau 
theory, close to the critical point we have
\begin{eqnarray}
|<\psi'>|^2 \sim |\tau|^{2 \beta}, 
\end{eqnarray}
where $\tau=(T-T_c)/T_c$. On the other hand, $\Upsilon_{\mu} \propto
\rho_{s \mu}$, where $\rho_{s \mu}$ is the superfluid density in the
$\mu$-direction. Using the Josephson scaling relation $\rho_{s \mu}
\sim \xi^{2-d} \sim |\tau|^{\nu(d-2)}$ \cite{Josephson:PPS67} along
with the scaling laws $\gamma=\nu(2-\eta)$ \cite{Fisher:PR69} and $2
\beta=2-\alpha-\gamma$ \cite{Rushbrooke:JCP63}, we find
\begin{eqnarray} 
\Upsilon_{\mu} \sim |\tau|^{2 \beta - \eta \nu}.
\end{eqnarray}
Here, $d$ is the dimensionality of the system, $\nu$ is the
correlation length exponent of the system, $\beta$ is the order
parameter exponent, $\gamma$ is the order parameter susceptibility
exponent, and $\eta$ is the anomalous dimension of the order parameter
two-point correlation function at the critical point.  Therefore,
although $|<\psi'>|^2$ and $\Upsilon_{\mu}$ are in principle
different, they may {\it appear} to be very close if the anomalous
dimension $\eta$ of the $\psi$-field is small, as indeed is the case
for the Ginzburg-Landau theory, where $\eta \approx 0.04$
\cite{Hasenbuch:cm99}. Note that for $\eta > 0$, the curve for
$\Upsilon_{\mu}$ should bend more sharply towards zero at the critical
point than $|<\psi'>|^2$. We will explicitly show by direct
calculations within the Ginzburg-Landau theory that $|<\!\psi'\!>|^2$
is very close to $\Upsilon_\mu$ both in zero field and finite magnetic
field. In zero magnetic field this is precisely what one would expect
based on the above, when $\eta << 1$ \cite{Weichman:pc}. For the
special case of $d=3$, we have $2 \beta - \eta \nu = \nu < 2 \beta$.
To high precision, we have for the $3DXY$-model, that $\nu = 0.673$
and $\eta = 0.038$ \cite{Hasenbuch:cm99}

\subsubsection{Vortex loop distribution $D(p)$}
To probe the typical perimeter $L_0 (T)$ and the effective
long-wavelength {\it vortex-line tension} $\varepsilon(T)$ (not to be
confused with the {\it flux-line tension}, which is always zero when
gauge-fluctuations are completely suppressed due to the absence of
tubes of confined magnetic flux), we define a vortex-loop distribution
function $D(p)$, which measures the ensemble-averaged number of vortex
loop in the system having a perimeter $p$
\cite{Li:B94,Nguyen:B98a,Chin:cm98,Williams:L98}. In order to compare
results from different system sizes, we normalize $D(p)$ with respect
to the system size.

We search for a vortex loop using the following procedure. Given a
vortex configuration, we start with a randomly chosen unit cell with
vortex segments penetrating its plaquettes. We follow the directed
vortex path and record the trace. When the directed vortex path
encounters a unit cell containing more than one outgoing direction, we
choose the outgoing direction {\it randomly}. When the vortex path
encounters a previously visited unit cell, i.e. when it crosses its
own trace, we have a closed vortex loop, its perimeter being $p$.  We
now delete the vortex loop from the vortex configuration, to prevent
double counting, and continue the search. The search is continued
until all vortex segments are deleted from the system.

Using a 3D non-interacting boson analogy to the vortex system, it can
be shown \cite{Hoye:JSP94} that the distribution-function $D(p)$ can be
fitted to the form \cite{Footnote:dp}
\begin{eqnarray}
  D(p) = A ~ p^{-\alpha} ~ \exp [-\frac{\varepsilon(T) p}{k_BT} ].
\label{Loop.Dstrb}
\end{eqnarray}
Here, A is a temperature independent constant, and the exponent
$\alpha \approx 5/2$ to a first approximation
\cite{Footnote:exponent}. When $\varepsilon(T)$ is finite, there
exists a typical length scale $L_0=k_BT/\varepsilon$ for the thermally
excited vortex loops. The probability of finding vortex loops with
much larger perimeter than $L_0$ is exponentially suppressed,
according to Eq.  \ref{Loop.Dstrb}. When $\varepsilon=0$, $D(p)$
decays algebraically, and the length scale of the problem\, $L_0 = k_B
T/\varepsilon(T)$, has diverged. As a consequence, configurational
entropy associated with topological phase-fluctuations is gained
without penalty in free energy. In zero field, there is only one
critical point, and in this case $L_0$ must be some power of the
superconducting coherence length $\xi(T)$.

\subsubsection{Probe of vortex-connectivity, $O_L$}
For probing the connectivity of a vortex tangle in a type-II
superconductor, in zero as well as finite magnetic field, we introduce
a quantity $O_L$, defined in zero magnetic field as the probability of
finding a vortex configuration that {\it can} have at least one
connected vortex path threading the entire system in any direction. In
the presence of a finite magnetic field, $O_L$ is defined as the
probability of finding a similarly connected vortex path in a
direction transverse to the field direction, {\em without using the
  PBC along the field direction}. In zero field, we use the same
procedure as in finite-field, namely searching for connected vortex
paths perpendicular to the $z$-direction, although in this case we
could just as well have used any direction. {\it Note that $O_L$ is
  very different from the winding number $W$ in the 2D boson analogy }
\cite{Nelson:B89,Nordborg:Thesis}. There, $W$ is proportional to the
number of vortex paths percolating the system transverse to the field
direction. However, in the calculation of $W$, the PBC along the field
direction is used many times.

In an attempt to probe``vortex-percolation", a slightly different
quantity than $O_L$ has been considered in the context of
high-temperature superconductors by others
\cite{Jagla:B96,Footnote:Jagla}. A crucial difference between our work
and that of Ref. \onlinecite{Jagla:B96}, is that Ref.
\onlinecite{Jagla:B96} allows periodic boundary conditions along the
field direction to be used several times before the vortex path winds
once around the $x$- or $y$-axis, as is easily seen from Fig. 2b of
Ref. \onlinecite{Jagla:B96}. This ultimately is the same as computing
the winding number of the $2D$ non-relativistic boson-analogy of the
vortex-system \cite{Nelson:L88}, as recently done in careful
Monte-Carlo simulations in Ref. \onlinecite{Nordborg:B98}.  It also
explains why the authors of Ref. \onlinecite{Jagla:B96} get
longitudinal dissipation at the onset of what they denote
``vortex-percolation", which is nothing but the temperature at which
the winding number becomes finite.

This is entirely consistent with a number of other Monte-Carlo
simulation results on the $3DXY$-model
\cite{Hu:L97,Nguyen:B98a,Nguyen:B98b,Nguyen:cm98,Chin:cm98} which all
show the loss of longitudinal phase-coherence and onset of
longitudinal dissipation precisely at the vortex lattice melting
transition. This is measured simply by the helicity modulus
$\Upsilon_z$, which is quite different from $O_L$.  To the contrary,
in our calculation of $O_L$, we do not allow for the use of periodic
boundary conditions in the $z$-direction to measure vortex-tangle
connectivity in the $x$- or $y$-directions, in other words the
``percolating" configurations of Fig. 2b of Ref.
\onlinecite{Jagla:B96} are not counted when computing $O_L$.

We have
\begin{eqnarray}
  O_{L} = \frac{N_{c}}{N_{total}}.
\label{Perc.Prob}
\end{eqnarray}
$N_{total}$ is the total number of independent vortex configurations
provided by the Monte Carlo simulation. Furthermore, $N_{c}$ is the
number of vortex configurations containing {\em at least one} directed
vortex path that traverses the entire system perpendicular to the
direction, without using the PBC along the field direction. For
convenience, we treat the zero field case as the limit $\lim_{B \to
  0}$ keeping the ``field direction'' intact.

We search for the {\it possibility} of finding a vortex path such as
described above by using the following procedure. Assume that the
magnetic induction points along the $z$-axis. We follow all paths of
directed vortex segments starting from all four boundary surfaces with
surface normal perpendicular to $\hat{z}$, and check whether at least
one of these vortex paths percolates the system and reaches the
opposite surface, {\em without applying the PBC in the z-direction}.
Note that when crossing vortex segments are encountered, the procedure
is to attempt to continue in a direction that will bring the path
closer to the opposite side of the system, rather than randomly
resolving the intersection. $O_L$ is therefore a {\it necessary}, but
not {\it sufficient} condition for finding an actual vortex-path
crossing the system. However, in zero field this procedure does not
make a difference to that of resolving the intersections randomly.
This is demonstrated by the correlations of the change in $O_L$ and
$D(p)$, to be detailed in the next section.

If a vortex path is actually found crossing the system in any
direction in zero field, or without using PBC in the field-direction
when a field is present, one may safely conclude that the vortex-line
tension has vanished. If it were finite, it would not be possible to
find such a path at all, either because all vortex-lines form closed
confined loops in zero field, or because the vortex-line fluctuations
along the field direction would be {\it diffusive} in finite field. In
zero-field, this is clear by the above mentioned correlation between
the change in $O_L$ and $D(p)$, cf. the results of the next section.
In this paper, we also investigate this in detail for the finite-field
case, by considering the position of the lowest temperature $T_L$
where we have $O_L =1 $ both as a function of system size and aspect
ratio $L_x/L_z = L_y/L_z$. {\it If vortex-line physics remains intact,
  $T_L$ should move monotonically up with system size, and should
  scale with $L_x/L_z$.} Instead, we will find that $T_L$ moves {\it
  down} slightly, and saturates with increasing system size at fixed
aspect ratio. In addition, we find that $T_L$ is virtually independent
of aspect ratio for large enough systems.

This contradicts expectations based on a lines-only
approximation to the vortex-liquid. It demonstrates that the {\it
connectivity of the vortex-tangle} undergoes a fundamental change
inside the vortex-liquid. {\it The above mentioned finite size scaling
analysis, suggests to us that this geometric transition is a
property that survives in the thermodynamic limit.} The issue is
whether the change in connectivity has anything to do with a
thermodynamic phase-transition. This will be investigated in detail
for zero magnetic field in Section IIIA, and for finite magnetic field
in Section IIIB. In particular, we look for a specific heat anomaly
scaling up with system size, at the putative transition point $T_L$.
This will reveal if the change in the geometric properties of the
vortex liquid is indeed associated with singular thermodynamics. In
any case, once the {\it geometric} transition has taken place, it is
no longer possible to model the vortex-liquid regime in terms of
field-induced flux lines only, with merely renormalized interactions
between them.

In the VLL phase $O_L = 0$, since the field induced flux lines are
well defined and do not ``touch'' each others, and the thermally
excited vortex loops are confined to sizes smaller than the magnetic
length \cite{Nguyen:B98a}. $O_L = 1$ in the normal phase above the
crossover region where the remnant of the zero field vortex loop
blowout takes place. Needless to say, it is a matter of interest to
investigate precisely where $O_L$ changes value  from zero to one.

{\it Note that $O_L$ itself is not a genuine thermodynamic order
  parameter}, although it may be said to {\it probe} an order-disorder
transition \cite{Nguyen:cm98}. However, by the transcription of the
vortex-content of the Ginzburg-Landau theory to the form Eq.
\ref{Sphiha} in Section IIF, it is brought out that probing the
vortex-tangle connectivity by considering $O_L$ is closely connected
to probing the two-point correlator of a local complex field
$\phi({\bf r})$, the dual field of the local vorticity-field
$m_{\mu}({\bf r})$ of the Villain-approximation and
London-approximation to the Ginzburg-Landau theory, Eqs. \ref{Shma}.
The two-point correlator
$<\phi^*({\bf r}) \phi({\bf r}')>$ is ultimately the probe of whether
or not the $\phi$-theory Eq. \ref{Sphiha} exhibits off-diagonal
long-range order and a broken $U(1)$-symmetry.  An entirely equivalent
interpretation of the change in $O_L$ was given in Ref.
\onlinecite{Nguyen:cm98} which did not involve a local field
$\phi({\bf r})$, but number conservation of vortex-lines threading the
entire superconductor. This number is conjugate to the phase-field of
the local complex field $\phi({\bf r})$. An advantage of the present
formulation involving Eq.  \ref{Sphiha} is that it directly relates he
change in $O_L$ to the long-distance part of a correlation for a local
field, and hence to a local order parameter $<\phi({\bf r})>$. This
connection makes it at least {\it plausible} that the change in
vortex-tangle connectivity, i.e.  a change in the geometry of the
vortex-tangle, may be related to a thermodynamic phase-transition.
{\it We emphasize that the present problem is very different from the
  percolation transition known to occur in the $3D$ Ising-model, and
  which has nothing to do with the thermodynamic phase-transition in
  that model} \cite{Footnote:Ising}.

\subsubsection{Structure function $S({\mathbf k})$}

To probe the structure of the VLL, we consider the structure function
for vortex segments directed along the field direction
\cite{Cavalcanti:E92,Nguyen:L96}. Given an applied field along
$z$-axis, the structure function $S({\mathbf k})$ is defined by
\begin{eqnarray}
  S({\mathbf k}) = \frac{<\mid \sum_{\mathbf r} n_z({\mathbf r}) \exp
    ~ [i {\mathbf k} \cdot {\mathbf r} ]\mid^2>} {(fV)^2}.
\label{Structure.Function}
\end{eqnarray}
Here, ${\mathbf k}$ is a reciprocal lattice vector, and the filling
fraction $f$ is defined in Eq. \ref{Filling.Fraction}. In the VLL
phase, $S({\mathbf k}_\perp,k_z \keq 0)$ has $\delta$-function Bragg
peaks at ${\mathbf k}_\perp \keq {\mathbf K}$, where ${\mathbf K}$ is
reciprocal lattice vector for the ordered VLL.  When the VLL melts,
the Bragg peaks are smeared out $S({\mathbf K},k_z \keq 0)$ drops to a
very small value. Thus, the VLL melting temperature can be defined as
the temperature where $S({\mathbf K},k_z=0)$ shows a sharp drop to a
value very close to zero. In a liquid where we have full rotational
invariance, we expect $S({\mathbf k}_\perp, k_z \keq 0)$ to exhibit a
ring pattern, with a very small amplitude.

\subsubsection{Extended Landau gauge:}
Periodic boundary conditions together with Landau gauge
\begin{eqnarray*}
  {\mathcal A}_y = 2\pi f x,
\end{eqnarray*}
give rise to a constraint, $L_x f = 1,2,3..$. Thus, for
given $L_x$, the smallest $f$ allowed is $f = 1/L_x$. To perform
simulations and finite size scaling of systems with very low filling
fractions, we define an ``extended'' Landau gauge,
\begin{eqnarray}
  {\mathcal A}_x=\frac{2 ~\pi ~y ~ m_y ~m ~ n}{L_x L_y};~~ {\mathcal
    A}_y=\frac{2~ \pi ~x ~ n_x ~m ~ n}{L_x L_y},
\label{E:Landau:Gauge}
\end{eqnarray}
where $n_x,n,m_y,m$ are positive integers satisfying $n_x ~ n = L_y$,
and $m_y ~ m = L_x$. The filling fraction $f$ is now given by
\begin{eqnarray*}
  f = \frac{n ~ m ~ [n_x-m_y]}{L_x L_y}.
\end{eqnarray*}
Hence, it is possible to choose systems with a filling fraction as low
as $f=1/L_xL_y$ \cite{Nguyen:cm98}.

\subsection{Details of the Monte-Carlo simulations}

The statistical mechanics of the 3DXY model and the FG model is
investigated by Monte-Carlo simulations on the effective Hamiltonians
Eq.  \ref{XY:Hamiltonian} and Eq.  \ref{FG:Hamiltonian}.

For the 3DXY model, a Monte-Carlo move is an attempt to replace a
phase angle at a given site $\theta({\mathbf r})$ with a new randomly
chosen phase angle $\theta' \in [0, 2\pi>$. For the FG model, a
Monte-Carlo move is an attempt to replace a complex number at a given
site $\psi({\mathbf r})$ with a new randomly chosen complex number
$\psi'$.  Here, $|\psi'| \in [0:1+\epsilon]$ and $\theta' \in [0,
2\pi>$.  We have introduced a small positive parameter $\epsilon$ to
allow the system to perform Gaussian fluctuations, around the extremal
field configuration $|\psi'|^2 = 1$, at very low temperature. Note
that we are letting the amplitude fluctuate around its mean value at
every temperature. The Monte-Carlo move is accepted or rejected
according to the standard Metropolis algorithm \cite{Plischke:Bo94}.
If the new phase angle causes a gauge invariant phase differences
$j_\mu = \Delta_\mu \theta - {\mathcal A}_\mu$ to fall outside the
primary interval $[-\pi,\pi>$, we take it back into the primary
interval. This compactization procedure creates a closed unit vortex
loop around the link where $j_\mu$ is changed. In this way, all the
vortex loops introduced into the system are closed, and the net
induction is always conserved.

A Monte-Carlo sweep consists of $L_x \times L_y \times L_z$
Monte-Carlo moves. Typical runs consist of $1.2 \times 10^5$ sweeps
per temperature, where the first $2 \times 10^4$ sweeps are discarded
for equilibration.  Near the phase transitions up to $2 \times 10^6$
sweeps per temperature is necessary to capture the correct physics.
For a given system, we always start the simulation by a cooling
sequence, where the starting temperature is significantly higher than
all temperatures associated with phase-transitions or crossovers the
model might exhibit. The results shown in this paper originate both
from cooling and heating sequences. Since these two methods give
essential identical results, we do not differentiate between them.

In order to resolve anomalies in the specific heat, we must in some
cases perform simulations on systems as large as $360^3$. To be able
to carry out simulation on such large systems, we must 1) write part
of the code in assembly and 2) carry out the simulations in a parallel
manner. Our systems are divided into ``black and white'' subsystems,
arranged in a 3D checkerboard pattern.  Each black subsystem has only
six white subsystems as its nearest neighbors, and visa versa. Since
the 3DXY and the FG model only have nearest neighbor interactions, all
subsystems with the same color can be updated simultaneously.

To be able to calculate a nonparallel-able routine as $O_L$ in an
effective manner, we divide the computer nodes in 2 groups, the large
main group takes care of the Monte Carlo simulation, and a small
subgroup carries out, simultaneously, the calculation of $O_L$.

\subsection{Model parameters}

\paragraph{System sizes:}
We put our coordinate (x,y,z)-axes along the crystal (a,b,c)-axes.
For the anisotropic cases, we assume uniaxial anisotropy, and use the
crystal $c$-axis as the anisotropy axis. We perform simulations on
tetragonal systems with dimensions $L_x, L_y, L_z$. The main part of
the simulations is done on cubic or nearly cubic systems. Nearly cubic
systems $L_x \sim L_y = L_z$ is some times necessary in order to
satisfy the boundary conditions enforce by the extended Landau gauge,
Eq.  \ref{E:Landau:Gauge}. To check for the finite size effect of
$O_L$, we carry out simulations on slab systems with the aspect ratios
$L_y/L_z = 1.00, 1.25, 1.5, 1.75, 2.00$. System sizes up to $360^3$
were used.

\paragraph{Cooper-pair chemical potential $\alpha(T)$:}.
We let the Cooper-pair chemical potential have the simple linear form
\begin{eqnarray*}
  \frac{\alpha(T)}{\alpha(0)} = \frac{T - T_{MF}}{T_{MF}}.
\end{eqnarray*}
We have also carried out simulations with other forms for
$\alpha(T)/\alpha(0)$, such as $\tanh([T-T_{MF}/T_{MF}]T_0)$. Here,
$T_0$ is a constant regulating the size of the region where
$\alpha(T)/\alpha(0)$ grows from -1 to 1. The results are, however,
qualitatively the same as for the linear case. The parameter $T_{MF}$
is the parameter effectively controlling the width of the critical
region in these calculations. In units of $J_0$, Eq.
\ref{Energy.Scale}, we write $T'_{MF}=k_BT_{MF}/J_0$. The values we
will use are $T'_{MF}=0.3,1.0$. An estimate for what temperatures
these values correpond to may readily be obtained by using $a_z=11$
{\AA}, $\lambda_{ab} = 1500$ {\AA}, implying that $T'_{MF}=1$
corresponds to $300K$, while using $\lambda_{ab}=2000$ {\AA} implies
that $T'_{MF}=1$ corresponds to $180K$. These are very reasonable
numbers. {\it With these parameters, it will be shown that amplitude
fluctuations of the local Ginzburg-Landau order parameter, when
included on an equal footing with the phase-fluctuations, are far
from being critical}.

\paragraph{Anisotropy parameter $\Gamma$:}
The anisotropy parameter $\Gamma$ is defined as
\begin{eqnarray*}
  \Gamma = \frac{a_z \xi_x}{a_x \xi_z} = \frac{\lambda_c
    a_z}{\lambda_{ab}a_x}.
\end{eqnarray*}
Note that $\Gamma > 1$ only when the layering of the superconductor to
be simulated is pronounced, i.e.  $d_\mu > \xi_\mu$ for at least one
direction $\mu$. In this article, we consider systems with the
anisotropy parameter $\Gamma = 1,3,7$.

\paragraph{Filling fraction $f$:}
The filling fraction along the $\mu$-direction, $f_\mu$, is defined as
\begin{eqnarray*}
  2\pi f_\mu = ({\mathbf \Delta} \times {\mathbf {\mathcal{A}} })_\mu.
\end{eqnarray*}
$f_\mu$ is a measure of the fraction of flux quanta of magnetic
induction penetrating a single plaquette with surface normal along
$\hat{\mu}$.  When the magnetic field is applied along $z$-axis, $f_x
= f_y = 0$, and
\begin{eqnarray}
  f \equiv f_z = \frac{Ba_xa_y}{\Phi_0} ~~~ \stackrel{\small 3DXY}{=}
  ~~~ \frac{B\xi_{ab}^2}{\Phi_0}.
\label{Filling.Fraction}
\end{eqnarray}
In this work, we consider filling fractions $f = 0, 1/20,.....,1/1560$.

\section{Monte-Carlo simulations, ${\bf B } = 0$}

In this subsection, we discuss the zero field superconducting-normal
(S-N) phase transition, both in terms of the usual Ginzburg-Landau
order field $\psi({\bf r})$, and in terms of the behavior of
topological excitations which can be tied to the formulation of the
transition using the disorder-field picture presented in Section IIF.
We compare our results obtained from the FG model to known simulation
results of the 3DXY model \cite{Chui:L88,Nguyen:B98b}, the London
model \cite{Dasgupta:L81}, and the Villain model
\cite{Janke:B90,Nguyen:B98a}.

Unless otherwise stated, in this subsection we show simulation results
for the FG model with the parameters: $f=0$, $\Gamma = 1$, $T'_{MF} =
0.3,1$, $a_\mu/\xi_\mu = 6$, and ${\mathcal{V}} = 60^3$.  We have
chosen $a_\mu/\xi_{\mu} = 6$ to slightly enhance the critical features
of the FG model. Simulations of the FG model using a smaller ratio
$a_\mu/\xi_\mu = 4$ leads to the same conclusions, but larger systems
and longer simulation times are required to obtain the same quality of
the data. What we will find is that the width of the regions where
phase-fluctuations dominate is controlled by the parameter $T_{MF}$,
increasing with $T_{MF}$. Our picture of the zero-field transition as
a vortex-loop unbinding is however borne out regardless of the value
of $T_{MF}$, close enough to the true critical temperature $T_c$.

\subsection{Order field}

First we present results for the S-N phase transition in terms of an
order field picture, i.e. in terms of the ordinary local pair-wave
function $\psi({\bf r})$ of the Ginzburg-Landau theory. We obtain
results of the full GL theory including amplitude and
phase-fluctuations. What will be shown, even when amplitude
fluctuations are included, is that we obtain a clear picture of the
S-N phase transition in terms of an Onsager-Feynman vortex loop
unbinding \cite{Hemmer:96,Onsager:1949,Feynman:PLTP55}, driven
exclusively by topological phase-fluctuations of the Ginzburg-Landau
order parameter.

\begin{figure}[htbp]
  \begin{picture}(0,400)(0,0)
     \put(-65,-60)
         {\includegraphics[angle=0,scale=0.6]
         {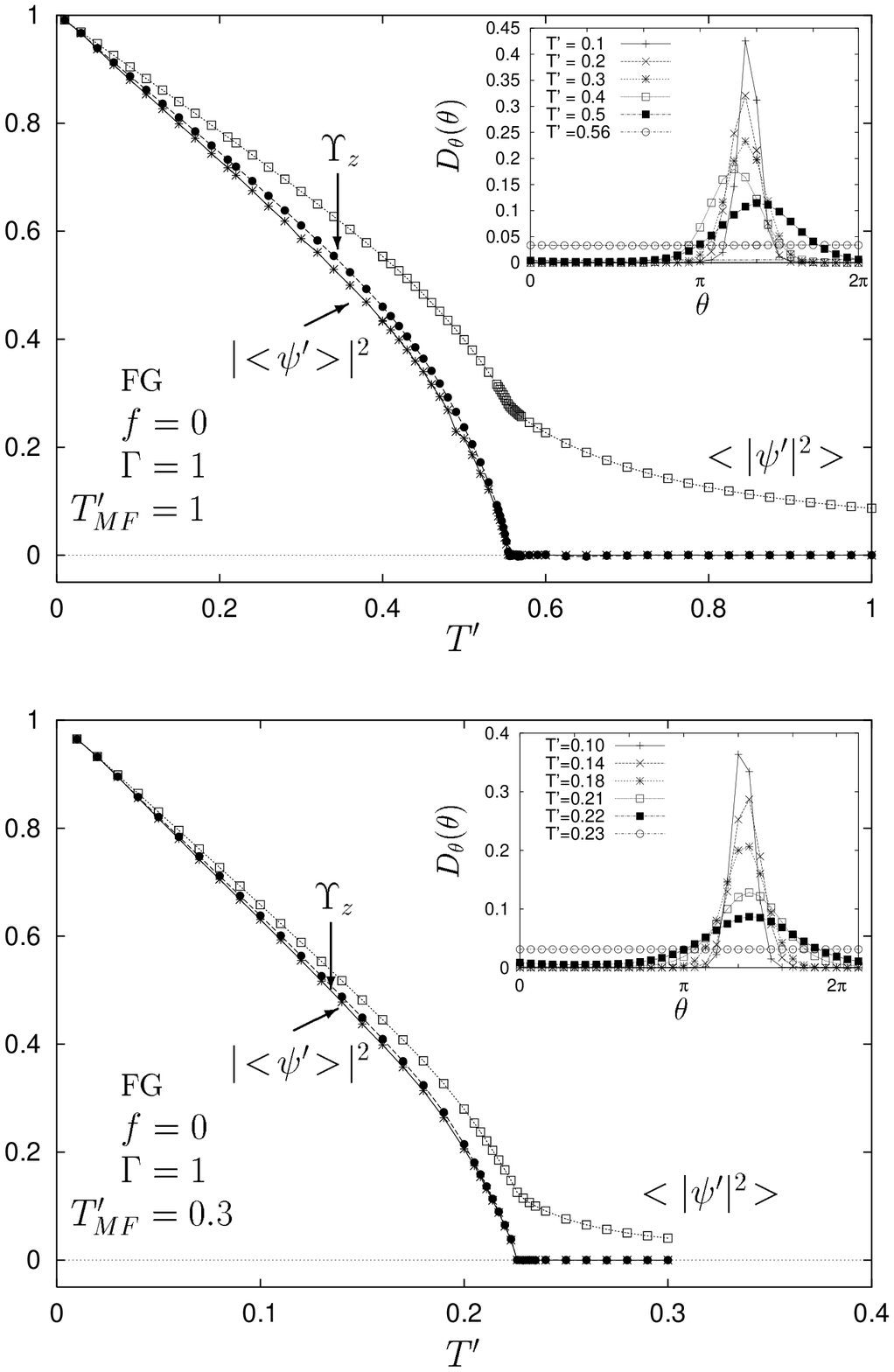}}
  \end{picture}
  { \small FIG. \ref{F1A1.YPsi.FG}. Helicity modulus $\Upsilon_z$,
    local Cooper-pair density $<|\psi'|^2 \! >$, and superfluid
    condensate density $|<\!\psi'\!>|^2$ as functions of temperature
    for the Ginzburg-Landau model in a frozen gauge approximation.
    Upper panel shows results for $f=0$, $\Gamma = 1$, $T'_{MF} =
    1.0$, $a_\mu/\xi_\mu = 6$, and ${\mathcal V} = 60^3$. Lines are
    guide to the eye. $\Upsilon_z$ and $|<\!\psi'\!>|^2$ develop
    finite expectation values for $T<T_c$, while $<|\psi'|^2\!>$ is
    finite both above and below $T_c$.  {\em Inset:} The distribution
    function of the phase angle of the order field $D_\theta(\theta)$
    as a function of $\theta$, for several temperatures. Below $T_c$,
    a preferred phase angle is chosen and the global $U(1)$ symmetry
    is spontaneously broken.  Note how the phase fluctuates around a
    mean value even in the ordered phase. Lower panel shows the same
    for $T'_{MF}=0.3$. The width of the critical region has decreased,
    but the amplitude fluctuations are still far from being critical.
    } \refstepcounter{figure}
\label{F1A1.YPsi.FG}
\end{figure}

In Fig. \ref{F1A1.YPsi.FG} we plot the helicity modulus $Y_z$, the
local density of Copper pairs $<|\psi'|^2\!>$, and the superfluid
condensate density $|<\!\psi'\!>|^2$ as functions of temperature. We
see that the condensate density $|<\!\psi'\!>|^2$ is zero above a
critical temperature and develops a finite expectation value below
$T_c$. In contrast to this, $<|\psi'|^2\!>$ is finite both above and
below $T_c$. Close to $T=T_c$ we have performed the simulations for a
very dense set of temperatures, and from the top panel of Fig.
\ref{F1A1.YPsi.FG}, we may however discern a kink in the curve and
hence a singular behavior of the temperature derivative of
$<|\psi'|^2>$. The top panel shows the results for $T_{MF}=1.0$, while
the lower panel shows the same results for $T_{MF}=0.3$. The
difference between the two panels is that since $T_{MF}$ has been
changed, the width of the critical region has changed, increasing upon
increasing $T_{MF}$. Had we chosen $T_{MF} =0.01$, an appropriate
value for conventional superconductors, the curves for
$|<\!\psi'\!>|^2$ and $<|\psi'|^2\!>$ would have been
indistinguishable, the conventional $BCS$ mean-field picture of the
superconducting transition would have been appropriate. The reason
that it is no longer the case in the high-$T_c$ cuprates is the large
energy scale for pairing, coupled with the fact that the
phase-stiffness is low.

Note how the curve for $\Upsilon_z \sim |\tau|^{2 \beta - \eta \nu}$
bends slightly more sharply towards zero than the curve for the
condensate density $|<\psi'>|^2 \sim |\tau|^{2 \beta}$, as expected
for a positive $\eta$, since in that case $2 \beta - \eta \nu < 2
\beta$. In fact, this provides a nice consistency-check on the
Monte-Carlo simulations.

We may therefore conclude that the existence of Cooper pairs does not
imply superconductivity, and the local Cooper pair density can not be
used as a probe of the superconducting phase. Rather, {\em the
  superconducting phase is characterized by a finite expectation value
  of the superfluid condensate density}, i.e $<\!\psi'({\mathbf r})\!>
\not =\! 0$.  Recall that $|<\!\psi'\!>|^2 \not = 0$ only if
$<\!\psi'({\mathbf r})\!> \not = 0$.  The slow decay of
$<|\psi'|^2\!>$ above the true superconducting transition temperature
$T_c$ is due to gaussian fluctuations in the amplitude of the order
parameter.

The results shown in Fig. \ref{F1A1.YPsi.FG} justify, quite clearly,
that the neglect of amplitude fluctuations in order to study the
critical behavior, is an entirely appropriate approach to this
problem. In the normal phase, the phase angle of the order field is
uniformly distributed, Fig. \ref{F1A1.YPsi.FG} inset, while for $T <
T_c$, the system spontaneously chooses a preferred phase angle giving
a peak in $D_\theta(\theta)$. Due to our finite set of discrete phase
angles, $D_\theta(\theta) = 1/N_\theta$ for $T > T_c$, and not zero as
in the continuum-$\theta$ limit.

Note the difference between this picture compared to the usual mean
field picture of the S-N phase transition. In the mean field picture,
the phase angle of the order field does not fluctuate.  Thus,
$<|\psi'|^2> = |<\psi'>|^2$, and $T_c$ can be defined as the highest
temperature where the local Cooper pair density $<|\psi'|^2>$ develops
a finite expectation value. This happens, at the mean field level
neglecting all fluctuations in $\psi'$, at $T_{MF}$ where the
Ginzburg-Landau parameter $\alpha (T) = 1 - T'/T'_{MF}$ becomes
negative. Including fluctuations $<|\psi'|^2> \not= |<\psi'>|^2$, and
$T_c$ is defined as the lowest temperature where the  superfluid
condensate density $|<\psi'>|^2$ still maintains a value of zero.

Given that the condensate density is non-zero below $T_c$, we
next focus on a global quantity, the long-wavelength limit of the
helicity modulus $\Upsilon_\mu$, or equivalently the superfluid
stiffness in the $\mu$-direction. In Fig. \ref{F1A1.YPsi.FG} we see
that $\Upsilon_z$ vanishes for temperatures $T \geq T_c$, and develops
an expectation value for $T < T_c$. Thus, the superconducting phase
exhibits global phase coherence, while the normal phase does not. We
have also calculated $\Upsilon_x$ and $\Upsilon_y$, and found (not
shown) that they show the same behavior as $\Upsilon_z$. Apart from
minor details, we see in Fig.  \ref{F1A1.YPsi.FG} that the helicity
modulus is proportional to the condensate  density \cite{Fisher:A73}.
We will also show that this equality also applies to the finite field
case.

At low temperature, $\Upsilon_\mu$ decreases linearly. This feature is
also obtained in the zero field 3DXY model \cite{Nguyen:B98b}, but not
in the zero field Villain model \cite{Nguyen:B98a}. In the Villain
model, the spin waves and the vortex loops can be analytically
decoupled.  Here, at low temperatures, spin wave excitations do not
affect the vortex loops excitations and the superfluid phase stiffness
should decay in an activated manner due to the excitation of vortex
loops. In the 3DXY model, the spin wave and the vortex loops are
coupled together. 
Whether or not the low-temperature features of $\Upsilon_{\mu}$ in
Fig. \ref{F1A1.YPsi.FG} can explain experimental data on the
temperature dependence of $1/\lambda_{\mu}^2$, see for instance Ref.
\onlinecite{Hardy:96}, is an interesting but so far unsettled issue,
see also the results of Refs.
\cite{Nguyen:B98b,Nguyen:cm98,Carlson:cm99}.  Within the anisotropic
$3DXY$-model, the helicity modulus $\Upsilon_z$ has a {\it larger},
but negative slope of its linear low-$T$ behavior compared to
$\Upsilon_x$ and $\Upsilon_y$. On the other hand, it is not entirely
trivial to connect $\Upsilon_z(T)/\Upsilon_z(0)$ to the
$T=0$-normalized superfluid density $\rho_{sz}$ \cite{SAK:pc}, which
is the quantity measured in the experiments of Hardy {\it et al.}
\cite{Hardy:96}. {\em However, our main point of emphasis is that the
  vanishing of the superconducting phase stiffness at $T=T_c$ is
  caused exclusively by an unbinding of large vortex loops}. Further
evidence for the connection between $\Upsilon_\mu$ and the vortex
loops can be found in simulations of the lattice London model, where
vortex-loops are the only degrees of freedom
\cite{Dasgupta:L81,Nguyen:L96,Chen:B97b}. Here, the normalized
helicity modulus $\lim_{k \rightarrow 0}
\Upsilon_\mu(T)/\Upsilon_\mu(T=0)$ is renormalized to zero at $T_c$
exclusively by the expansion of vortex loops.

The critical behavior of the $3DXY$-model including coupling between
spin-waves and vortices, is the same as for the vortex-content of the
same theory, but taken in the Villain-approximation, where no coupling
between vortices and spin-waves exist. The only excitations in the
$U(1)$-symmetric gauge-theories in the frozen gauge- and
amlpitude-approximation, are longitudinal and transverse
phase-fluctuations, i.e. spin-waves and vortex-loops.

{\it The above considerations and results provide an overwhelming
  amount of evidence in favor of the proposition that unbound
  vortex-loops are precisely the critical fluctuations of an extreme
  type-II superconductor.}

Approaching $T_c$ from below, $\Upsilon_\mu(T)$ decays to zero with an
exponent consistent with $\upsilon = 2 \beta - \eta \nu$, and $\eta
\approx 0.04$ as discussed in Section III A.5, see Fig.
\ref{F1A1.YPsi.FG}.  For the special case of $d=3$, which we consider,
we have $\rho_{s \mu} \sim \xi^{-1} \sim |\tau|^{\nu}$, and hence we
find $2 \beta - \eta \nu = \nu < 2 \beta$.
We will show in the next section that this renormalization of 
$\Upsilon_\mu$ is nicely explained by the expansion and blowout of 
thermally excited vortex loops.

\subsection{Topological excitations and vortex-line tension}

We now discuss the S-N phase transition in terms of the behavior of
topological objects of the FG model, i.e. closed vortex loops and
vortex lines.

The first figure of this section shows our probe of the vortex-tangle
connectivity, $O_L$, as a function of temperature, for various system
sizes and $\Gamma=1$ and $7$. We reemphasize that $O_L$ is not
suggested as an order parameter of the transition, but that it
nevertheless can be tied to a local order parameter via the
formulation of the theory given in Section IIF, unlike the known
spin-percolation transition in the $3D$ Ising-model
\cite{Footnote:Ising}.

\begin{figure}
  \begin{picture}(0,400)(0,0)
     \put(-65,-60)
         {\includegraphics[angle=0,scale=0.6]
         {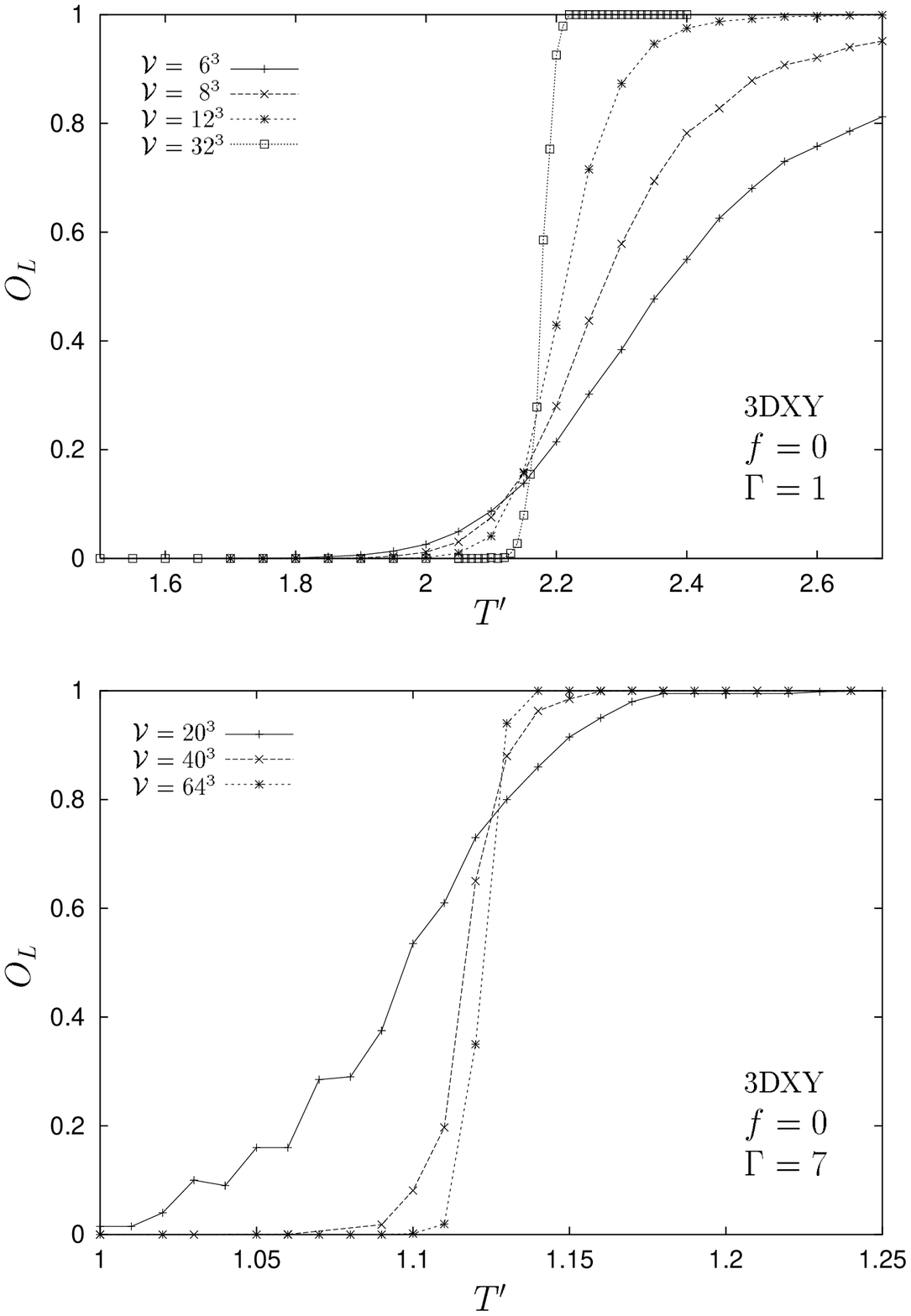}}
  \end{picture}
  { \small FIG. \ref{F1A1.Ol.FG}.  $O_L$ as a function of temperature
    for several system sizes for the 3DXY model with f=0. Lines are
    guide to the eye. Top panel: $\Gamma=1$, bottom panel: $\Gamma=7$.
    Note the finite-size effect in $O_L$, with the crossings of the
    curve approximately at the same temperature.  }
  \refstepcounter{figure}
\label{F1A1.Ol.FG}
\end{figure}

In Fig. \ref{F1A1.Ol.FG}, we show the probability of finding a
connected vortex-tangle across the system in zero magnetic field, for
$\Gamma=1,7$, and system sizes $L^3$, with $L=6,..64$. Notice how the
curves cross at approximately the same temperature and get
progressively sharper. Similar results were seen for considerably
smaller system sizes $L=4,6,8$ in Ref.  \onlinecite{Akao:B96}.  Below,
we will also give results for much larger system sizes, confirming
that the crossing temperature in Fig. \ref{F1A1.Ol.FG} gives a good
estimate for the threshold temperature for vortex-loop unbinding
throughout the sample. As pointed out in Ref.  \onlinecite{Akao:B96},
such a finite-size effect indicates that a percolation threshold
exists for the vortex tangle in the thermodynamic limit
\cite{Aharony:Bo94}. {\it A precisely similar finite-size effect in
  $O_L$ will be seen in finite magnetic field}, to be considered in
Section VB.  This will happen inside the vortex liquid phase at
elevated magnetic fields, but will coincide with VLL melting at low
fields, and suggests the revision of the picture of the molten phase
of the Abrikosov vortex system purely in terms of a vortex-{\it line}
liquid.

\begin{figure}
  \begin{picture}(0,410)(0,0)
     \put(-65,-50)
         {\includegraphics[angle=0,scale=0.6]
         {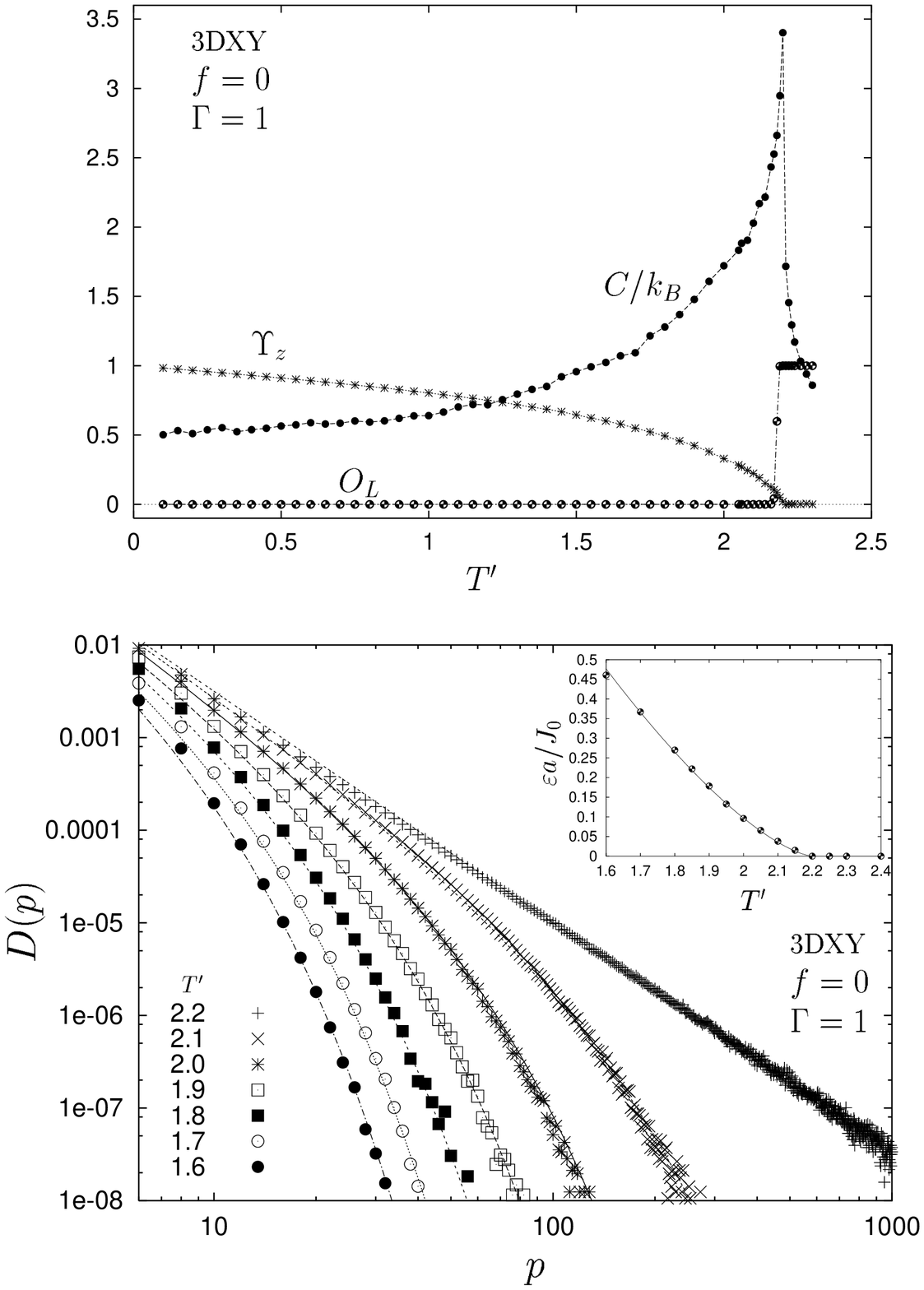}}
  \end{picture}
  { \small FIG. \ref{F1A1.XY.FG}. Top panel: Specific heat $C$,
    helicity modulus $\Upsilon_z$, and $O_L$ as functions of
    temperature for the 3DXY model with $f=0$, $\Gamma=1$ and
    ${\mathcal V}=120^3$. Lines are guide to the eye.\\  
    Bottom panel: Vortex-loop distribution function $D(p)$ as a
    function of loop-perimeter $p$ for various temperatures. Lines are
    fits using $D(p) = p^{-5/2} \exp(-\varepsilon(T) p/k_BT)$.  At
    $T=T_c$ the decay changes from exponential to algebraic implying
    that the vortex-line tension $\varepsilon$ vanishes.  Inset of
    bottom panel shows $\varepsilon(T)$.  Solid line is a fit using
    $|T'-T'_c|^{\gamma}$, with $\gamma = 1.45 \pm 0.05$.  }
  \refstepcounter{figure}
\label{F1A1.XY.FG}
\end{figure}

We next proceed to correlate the change in $O_L$ with the unbinding of
large-vortex loops and the loss of vortex-line tension at $T=T_c$, by
correlating its abrupt change with the characteristics developing in
$D(p)$, which probes the typical size of thermally induced vortex
loops in the system. We first consider the case of the $3DXY$-model,
for which the results are shown in Fig. \ref{F1A1.XY.FG}. The top
panel shows specific heat, $O_L$ and helicity modulus, while the
bottom panel shows the vortex-loop distribution function $D(p)$ as a
function of perimeter $p$ for a number of temperatures $T \leq T_c$,
while the inset of the bottom panel shows the temperature dependence
of the long-wavelength vortex-line tension $\varepsilon(T)$.  In the
top panel it is clear that the loss of helicity modulus, the anomaly
in specific heat, and the abrupt change in $O_L$ all occur at
precisely the same temperature. The change in the decay of $D(p)$ also
occurs at the same temperature, $T_c$.

For $T < T_c$, $O_L = 0$, and all vortex loops are confined, with
typical size given by $\L_0(T) = k_B T/\varepsilon(T)$, where
\begin{eqnarray}
L_0(T) = \frac{k_B T}{\varepsilon(T)},
\label{loopsize}
\end{eqnarray}
where $\varepsilon(T)$ is the effective long-wavelength vortex line
tension, equivalently the free energy per unit length of vortex-lines.
These objects, present also in the low-temperature phase, cause only a
local perturbation of the order parameter in the system, and may
simply be ``coarse grained" away.  The low energy physics of the model
is therefore described essentially by the physics of the zero
temperature fixed point. At and above $T_c$, $O_L=1$, and vortex loops
with infinite size always exist. The length scale $L_0(T)$ has
diverged, showing that there are vortex-loops on all length-scales
with a power-law tail in the distribution. Such loops cannot be coarse
grained away and taken into account by any ``appropriate
renormalization" of the zero-temperature theory. Thus, {\em the S-N
  phase transition can be viewed as a blowout out of thermally induced
  vortex loops}. Above $T_c$, free thermally induced ``vortex lines''
exist in all direction, and any infinitesimal applied current will
move these thermally induced ``vortex lines'' and dissipate energy
\cite{Footnote:ffh}.  Thus, the system is in the normal phase.

\begin{figure}
  \begin{picture}(0,410)(0,0)
     \put(-50,-50)
         {\includegraphics[angle=0,scale=0.6]
         {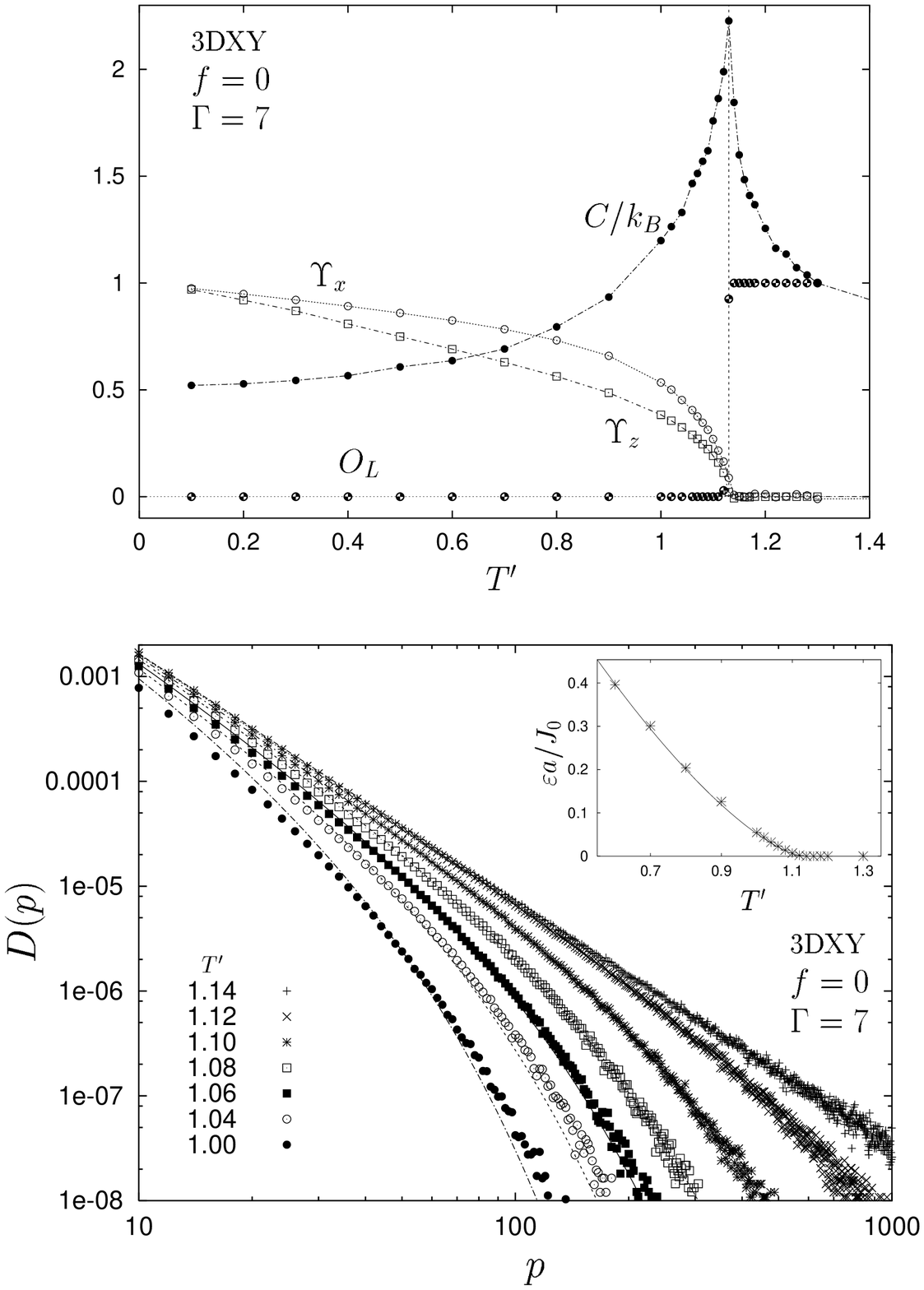}}
  \end{picture}
  { \small FIG. \ref{F1A7.XY.FG}. Top panel: Specific heat $C$,
    helicity moduli $\Upsilon_x, \Upsilon_z$, and $O_L$ as functions
    of temperature for the 3DXY model with $f=0$, $\Gamma=7$, and
    ${\mathcal V} = 140^3$. Lines are guide to the eye. \\
    Bottom panel: Vortex-loop distribution function $D(p)$ as a
    function of loop-perimeter $p$ at various temperatures. Lines are
    fits using $D(p) = 0.37p^{-2.35} \exp(-\varepsilon(T) p/k_BT)$.
    Inset: Vortex-line tension $\varepsilon(T)$ as a function of
    temperature.  Solid line is a fit using $|T'-T'_c|^{\gamma}$, with
    $\gamma = 1.45 \pm 0.05$.  } \refstepcounter{figure}
\label{F1A7.XY.FG}
\end{figure}

\begin{figure}
  \begin{picture}(0,410)(0,0)
     \put(-65,-50)
         {\includegraphics[angle=0,scale=0.6]
         {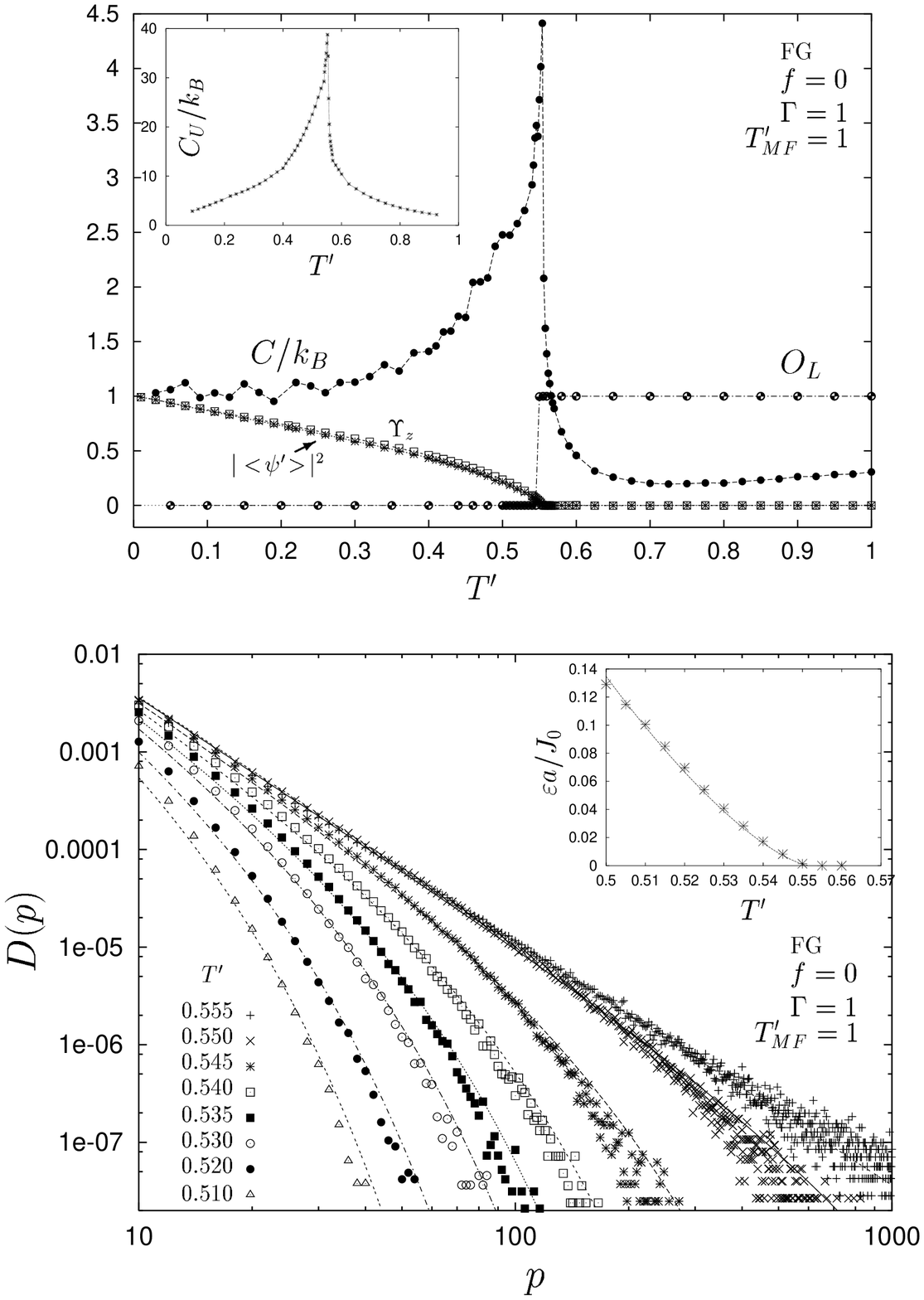}}
  \end{picture}
  { \small FIG. \ref{F1A1.1.0.FG}. Top panel: Specific heat $C$
    calculated using Eq. \ref{Spes.Heat.Fluc}, $O_L$, helicity modulus
    $\Upsilon_z$, and superfluid density $|<\psi'>|^2$ as
    functions of temperature for the Ginzburg-Landau model in a frozen
    gauge approximation with $f=0$, $\Gamma = 1$, $T'_{MF} = 1$,
    $a_\mu/\xi_\mu = 6$, and size ${\mathcal{V}} = 60^3$. Lines are
    guide to the eye. Inset shows the specific heat calculated using
    Eq. \ref{Spes.Heat.Diff}.\\
    Bottom panel: Distribution of vortex loops $D(p)$ as a function of
    the loop-perimeter $p$ for several temperatures. Lines are fits
    using $D(p) = 1.15p^{-5/2} \exp(-\varepsilon(T) p/k_BT)$. Inset:
    Vortex-line tension $\varepsilon(T)$ as a function of temperature.
    Dotted line is a fit using $|T'-T'_c|^{\gamma}$, with $\gamma = 1.45
    \pm 0.05$.  } \refstepcounter{figure}
\label{F1A1.1.0.FG}
\end{figure}

In Fig. \ref{F1A7.XY.FG}, we show the specific heat anomaly, the
helicity moduli $\Upsilon_x$ and $\Upsilon_z$, as well as $O_L$ for
the $3DXY$-model, with $\Gamma=7$. The correlation noted above in
connection with Fig.  \ref{F1A1.XY.FG} is again perfect, the only
difference being that the specific heat anomaly has become more
symmetric due to the increased anisotropy, $\Gamma=7$. Although the
amplitude of $\Upsilon_x$ is larger than the amplitude of $\Upsilon_z$
due to the uniaxial anisotropy along the $z$-axis, the temperature at
which they vanish, and the power law with which they vanish, are the
same. Note also the sharpness of the manner in which the moduli
$\Upsilon_{\mu}$ approach zero at $T_c$, there is no high-temperature
tail as one would have found in too small systems. This in fact serves
as a highly non-trivial benchmark on the quality of the Monte-Carlo
simulations.

\begin{figure}
  \begin{picture}(0,410)(0,0)
     \put(-50,-50)
         {\includegraphics[angle=0,scale=0.6]
         {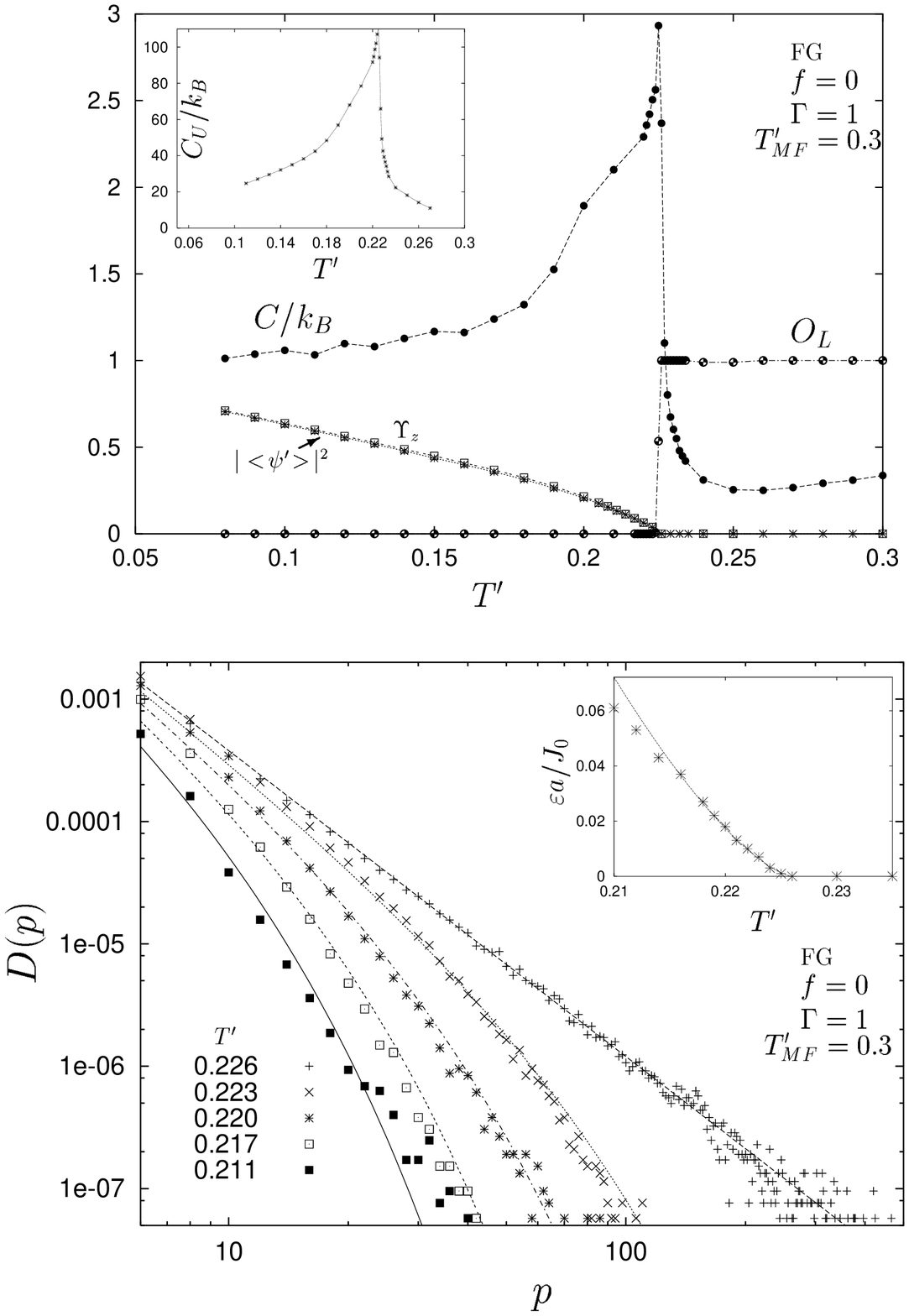}}
  \end{picture}
  { \small FIG. \ref{F1A1.0.3.FG}. Same as Fig. \ref{F1A1.1.0.FG},
    with $T'_{MF}=0.3$, $D(p)=0.12p^{-5/2}\exp(-\varepsilon(T)
    p/k_BT)$, and $\varepsilon(T)a/J_0 = 29|T'-T'_c|^{1.45}$.}
  \refstepcounter{figure}
\label{F1A1.0.3.FG}
\end{figure}

Figs. \ref{F1A1.1.0.FG} and \ref{F1A1.0.3.FG} show essentially the
same as Fig.  \ref{F1A7.XY.FG}, but now for the FG model, {\it i.e.
  including amplitude fluctuations on an equal footing with the
  phase-fluctuations}. Clearly, the picture that it is the topological
phase-fluctuations, or the vortex-loop unbinding, that drives the
superconductor normal-fluid transition, is not at all altered by the
fact that amplitude fluctuations are included. This is a
reconfirmation of the results obtained in Section IVA, showing that
amplitude fluctuations of the local Ginzburg-Landau order parameter
have a large mass at the critical temperature where the superfluid
density vanishes.  {\it The vortex-loop unbinding, which is the
  microscopic mechanism driving the transition, cannot be
  reparametrized in terms of critical amplitude fluctuations of the
  local order parameter of the theory, since the latter are clearly
  nowhere close to being critical.}

For a more detailed study of the properties of thermally induced
vortex loops, we now focus on the vortex-loop distribution function
$D(p)$ as a function of vortex-loop perimeter $p$ at various
temperatures.  These are shown in Figs. \ref{F1A1.XY.FG},
\ref{F1A7.XY.FG}, and \ref{F1A1.1.0.FG} and are clearly well
approximated by the form $D(p) = A p^{-5/2} ~ e^{-\varepsilon(T)
  p/k_BT}$ for all temperatures considered. {\em Note that
  $\varepsilon(T)$ is the only temperature-dependent fitting parameter
  in all plots}. The effective long-wavelength linetension of vortex
loops is finite below $T_c$, and vanishes for $T \geq T_c$. The
physical picture of this phase transition is as follows. Below $T_c$,
$\varepsilon(T)$ is finite defining a typical length scale for the
vortex loops, $L_0 = k_BT/\varepsilon(T)$. Here, $D(p)$ is dominated
by an exponential decay and vortex loops with much larger perimeter
$p$ than $L_0$, are exponentially suppressed. Thus, the topological
excitations that are present in the system may be coarse grained away.
At and above $T_c$, $\varepsilon(T) = 0$ and no typical length scale
for the vortex loops exist; the length scale $L_0$ has diverged.
Here, $D(p)$ is purely algebraic, and vortex loops of all sizes
including infinite size, exist. Thus, the S-N phase transition at
$T_c$ is triggered by an unbinding of large vortex loops, analagous to
the Onsager-Feynman mechanism, \cite{Nguyen:B98a} suggested for the
superfluid-normalfluid transition in $^4He$ \cite{Onsager:1949}.

In the insets in the bottom panels of Figs. \ref{F1A1.XY.FG},
\ref{F1A7.XY.FG}, \ref{F1A1.1.0.FG}, and \ref{F1A1.0.3.FG}, we show
the vortex line tension $\varepsilon(T)$ extracted from the vortex
loop distribution function $D(p)$. Regardless of whether the $3DXY$-
or the FG-models are used, we find that the long-wavelength
vortex-line tension vanishes as
\begin{eqnarray}
\varepsilon(T) \sim |T-T_c|^{\gamma}; ~~ \gamma = 1.45 \pm 0.05.
\end{eqnarray}
The numerical value of the exponent $\gamma$ has been extracted from
the systems with the largest critical regions, i.e. Figs.
\ref{F1A1.XY.FG}, \ref{F1A7.XY.FG}, and \ref{F1A1.1.0.FG}. The system
shown in Fig. \ref{F1A1.0.3.FG} does not allow a very precise value
for $\gamma$ to be obtained, although the qualitative aspects of the
results are clearly precisely the same as those for the $3DXY$-model
and the FG-approximation of the GL-model with $T'_{MF}=1.0$.  This
implies that the typical vortex-loop perimeter diverges when $T_c$ is
approached from below, using Eq. \ref{loopsize}, as
\begin{eqnarray}
L_0(T) \sim |T-T_c|^{-\gamma},
\end{eqnarray}
such that $L_0(T)$ is a power of the correlation length $\xi$ of the
$3DXY$ model.

\subsection{Anomalous dimension of the dual field}

We next connect the result for $\varepsilon(T)$ to the anomalous
dimension of the dual field $\phi$.  It is natural, within the
formulation of the problem given in Section IIF, to associate the
proliferation of unbound vortex-loops with a vortex-loop
susceptibility, or equivalently a susceptibility for the $\phi$-field
of Section IIF. This is seen as follows. The proliferation of
unbounded vortex loops as the temperature of the superconductor is
increased, is associated with the development of long-range
correlations in the two-point correlation function of the dual field,
$G(x) \equiv <\phi^*(x) \phi(0)>$, where on the low-temperature side
the dual order parameter has zero expectation value, $<\phi>=0$. A
scaling Ansatz for $G(x)$ reads
\begin{eqnarray}
G(x)  = \frac{1}{|x|^{d-2 + \eta_{\phi}}} ~~{\cal G}(x/\xi),
\end{eqnarray}
where $\eta_{\phi}$ is the anomalous dimension of the dual field
$\phi$, $\xi$ is its correlation length, and ${\cal G}(x/\xi)$ is some
scaling function. The square of the mass of the dual field,
$m_{\phi}^2$, is therefore naturally mapped to the line-tension
$\varepsilon(T)$ of the vortex-loops. This follows from the
observation that the dual boson system of which the $\phi$-theory is a
field-theory description, has a chemical potential $m^2_{\phi}$ which
in turn is nothing but the line-tension $\varepsilon(T)$ of the
vortex-loop system, when the density distribution $D(p)$ is viewed as
a partial density in a fugacity expansion for the density of the dual
bose-system \cite{Hoye:JSP94}.  The Fourier-transform $\tilde
G(k)=<\phi^*(k) \phi(-k)>$ of $G(x)$ may be written on form
\begin{eqnarray}
\tilde G(k) = \xi^{2 - \eta_{\phi}}~~{\cal F}(k \xi),
\end{eqnarray}
where ${\cal F}(k \xi)$ is some new scaling function. The $k \to 0$
limit of this is the static uniform susceptibility $\chi_{\phi}$ of
the dual field on the low-temperature side, where $<\phi>=0$. On the
other hand, as long as the dual field is massive, which it is on the
low-temperature side, we must have $\lim_{k \to 0} \tilde G(k) =
m^{-2}_{\phi}$. Hence, we obtain
\begin{eqnarray}
\chi_{\phi} \sim \frac{1}{m^2_{\phi}} 
            \sim \frac{1}{\varepsilon} 
            \sim \xi^{2 - \eta_{\phi}}
            \sim |\tau|^{-\nu_{\phi}(2 - \eta_{\phi})}.
\end{eqnarray}
The field $\phi$ has a correlation length exponent given by
$\nu_{\phi}=2/3$, the same as for the $3DXY$-model
\cite{Hasenbuch:cm99}. This follows from the fact that it is a
thermodynamic exponent describing the divergence of one and the same
length in the Ginzburg-Landau theory and dual theory. Very
importantly, it {\it must} be equal both for the dual model and its
Ginzburg-Landau counterpart by ``strong" duality
\cite{Herbut:L96,Olsson:L98,Hove:99}.  Were this {\it not} to hold,
the dual of the {\it dual} theory would not be the original theory, as
it ought to be. The above of course precisely amounts to the Fisher
scaling-law \cite{Fisher:PR69} relating the susceptibility exponent of
the dual field $\gamma_{\phi}$ to $\nu_{\phi}$ and $\eta_{\phi}$
\begin{eqnarray}
\gamma_{\phi} = \nu_{\phi} ~ (2-\eta_{\phi}).
\end{eqnarray}
Using our estimate $\gamma_{\phi} = 1.45 \pm 0.05$ with
$\nu_{\phi}=2/3$ gives $\eta_{\phi}=-0.18 \mp 0.07$ in close agreement
with previous renormalization group calculations \cite{Herbut:L96},
who found $\eta_{\phi}=-0.20$ to one-loop order.

The result $\eta_{\phi} = -0.18 \mp 0.07$ obtained directly from
computing the statistics of the loop-excitations of the $3DXY$-model
is a truly noteworthy result, when viewed juxtaposed to the
RG-calculations of Ref.  \onlinecite{Herbut:L96}. In Ref.
\onlinecite{Herbut:L96}, the RG-result for the anomalous dimension of
the dual field was obtained directly from the dual theory. On the
other hand, our numerical result is obtained directly from the
phase-only approximation to the original Ginzburg-Landau theory. The
agreement shows conclusively, and to our knowledge for the first time,
that viewing the zero-field transition of the $3D$ Ginzburg-Landau
theory as a vortex-loop unbinding, which {\it is} the phase-transition
of the dual theory, is {\it precisely} correct, not only
qualitatively, but {\it quantitatively} \cite{ZBT:pc}.

At and below $T_c$, the order field $<\psi'({\mathbf r})>$ develops an
expectation value, and explicitly breaks the global $U(1)$ symmetry of
the GL-theory. In contrast to the order field picture, in a
description using only topological excitations, {\em the global $U(1)$
  symmetry is hidden}. There does not appear to be any symmetry
operation involving the phase of a local field, that will leave the
effective action Eq. \ref{Shma} invariant. Therefore, there is also no
obvious local quantity that develops an expectation value in the
non-symmetric phase. Nevertheless, it is possible to define a global
quantity that implicitly probes the breaking of the global $U(1)$
symmetry, namely $O_L$. Let $N_\mu$ denote the number of ``vortex
lines'' (percolating directed vortex paths without using PBC) along
the $\mu$-direction. Below $T_c$, $N_\mu$ is fixed to zero and
$O_L=0$.  Concomitant with the conservation of the global quantity
$N_\mu$, the system must exhibit a global $U(1)$ symmetry.  At and
above $T_c$, $N_\mu$ develops an expectation value and $O_L \neq 0$.
This leads to a broken $U(1)$ symmetry.

\section{Monte-Carlo simulations, ${\bf B} \neq 0$ }

We next discuss the indications we have of phase-transitions in the
vortex-system in a finite magnetic field. In addition to the first
order VLL melting transition line $T_m(B)$ which we map out for a
large range of filling fractions, we find indications for a new phase
transition in the vortex liquid. {\it We emphasize that in all
  simulations performed in finite magnetic field, the filling fraction
  is low enough to ensure that there is zero transverse Meissner
  effect at any temperature of interest.} That is to say, the
vortex-line lattice is depinned from the numerical lattice at much
lower temperature than the temperatures where the Bragg-peaks in the
structure function of the VLL vanishes. Therefore, commensuration
effects due to defining the theory on a lattice effectively have been
eliminated at the temperatures of interest.

Before entering into the discussion, a clarifying remark is
appropriate.  {\it Note that the phase-transition that we suggest may
  be taking place inside the vortex liquid, is not a transition from a
  disentangled low-temperature vortex liquid to an entangled
  high-temperature vortex liquid, as discussed by numerous previous
  authors}. Such a transition would have has its hallmark that the
superfluid stiffness along the magnetic field, or equivalently the
helicity modulus $\Upsilon_z$, would vanish {\it inside} the
vortex-liquid phase. This has now been conclusively demonstrated not
to be the case \cite{Hu:L97,Nguyen:B98b,Chin:cm98,Olsson:cm98}.

At the phase transition we propose inside the vortex liquid, a global
$U(1)$ symmetry is spontaneously broken. {\em The probe for this new
  phase transition is, as for the zero field case, a sudden change in
  the connectivity of the vortex system, probed by $O_L$}. As defined
in the zero-field section, we have found that the vortex-loop
distribution function $D(p)$ is not useful as a probe for the vortex
line tension in finite field. Fortunately, non-local quantities with
directions built-in as the helicity modulus, and the percolation
probability still apply in finite field. To probe the VLL melting, we
must define a new direction dependent non-local quantity, namely the
structure function for vortex segments parallel to the field,
$S({\mathbf k})$. Quantities probing local properties as $<|\psi'|>$
and $|<\psi'>|$ are still useful in finite field.

\subsection{First order VLL melting transition}

In this sub-subsection, we show and discuss our results with the focus
on the first-order VLL melting transition.

\begin{figure}
  \begin{picture}(0,410)(0,0)
     \put(-65,-50)
         {\includegraphics[angle=0,scale=0.6]
         {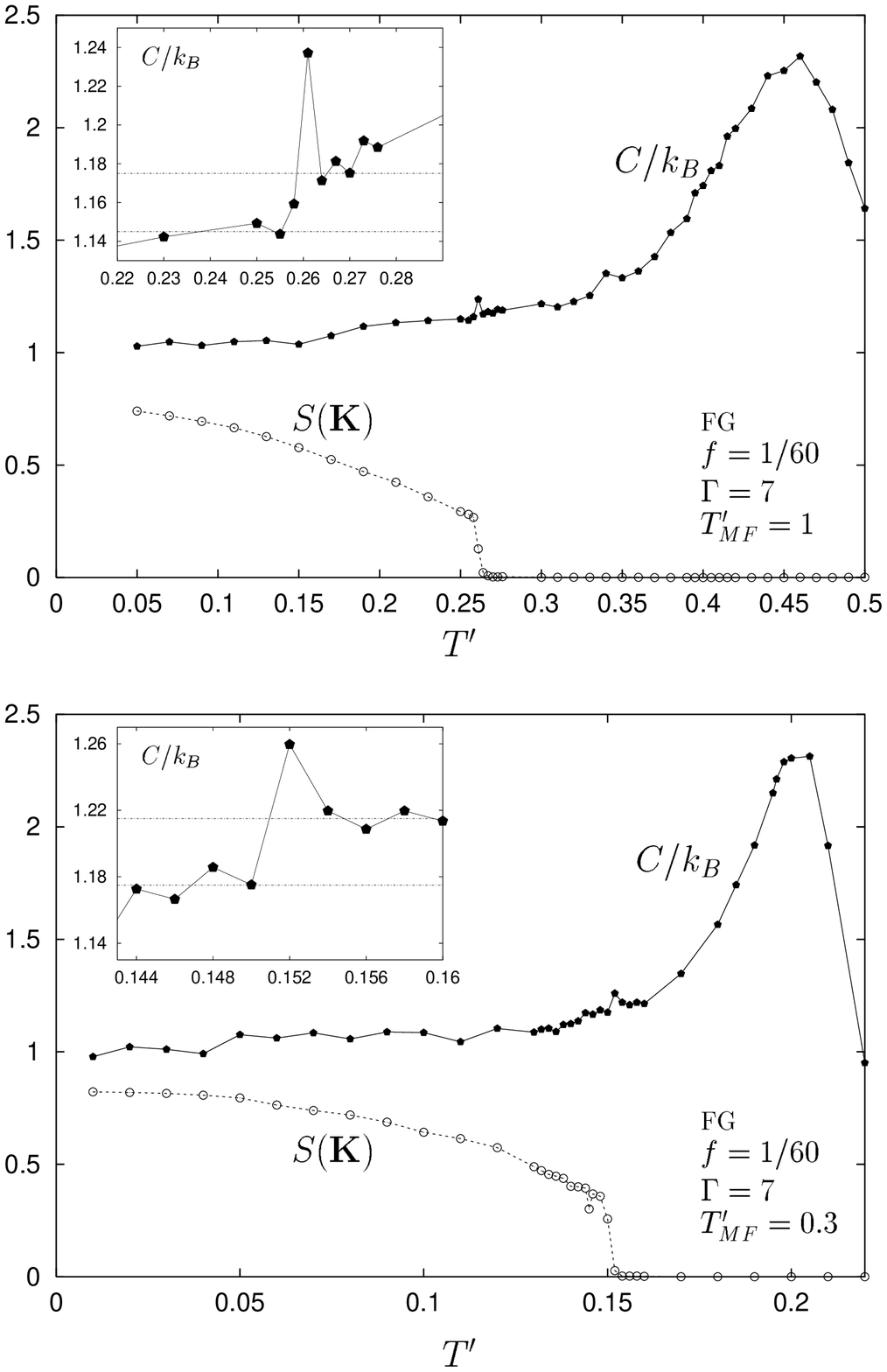}}
  \end{picture}
  {\small FIG. \ref{F60A7.SC.FG}. Top panel: Structure function for
    vortex segments along the field direction $S({\mathbf
      K}=[\pi/15,4\pi/15,0])$ and specific heat $C$ as functions of
    temperature for the Ginzburg-Landau model in a frozen gauge
    approximation with $f=1/60$, $\Gamma = 7$, $T'_{MF} = 1$,
    $a_\mu/\xi_\mu = 6$, and ${\mathcal V} = 60^3$. $S({\mathbf K})$
    shows a sharp drop at $k_BT_m/J_0 = 0.26$ indicating a first order
    VLL melting transition. $C$ shows a delta-function like anomaly at
    $T_m$, the VLL melting transition, detailed in the inset. \\
    Bottom panel: Same as top panel, but for $T'_{MF}=0.3$. }
  \refstepcounter{figure}
\label{F60A7.SC.FG}
\end{figure}

We show in Fig. \ref{F60A7.SC.FG} the structure function for vortex
segments along the field direction $S({\mathbf K}=[\pi/15,4\pi/15,0])$
and the fluctuation specific heat $C$ as functions of temperature for
the FG model. Here, ${\mathbf K}$ is a reciprocal lattice vector of
the VLL.

We see in Fig. \ref{F60A7.SC.FG} that $S({\mathbf K})$ shows a sharp
drop at $k_BT_m/J_0 = 0.26$ from $\sim 0.2$ to $0$, indicating a first
order VLL melting transition. The Lindemann number for this melting
transition estimated from the Debye-Waller factor at the melting
temperature, is 0.25, consistent with previous estimates
\cite{Houghton:B89,Nordborg:Thesis,Nguyen:B98b}. This abrupt
disappearance of the VLL structure function has also been
experimentally found in BSCCO \cite{Cubitt:N93,Oral:L98}. {\it In the
  high-field regime, the position of the melting line is well
  estimated by the Lindemann-criterion}
\cite{Houghton:B89,Nguyen:cm98}.

Density plots (not shown) of $S({\mathbf k}_\perp,0)$ show Bragg spots
for $T$ below $T_m$, and rings pattern for $T$ just above $T_m$.
Here, ${\mathbf k}_\perp = [k_x,k_y,0]$ and $(k_x,k_y) \in
[-\pi:\pi]$. This expected peak-to-ring feature is also found at $T_m$
in simulations on the 3DXY model \cite{Nguyen:B98b} and in the Villain
model \cite{Li:B93,Nguyen:B98a}. Below $T_m$, the vortex lines perform
small Gaussian fluctuations around their equilibrium positions, and
renormalize $S({\mathbf K};T)$ to a smaller value compared to
$S({\mathbf K};T=0)$.  Precisely at $T_m$ the specific heat shows a
delta-function like anomaly, Fig. \ref{F60A7.SC.FG}, indicating the
well-established first-order character of the VLL-melting transition
in clean systems \cite{Hetzel:L92,Safar:L92}.  Note that $T_m/T_{Bc2}$
increases for decreasing $T_{MF}$. For very small $T_{MF}$, where the
mean field approximation applies, $T_m \simeq T_{Bc2}$.

\begin{figure}
  \begin{picture}(0,410)(0,0)
     \put(-65,-50)
         {\includegraphics[angle=0,scale=0.6]
         {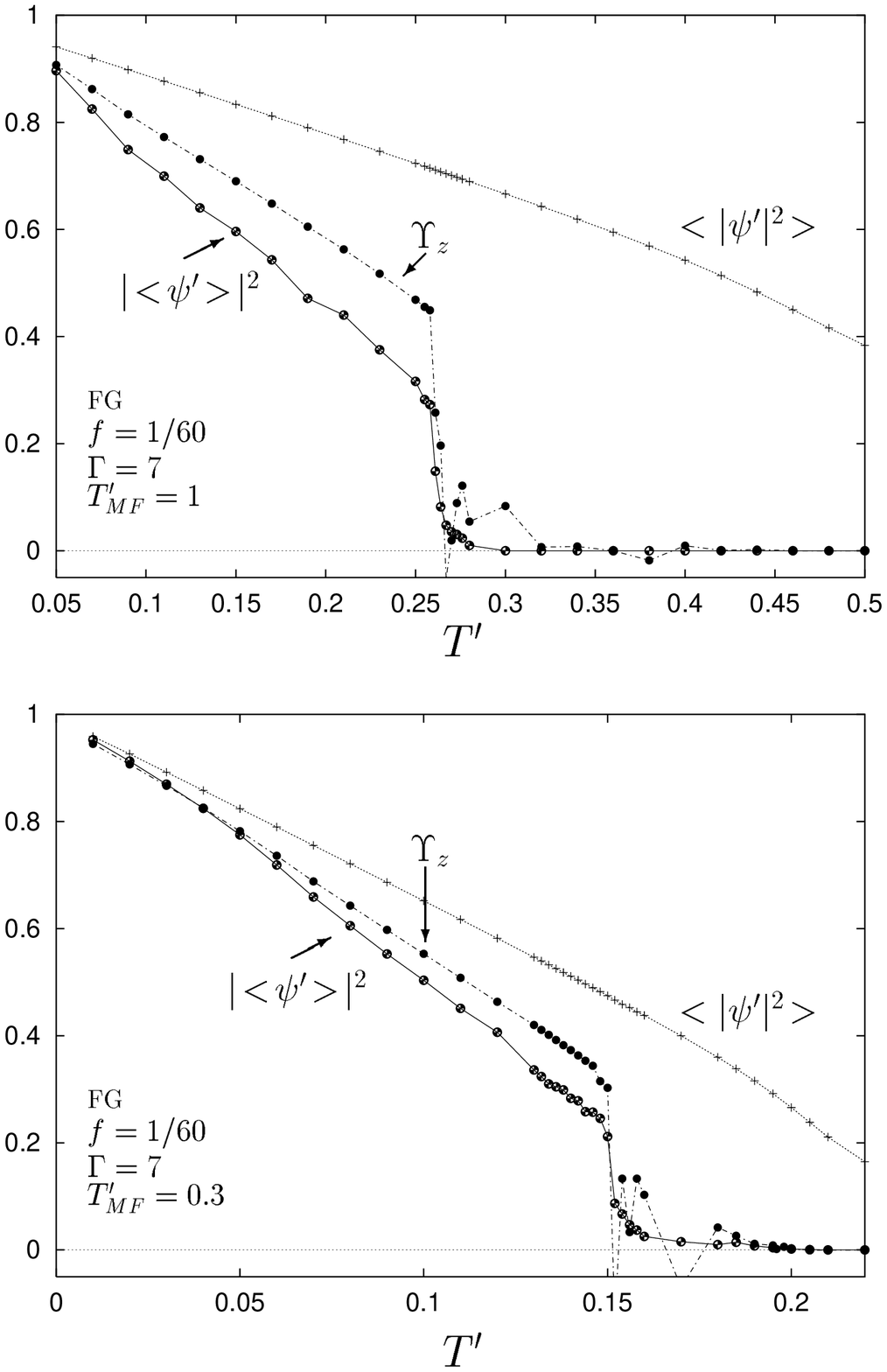}}
  \end{picture}
  {\small FIG. \ref{F60A7.PsiY.FG}. Top panel: Local Cooper-pair
    density $<|\psi'|^2>$, local condensate density $|<\psi'>|^2$ and
    helicity modulus along the field direction $\Upsilon_z$ as
    functions of temperature for the Ginzburg-Landau model in a frozen
    gauge approximation with $f=1/60$, $\Gamma = 7$, $T'_{MF} = 1$,
    $a_\mu/\xi_\mu = 6$, and ${\mathcal V} = 60^3$. $|<\psi'>|^2$ and
    $\Upsilon_z$ show a sharp drop at $T_m$, implying zero
    condensate density and lack of global phase coherence in the vortex
    liquid. \\
    Bottom panel: Same as top panel, but with $T'_{MF}=0.3$}
  \refstepcounter{figure}
\label{F60A7.PsiY.FG}
\end{figure}

In Fig. \ref{F60A7.PsiY.FG}, we show the local Cooper-pair density
$<|\psi'|^2>$, the local condensate density $|<\psi'>|^2$ and the
helicity modulus along the field direction $\Upsilon_z$ as functions
of temperature for the FG model.

We see in Fig. \ref{F60A7.PsiY.FG} that $<|\psi'|^2>$ does not show
any particular anomalous feature at the melting temperature $T_m$, it
is completely smooth. In contrast to this, $|<\!\psi'\!>|^2$ shows a
sharp drop to zero at $T_m$, which indicates a first order phase
transition at $T_m$. $<|\psi'|^2>$ shows a small tail above $T_m$.
However, note that this tail decreases towards zero for increasing
simulation length. {\em Thus, the superfluid condensate density is
  finite in the VLL phase, jumps discontinuously to zero at $T_m$, and
  is zero in the entire vortex liquid phase}.  The vortex-liquid phase
is therefore phase-incoherent, and we denote the phase as an {\it
  incoherent vortex liquid}. The behavior of the helicity modulus
shows that there is no signature of any transition from a disentangled
to an entangled {\it vortex liquid phase}, in agreement with previous
results \cite{Hu:L97,Nguyen:B98b}. In the language of the
non-relativistic $2D$ boson-analogy commonly invoked in the study of
vortex-liquids, there is no transition from a normal fluid to
superfluid bose-system at $T=0$ of the bose-system, in agreement with
general arguments given by Landau. In the absence of disorder, a
normal bose-fluid cannot exist at $T=0$ in the thermodynamic limit,
the only possible phases are bose-crystals or superfluids and hence
there should be a direct transition from an insulating bose-system
(corresponding to the Abrikosov vortex lattice phase) to a superfluid
bose-system, corresponding to an incoherent vortex liquid.

This is precisely what is observed in all our simulations, {\it and
  demonstrates that all our systems are sufficiently large in the
  $z$-direction to capture this physics correctly.} In Fig.
\ref{F60A7.PsiY.FG}, the helicity modulus along the field direction
$\Upsilon_z$ also shows a sharp drop to zero at $T_m$.  Thus, the
vortex liquid is incoherent and can not carry any supercurrent. Note
that $\Upsilon_z$ is a global quantity and says nothing about local
(in space and time) phase coherence. Thus, even when $\Upsilon_\mu =
0$, the system can exhibit local superconductivity, and diamagnetic
response is expected in the incoherent vortex liquid.

It is interesting to note that while this comes out fairly
straightforwardly in simulations on the $3DXY$ model
\cite{Hu:L97,Nguyen:B98b}, it is much more difficult to obtain using
Quantum Monte Carlo simulations of the $2D$ non-relativistic
boson-analogy of the lines-only approximation to the vortex-liquid
\cite{Nordborg:Thesis,Nordborg:B98}. This is due to the fact that the
so-called winding number, which is the appropriate quantity to
measure, is essentially inaccessible for large systems. One has to
resort to computing quantities which are not the same as the winding
number, but hoping that it is a representative of it. For a nice
discussion of this point, see Refs.
\onlinecite{Nordborg:Thesis,Nordborg:B98}. It still has not been very
well established precisely which fluctuations are responsible for
destroying longitudinal phase-coherence at the VLL melting transition.

The loss of global phase coherence in the incoherent vortex liquid
does not mean that the layers are decoupled. The correlation length
$\xi_\mu(T)$, for all directions $\mu$, jumps from infinity to a
finite value at $T_m$, and further decreases for increasing
temperature. Estimates for various correlation lengths along the field
direction can be found in Refs.
\onlinecite{Nordborg:Thesis,Olsson:cm98}.  This picture nicely
explains the experimental results found in BSCCO in Ref.
\cite{Enriquez:cm98}. Here, the difference in the high frequency
resistivity, for different sample configurations, is used to probe the
VLL melting transition. In the VLL phase $\xi_\mu$ is infinite in all
directions, and the ``rigid'' VLL does not contribute to the
dissipation when a high-frequency circulating current is applied (Fig.
2 in Ref. \cite{Enriquez:cm98}) is applied. In the vortex liquid
however, this dissipation depends on the correlation length in the
different directions. Due to anisotropy and the applied field along
the crystal's $c$-axis, $\xi_c \not = \xi_{ab}$, and the
high-frequency microwave dissipation is different for different sample
orientation. For increasing temperature, $\xi_\mu(T)$ decreases
further, and thus the dissipation rate increases as found in Ref.
\cite{Enriquez:cm98}.

In the past there have been simulations on the 3DXY model
\cite{Chen:B97b,Ryu:B98}, the Villain model \cite{Li:B93}, and the
London model \cite{Chen:L94,Nguyen:L96} that suggest a two stage
``melting transition''. Here, the VLL melts at $T_m$ into a coherent
(disentangled) vortex liquid phase, where the free vortex lines are
straight a well defined. In this phase, phase coherence along the
field direction is still intact, and does not disappear before a new
``entangled'' phase transition at $T_e > T_m$. More recently and in
this work, by longer simulation time on larger systems, it is found
\cite{Nguyen:B98a,Hu:L97,Nguyen:B98b,Koshelev:B97} that $T_e=T_m$, and
the VLL melts directly into the incoherent vortex liquid.

\subsection{Change in vortex-tangle connectivity}
We next discuss in some detail the results obtained for the quantity
$O_L$, which probes the connectivity of the vortex-tangle in extreme
type-II superconductors. We will make the following point: as for the
zero field case, the increasingly sharp change in $O_L$ from zero to
one in finite field, with increasing system size, also denotes a phase
transition where a global $U(1)$-symmetry is broken. This refers to a
$U(1)$-symmetry associated with the vortex-content of the
Ginzburg-Landau theory. As argued in Section IIF, this symmetry of the
vortex content of the theory is seen explicitly when rewriting it to a
gauge-theory involving a local complex matter-field, see Section IIF.
We also discuss the finite size effects of $O_L$, in systems with slab
geometry, i.e. where $L_x/L_z \approx L_y/L_z > 1$, as well as in
cubic systems.

\begin{figure}
  \begin{picture}(0,410)(0,0)
     \put(-65,-50)
         {\includegraphics[angle=0,scale=0.6]
         {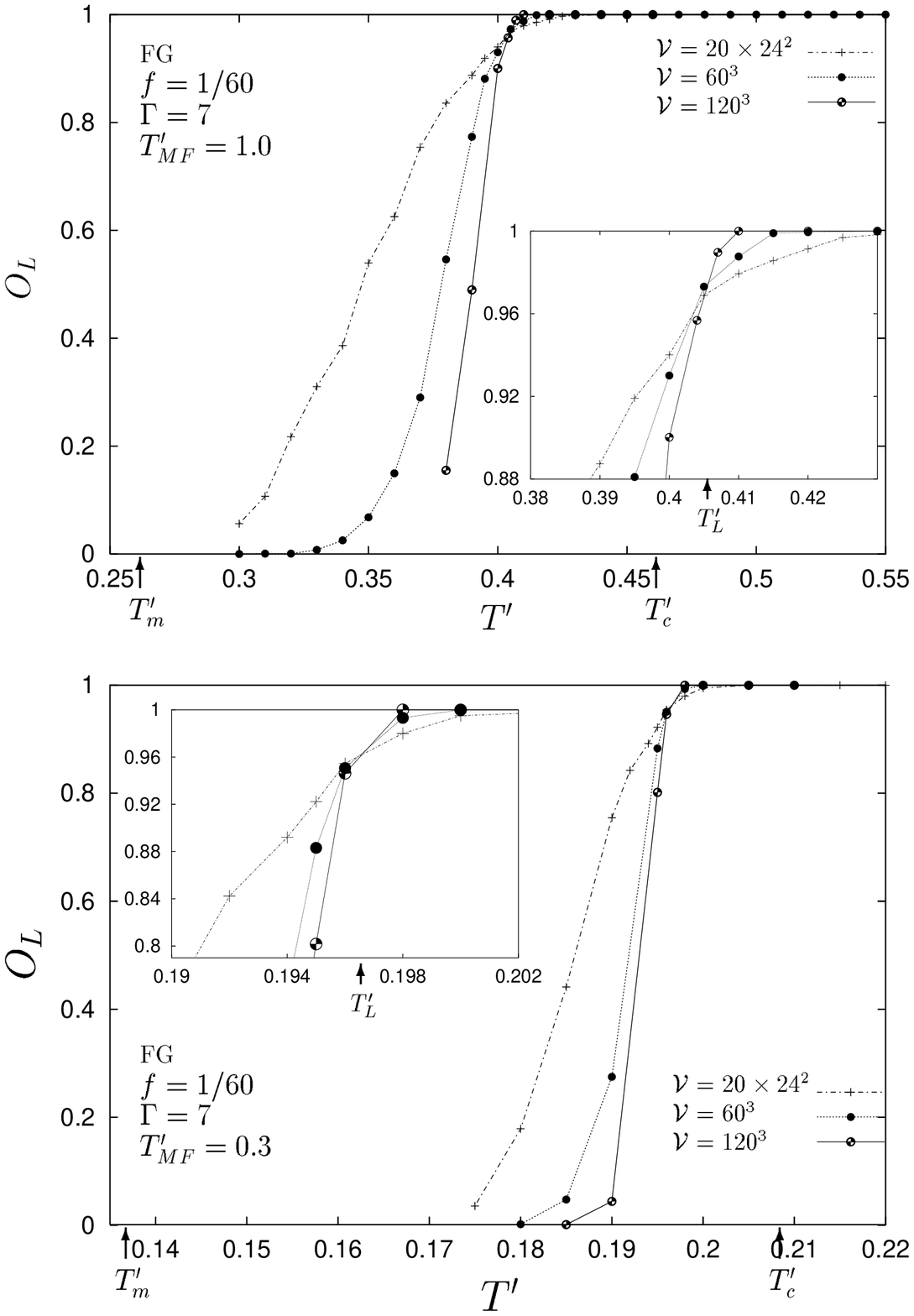}}
  \end{picture}
  {\small FIG. \ref{F60A7.OL.FG}.  Top panel: $O_L$ as a function of
    temperature for the Ginzburg-Landau model in a frozen gauge
    approximation with $f=1/60$, $\Gamma = 7$, $T'_{MF} = 1$,
    $a_\mu/\xi_\mu = 6$, and ${\mathcal V} = 20^3,60^3,120^3$. For
    increasing system size the largest temperature where $O_L=0$
    increases and the smallest temperature where $O_L = 1$ decreases.
    Thus, in the thermodynamical limit, there exists a well defined
    temperature $T_L$ where $O_L$ jump sharply from zero to one.  If
    we use the criteria $O_L \sim 0.9$, to determine $T_L$, we find
    that $T_L$ {\em monotonically decreases to a limiting value for
      increasing system sizes}. The inset shows the details of $O_L$
    close to $T_L$. Note how the curves for $O_L$ all cross at the
    same temperature with increasing $L$. Note also how the lowest $T$
    at which $O_L=1$ actually {\it decreases} with $L$. \\
    Bottom panel: Same as for top panel, but with $T'_{MF}=0.3$.}
  \refstepcounter{figure}
\label{F60A7.OL.FG}
\end{figure}

In Fig. \ref{F60A7.OL.FG} we show $O_L$ as a function of temperature
for several system sizes, ${\mathcal V} = 20^3, 60^3, 120^3$. {\it We
  see that for increasing system sizes, the largest temperature where
  $O_L = 0$ increases, while the smallest temperature where $O_L = 1$
  decreases}.

This finite-size effect in the behavior of $O_L$ can not be explained
by the $2D$ non-relativistic boson-analogy picture. In this picture
the transverse wandering $u(z)$ of a vortex lines along the field
direction ($\hat{z}$ )is given by \cite{Nelson:B89,Nordborg:Thesis}
\begin{eqnarray}
  <(u(z)-u(0))^2> = 2 D z; ~~ D = \frac{\Gamma^2 k_BT}{\varepsilon}.
\label{Diffusion}
\end{eqnarray}
We see in Eq. \ref{Diffusion} that the probability of finding {\it a
  vortex line with finite line-tension} traversing the system in a
direction perpendicular to the magnetic field, without using PBC along
the field direction, should decrease with increasing system size.
Thus, if the vortex liquid regime is always describable as a liquid of
vortex lines, then an inescapable consequence of this picture would be
that $T_L$ should shift to higher temperature with increasing system
size. This is in clear contrast to the finite size effect of $O_L$
shown in Fig. \ref{F60A7.OL.FG}. In the zero field case, $O_L=0$
indicates that the line tension of vortex loops is finite, while $O=1$
indicates that the line tension of vortex loops is zero. The
temperature where $O_L$ jumps from zero to one is the critical
temperature for the S-N phase transition.

We now focus on the inset of Fig. \ref{F60A7.OL.FG}. Note how the
curves for $O_L$ cross, {\it and reach a value $O_L=1$ for
  progressively lower temperatures as $L$ increases.} If a picture of
the vortex-liquid in terms of well-defined vortex lines with non-zero
linetension were applicable to this point, one would expect the point
$T_L$ at which $O_L$ reaches the value $1$, to move {\it monotonically
  up} with $L$.  The crossings of the curves for $O_L$ observed in the
inset of Fig.  \ref{F60A7.OL.FG} simply would not occur. {\it Note
  also the similarity of this finite-size effect, and the ones
  observed in Fig.  \ref{F1A1.Ol.FG} for the zero-field case}. There,
it was argued that such a finite-size effect was strongly indicative
of a percolation threshold for thermally induced unbound vortex-loops
in the thermodynamic limit \cite{Akao:B96,Aharony:Bo94}. The crossing
point $T_{\rm{cross}}$ seems a likely candidate for the limiting value
of $T_L$ as $L \to \infty$, see Fig. \ref{TL.L.FG} and the more
detailed discussion below. This, in our view, provides strong
numerical evidence that the progressively more abrupt change in the
connectivity of the vortex-tangle as $L \to \infty$, is a real feature
of the vortex system that survives in the thermodynamic limit, {\it
  also at a finite magnetic field}. In other words, the {\it geometric
  transition} signalled by the change in $O_L$ seems to be a real
feature and not an artifact of small systems. Whether or not it also
corresponds to a {\it finite-field thermodynamic} phase-transition
will be discussed below.

In the vortex representation, Eq. \ref{Shma} the
$U(1)$-symmetry to be broken is hidden, and can only be explored
implicitly using the conservation of $N_\mu$.  The connection is made
explicit by rewriting the vortex Hamiltonian in the disorder-field
language, see Eq.\ref{Sphiha} of Section IIF. Below $T_L$, only field
induced vortex line percolate the system. Thus, $N_x=N_y=0$ and
$N_z=fL_xL_y$. Here, $fL_xL_y$ is the number of field induced vortex
lines.  For $T > T_L$, in addition to the field induced vortex lines,
thermally excited ``vortex lines'' also exist. Thus, above $T_L$,
$N_\mu$ is not a conserved quantity and the global $U(1)$ symmetry is
broken, as for the zero field case.

In Ref. \onlinecite{Nguyen:B98b} it was claimed that because the
longitudinal superfluid density vanished precisely at the melting
line, as now found by several authors
\cite{Hu:L97,Nguyen:B98b,Chin:cm98,Olsson:cm98} including the
isotropic case, the vortex lines could not be considered well defined
in the vortex liquid phase. By itself, this is not a tenable
conclusion. Nor does it follow automatically that the vortex-lines are
entangled and that the mechanism for VLL melting is entanglement
\cite{Nonomura:cm98}. To substantiate such a claim one has to
investigate in more detail the geometric properties of the vortex
tangle in the liquid phase, as done above and in Refs.
\onlinecite{Nguyen:cm98,Chin:cm98}.  Even if it should turn out that
the loss of longitudinal superfluid density is entanglement it is
probably more appropriate to view the entanglement as triggered by VLL
melting transition rather than the converse. However, it is worth
while pointing out at this stage that there is now consensus on the
fact that at intermediate fields and above, the VLL melts into an
incoherent vortex-liquid and that there does exist a regime where the
molten phase consists of intact vortex lines, remarks to the contrary
in Ref. \onlinecite{Nguyen:B98b} not withstanding. Moreover, various
Monte-Carlo simulations agree that the Lindemann-criterion for VLL
melting applies in this regime
\cite{Nguyen:B98b,Nordborg:B98,Nguyen:cm98,Nonomura:cm98}. In the
low-field regime, far less consensus has so far been reached.
Therefore, the question of whether vortex loops influence VLL melting
or not, and whether there exists a genuine transition line $T_L(B)$
inside the vortex liquid, are two separate issues.

\begin{figure}
  \begin{picture}(0,350)(0,0)
     \put(-35,-20)
         {\includegraphics[angle=0,scale=0.47]
         {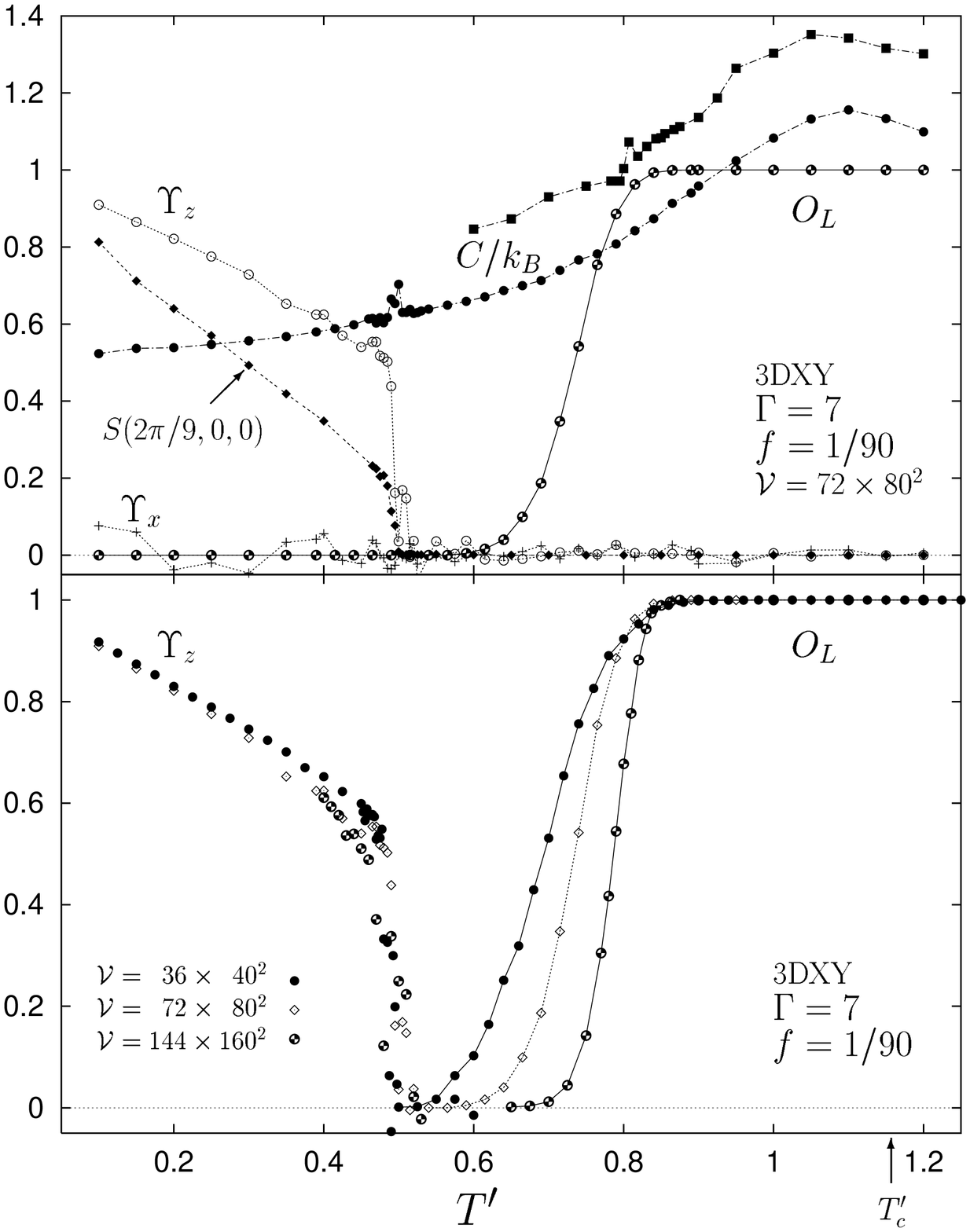}}
  \end{picture}
  {\small FIG. \ref{F90A7.Combine.FG}.  Top panel: Helicity moduli
    $\Upsilon_x$ and $\Upsilon_z$, structure function $S({\mathbf
      K},k_z \keq 0)$ and $O_L$ for the $3DXY$-model as a function of
    temperature for $f=1/90$, $\Gamma = 7$, and system size ${\mathcal
      V} = 72 \times 80 \times 80$.\\
    Bottom panel: $\Upsilon_z$ and $O_L$ for increasing system sizes.
    For increasing system size the largest temperature where $O_L=0$
    increases and the smallest temperature where $O_L = 1$ decreases.
    Thus, in the thermodynamical limit, there exists a well defined
    temperature $T_L$ where $O_L$ jumps sharply from zero to one,
    precisely as seen in the zero-field case.  Shown is also specific
    heat for a system of size ${\mathcal{V}}=360^3$ (shifted up by
    $0.2k_B$ for clarity).  } \refstepcounter{figure}
\label{F90A7.Combine.FG}
\end{figure}

We hereafter focus on simulation results obtained for the
$3DXY$-model.  There is no qualitative difference between the results
for this model, and the Ginzburg-Landau model. In Fig.
\ref{F90A7.Combine.FG} we show results for the $3DXY$ model $f=1/90$,
$\Gamma=7$. The top panel shows structure factor, superfluid density
along the field, specific heat and $O_L$ for a system of size
$72 \times 80 \times 80$. The bottom panel shows a sharpening of $O_L$
for increasing system sizes. The trend in the change in the
vortex-tangle connectivity is precisely the same as that seen for
$f=1/60$ within the Ginzburg-Landau model including amplitude
fluctuations.  The lowest temperature at which $O_L$ rises from zero,
increases with system size, {\it but the highest temperature at which
  it reaches the value $O_L=1$ decreases with system size}. Again, we
find a feature which indicates that a change in the vortex-tangle
connectivity is undergoing a change.

\subsection{Effect of varying system aspect ratio}

According to the $2D$ non-relativistic boson-analogy of the
vortex-liquid, $T_L$ should be proportional to the aspect ratio
$L_x/L_z$ \cite{Nordborg:pc98}.

\begin{figure}
  \begin{picture}(0,410)(0,0)
     \put(-65,-50)
         {\includegraphics[angle=0,scale=0.6]
         {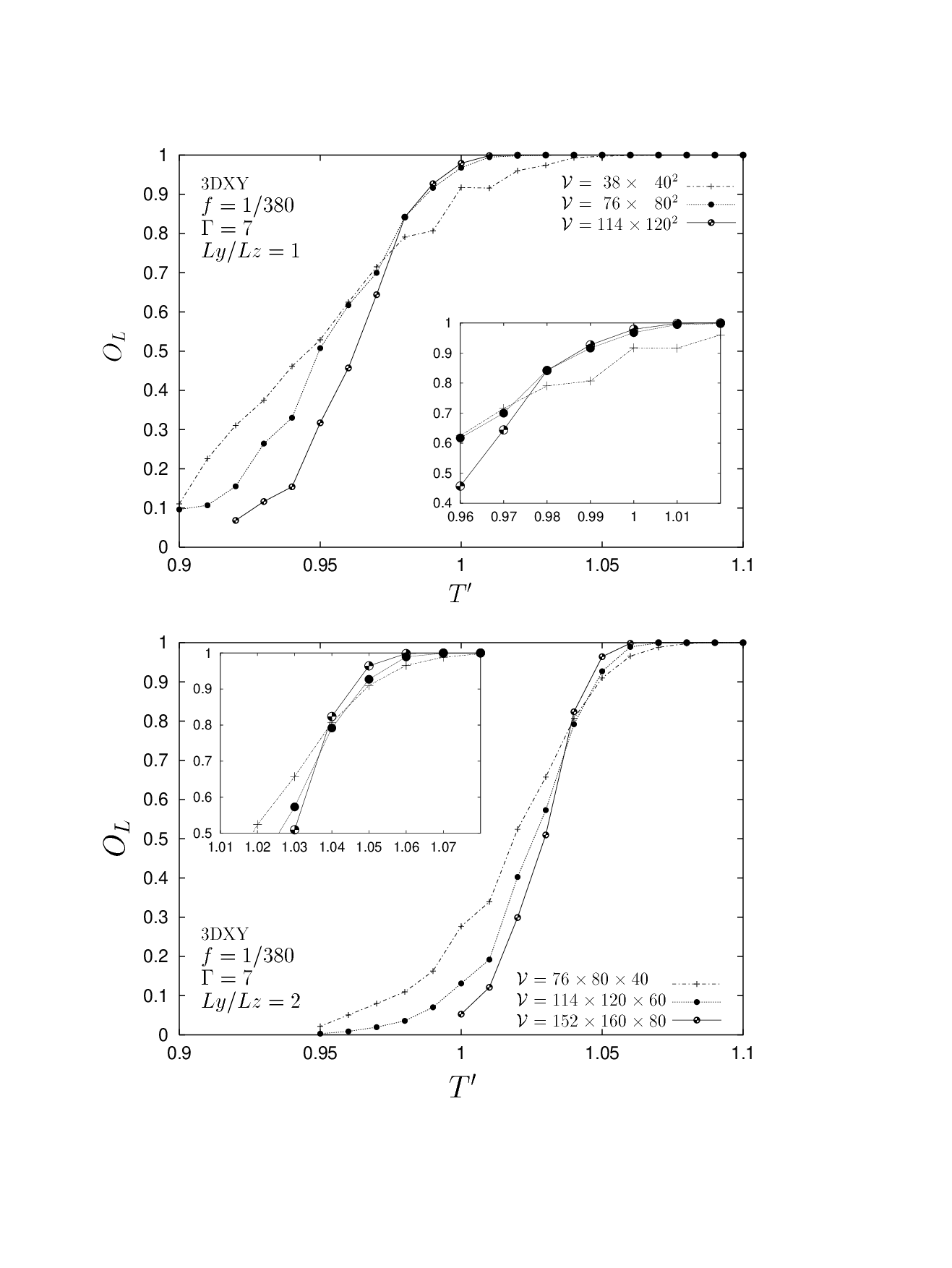}}
  \end{picture}
  {\small FIG. \ref{F380A7.Ol.FG}.  $O_L$ as a function of system
    temperature, obtained within the $3DXY$-model for $f=1/380$,
    $\Gamma=7$, for various system sizes. Top panel: Aspect ratio
    $L_y/L_z=1$. \\
    Bottom panel: aspect ratio $L_y/L_z=2$. Insets show details of the
    curve-crossings close to $T_L$. Note that while the lines-only
    approximation would predict a change in the crossing temperature
    of roughly a factor $2$, they only change by about $5\%$, which is
    within the uncertainty of the estimate for the crossing
    temperature.  } \refstepcounter{figure}
\label{F380A7.Ol.FG}
\end{figure}

To further investigate the possibility of a breakdown of vortex-line
physics inside the vortex-liquid regime, we consider the crossing
feature found in $O_L$ in more detail for various aspect ratios
$L_x/L_z$ of the systems on which the simulations are done. In this
paragraph, we carry out the simulations using the $3DXY$-model with
the parameters $f=1/380$ and $\Gamma=7$. We have varied the field to
illustrate that the features of $O_L$ are the same as for the higher
fields, but do become sharper. Furthermore, comparing the results
obtained from the FG model to the results obtained from the
$3DXY$-model, we again find that these models give qualitatively the
same results when parameters are comparable.

\begin{figure}
  \begin{picture}(0,410)(0,0)
     \put(-65,-50)
         {\includegraphics[angle=0,scale=0.6]
         {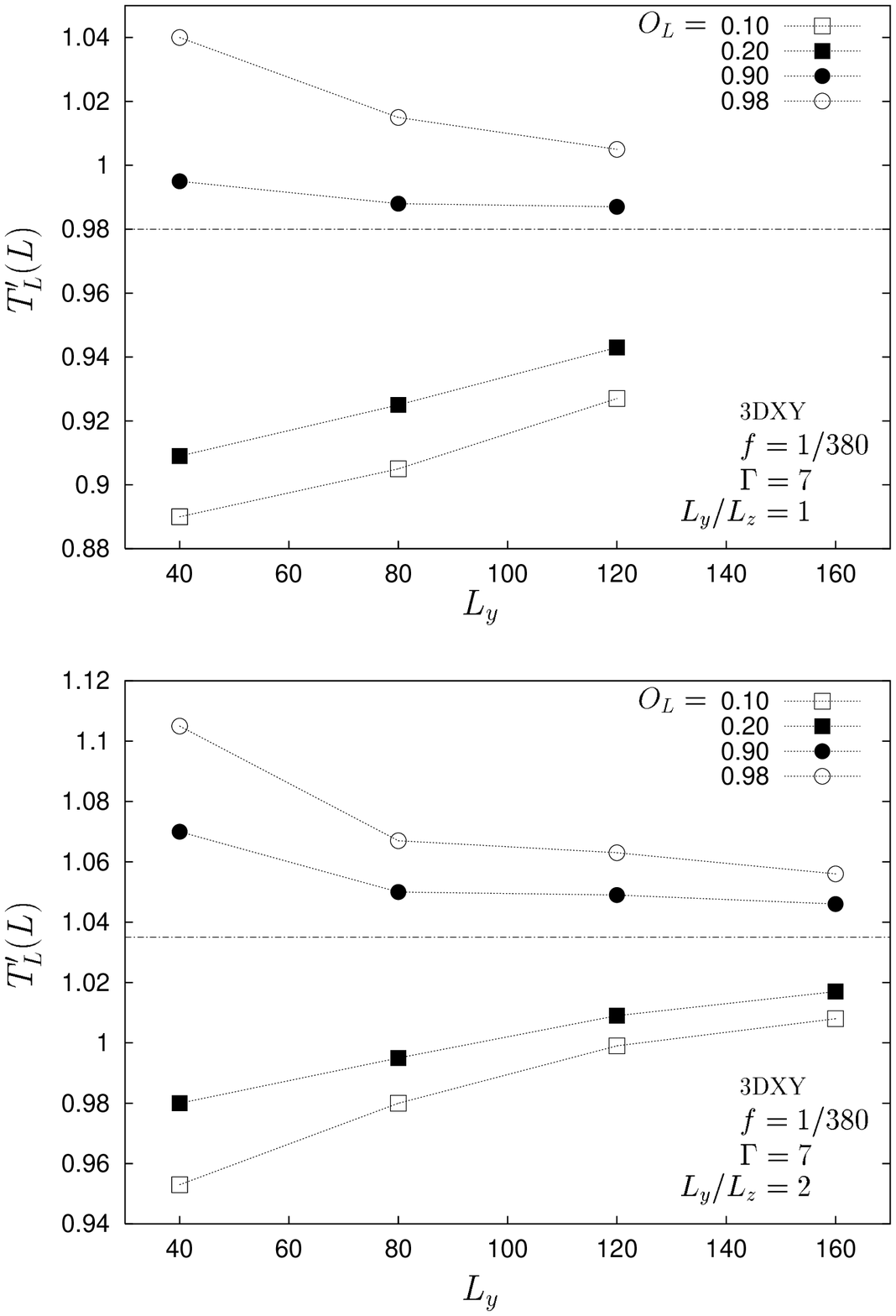}}
  \end{picture}
  {\small FIG. \ref{TL.L.FG}.  Finite-size effect in $T_L$ as obtained
    for the $3DXY$-model using four sets of defining criteria for
    $T_L$, $O_L=(0.10,0.20,0.90,0.95)$. Each of the criteria gives
    converging curves for $T_L(L)$, whose limiting values are
    estimated by the crossing temperatures in the insets of Fig.
    \ref{F380A7.Ol.FG}. Top panel: aspect ratio $L_y/L_z=1$. \\
    Bottom panel: aspect ratio $L_y/L_z=2$. Note that the limiting
    values for $L_y/L_z=1$ and $2$ differ by about $5\%$, whereas
    according to a vortex-line liquid picture, they should differ by
    about a factor of $2$.  } \refstepcounter{figure}
\label{TL.L.FG}
\end{figure}

In Fig. \ref{F380A7.Ol.FG}, we show $O_L$ as a function of $T$ for the
$3DXY$-model with parameters $f=1/380$, $\Gamma=7$, for a system of
size $L_x \sim L_y = L_z$ in one case, and $L_x \approx L_y$ and
$L_z=0.5L_y$ in the other case. Using the crossing temperature in the
insets as an estimate for the temperature $T_L$ in the thermodynamic
limit, as for the zero-field case, we see that this temperature
changes very little when changing the aspect ratio by a factor of $2$.
This indicates that in the thermodynamic limit there is only one
$T_L$, regardless of the system shape. What these results indicate, is
that the expectation one has based on the 2D non-relativistic
bose-analogy of the vortex liquid, namely that $T_L$ should scale with
$L_x/L_z$, is not borne out. Note that the present case is very
different from the situation encountered in the $3D$ Ising-model where
a percolation threshold for overturned spins in an ordered spin state
is found at a temperature which is lower than the critical temperature
\cite{Footnote:Ising}.

This may be further illustrated by considering the finite-size effect
of $T_L$, for two different aspect ratios $L_x/L_z = 1$ and
$L_x/L_z=2$. We investigate this by defining $T_L$ by four sets of
criteria, namely the temperature at which $O_L=(0.10,0.20,0.90,0.95)$.
If the curves for $O_L$ sharpen up, as seen in the above results, it
is ultimately immaterial what sets of criteria are being used. The
sets will give converging curves for $T_L(L)$, one coming up from
below and one coming down from above, see Fig. \ref{TL.L.FG}. We may
use the best estimate for the crossing temperatures in Fig.
\ref{F380A7.Ol.FG} as an estimate for what the limiting value of $T_L$
will be in the thermodynamic limit. These results illustrate two
important points, namely i) $T_L$ does not move up monotonically with
system size, but saturates at a specific value as $L \to \infty$
precisely as for the zero-field case, and ii) the limiting value of
$T_L$ is independent of aspect ratio. Both of these two points
contradict expectations based on a vortex-{\it line} liquid picture of
the molten phase of the Abrikosov VLL.

\subsection{Scaling of the melting line $T_m(B)$}

\begin{figure}
  \begin{picture}(0,410)(0,0)
     \put(-65,-50)
         {\includegraphics[angle=0,scale=0.6]
         {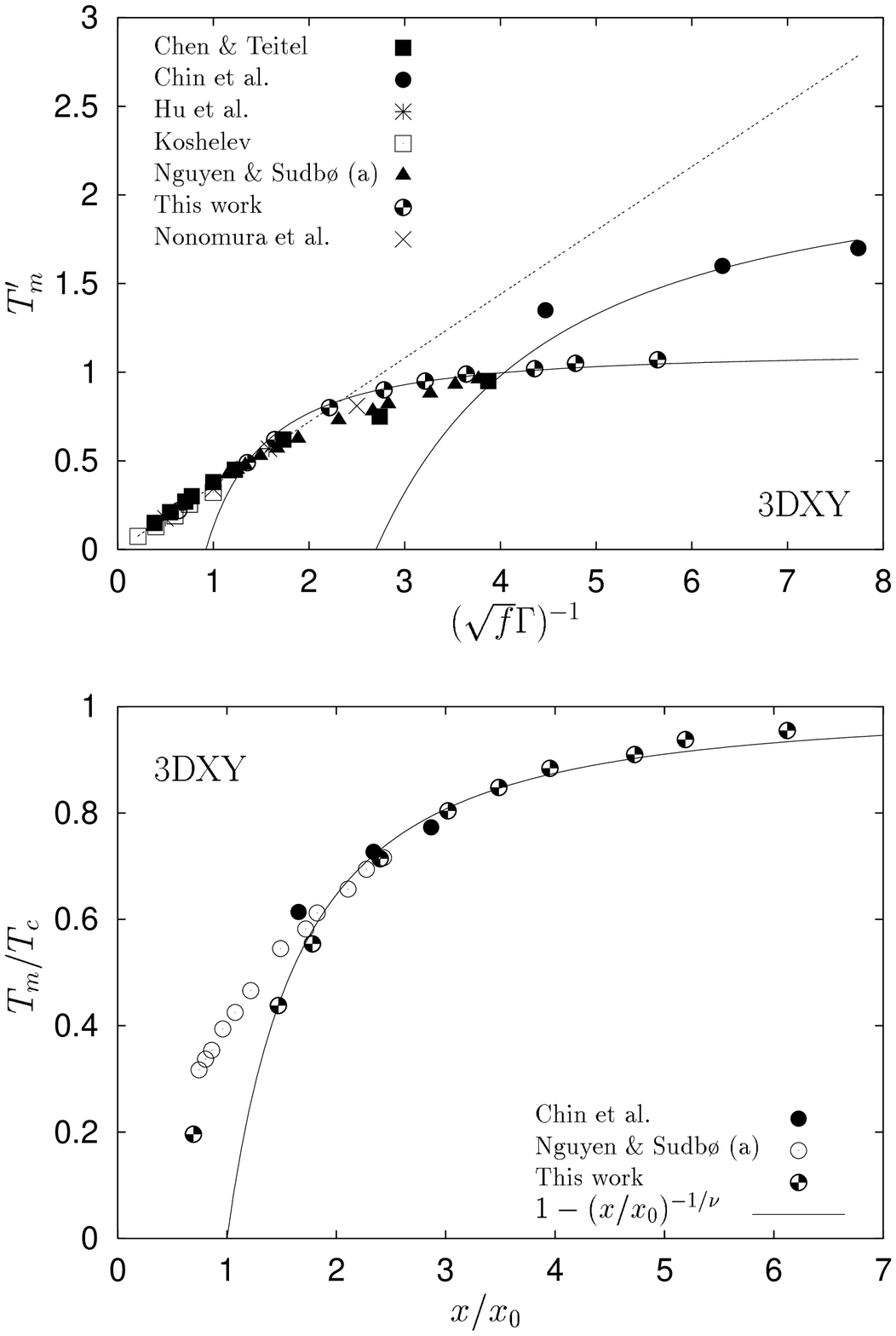}}
  \end{picture}
  {\small FIG. \ref{Scaling.Tm.FG} Top panel: Melting temperature
    $T'_m(B)$ of the vortex lattice as a function of $y=1/\sqrt{f}
    \Gamma$. At large enough filling fractions, $y<2.0$, $T'_m(B)$
    obtained from various simulations on the $3DXY$-model and
    boson-analogues of the vortex system, agree and is well described
    by $T'_m(B)=0.41y$, the dotted line.  At low filling fractions,
    $y>2.0$, there is a crossover to $3DXY$ critical-scaling of
    $T_m(B)$. The solid lines through the two sets of data points are
    $3DXY$ critical-scaling functions, described in the text.\\  
    Bottom panel: Normalized melting temperature $T'_m(B)/T'_c$ as a
    function of the varibale $x/x_0$, where $x=1/\sqrt{f}$ and
    $x_0$ is an anisotropy dependent fitting parameter. Solid line is
    the $3DXY$ scaling function $h(x) = 1-(x/x_0)^{-1/\nu}$, where
    $\nu=0.67$.}
  \refstepcounter{figure}
\label{Scaling.Tm.FG}
\end{figure}

In Fig. \ref{Scaling.Tm.FG}, we show data from various simulations, of
the vortex lattice melting line $T_m(B)$. We want to emphasize the
fact that there are {\it two} distinct scaling regimes for the melting
line $T_m(B)$, one at high fields which we somewhat arbitrarily denote
high-field scaling regime \cite{Koshelev:B99}, and one at low magnetic
fields which we identify to be $3DXY$-scaling.

The dotted straight line is the curve given by \cite{Koshelev:B99}
$k_B T_m(B)/J_0 = 0.41 ~ y; ~~y^2 < 4$, where $y = 1/\sqrt{f} \Gamma$.
It describes the published numerically obtained melting lines for
large enough filling fractions $y^2 < 4$ or so well, in our case with
$\Gamma=7$ corresponding to approximately $f > 1/200$. On the other
hand, for $y^2 > 4$, clear deviations from linear behavior is seen.

The melting curves obtained for $\Gamma=1$ in Ref.
\onlinecite{Chin:cm98} shown by the filled circles, and for $\Gamma=7$
in Ref. \onlinecite{Nguyen:B98b} shown by the half-filled circles
saturate at low filling fractions $f$ to the values given by the
zero-field critical temperature, $T_c$. For $\Gamma = 1$, we have $k_B
T_c/J_0 =2.2$, while for $\Gamma=7$ we have $k_B T_c/J_0=1.12$
\cite{Nguyen:cm98}. The data given by the filled triangles
\cite{Nguyen:B98a} are obtained on the $3DXY$-model with an anisotropy
parameter $\Gamma=3$. As $\Gamma$ increases from $1$, the zero-field
transition temperature $T_c$ rapidly approaches its $2D$ value,
although the transition is always $3D$ in character for finite
anisotropy.  Hence, the results from the anisotropic $3DXY$-model with
$\Gamma=3$ \cite{Nguyen:B98a} are very close to those of the
$3DXY$-model with $\Gamma=7$, see Ref.  \onlinecite{Nguyen:B98b}.

The results of Ref. \onlinecite{Chen:B97a}, obtained by fixing
$f=1/15$ and varying $\Gamma \in (1,..10)$, agree entirely with our
results of Refs.  \onlinecite{Nguyen:B98b,Chin:cm98,Nguyen:cm98} in
the low-field regime $1/\sqrt{f} \Gamma > 2$.  The significance of all
these three sets of points is that they fall significantly below the
straight line obtained from the high-field scaling of the melting
line.

Note also that even if we normalize the melting line $T_m(B)$, quite
arbitrarily, with a factor $1/(1-T_m/T_c)^{2 \nu}$
\cite{Koshelev:B99}, this might take out the strong downward curvature
of the data in the top panel of Fig. \ref{Scaling.Tm.FG}, but there is
absolutely no reason for why the slope of the resulting curve in the
low field regime, which would be a straight line, should be the same
as in the high-field regime.

Assuming $3DXY$-scaling for the melting line when $y >> 2$, i.e.
$B/|1-T/T_c|^{2 \nu} = B_0$ where $B_0$ is a field-scale that depends
on anisotropy, we find $k_B T_m(B)/J_0=(k_B T_c/J_0) ~ [1-(x_0/ \Gamma
y)^{1/\nu}]$ on the melting line, and where the last term is
negligible for low fields. Hence, we find that the melting line
saturates to the true critical temperature $T_c$, as it obviously
must. The dotted straight line $k_B T_m/J_0=0.41 y$, overshoots $T_c$
as the field is lowered. The Monte-Carlo results follow this line at
large fields, {\it but are however starting to be arrested in their
  tracks by the zero-field vortex-loop critical fluctuations already
  at around} $y=2$, thus crossing over to $3DXY$ critical scaling, as
our Monte-Carlo simulations results show.

In the top panel of Fig. \ref{Scaling.Tm.FG}, we have drawn the
function $T_m(B)/J_0 = (T_c/J_0) [1 - (x_0/y
\Gamma)^{1/\nu}]$ through the two sets of points obtained from
Monte-Carlo simulations for $y > 4$ and $\Gamma=1,7$, given by filled
and half-filled circles, respectively. Using $x_0=2.70$ for $\Gamma=1$
and $x_0=6.45$ for $\Gamma=7$, we find that the $3DXY$ scaling
function given above fits the Monte-Carlo data well for $y > 4$, while
the high-field scaling is excellent for $y < 2$. Note how vastly
different the scaling of $T_m(B)$ in the two regimes $y < 2$ and $y >
4$ is.

The bottom panel of Fig. \ref{Scaling.Tm.FG} shows the low-field
melting line $T_m(B)$ normalized by the zero-field critical
temperature, obtained from simulations of the $3DXY$ model with
$\Gamma=1, k_B T_c/J_0=2.2$ in Ref. \onlinecite{Chin:cm98}, $\Gamma=3,
k_B T_c/J_0=1.34$ in Ref. \onlinecite{Nguyen:B98b}, and $\Gamma=7, k_B
T_c/J_0=1.12$ in this work and in Ref. \onlinecite{Nguyen:cm98},
plotted in terms of the variable $x/x_0$, where $x = 1/\sqrt{f} $ and
$x_0$ is a fitting parameter for each $\Gamma$.  The corresponding
values of $\Gamma$ and $x_0$ are $(1,2.70)$, $(3, 4.65)$, and
$(7,6.45)$.  For $x/x_0 \approx 2$ or less, i.e. at large enough
fields, we see that deviations from $3DX$-scaling occur. For $\Gamma
=(1,3,7)$ this corresponds to $1/f=(30,90,160)$, respectively. The
line through the low-field data, is the $3DXY$ scaling function
$1-(x/x_0)^{-1/\nu}$. Notice the sharp bending of the $3DXY$-scaling
function as $x/x_0$ increases beyond the value $3$, and how the
available numerically obtained melting curves follow this line. This,
in our view, provides strong numerical support for the notion that at
low filling fractions $f \Gamma^2 << 1$, the melting line $T_m(B)$
obeys $3DXY$ critical scaling, while it follows follows a quite
different ``mean-field'' type of scaling $T_m(B) \sim 1/\sqrt{B}$ at
large fields \cite{Koshelev:B99}.

The field above which deviations from $3DXY$ critical scaling is seen,
increases as the anisotropy goes down. This is due to the fact that
with increasing $\Gamma$, the melting curve becomes flatter at low
fields, leaving the critical region more rapidly as $\Gamma$ increases.
The width of the critical region appears to widen only slightly with
increasing anisotropy \cite{Nguyen:B98b}, whereas the melting line
rapidly becomes flatter at low fields, such that this is the dominant
effect in determining the field at which the melting line enters the
critical region.

\subsection{Phase diagram, clean limit}
\begin{figure}
  \begin{picture}(0,230)(0,0)
     \put(-20,-105)
         {\includegraphics[angle=0,scale=0.43]
         {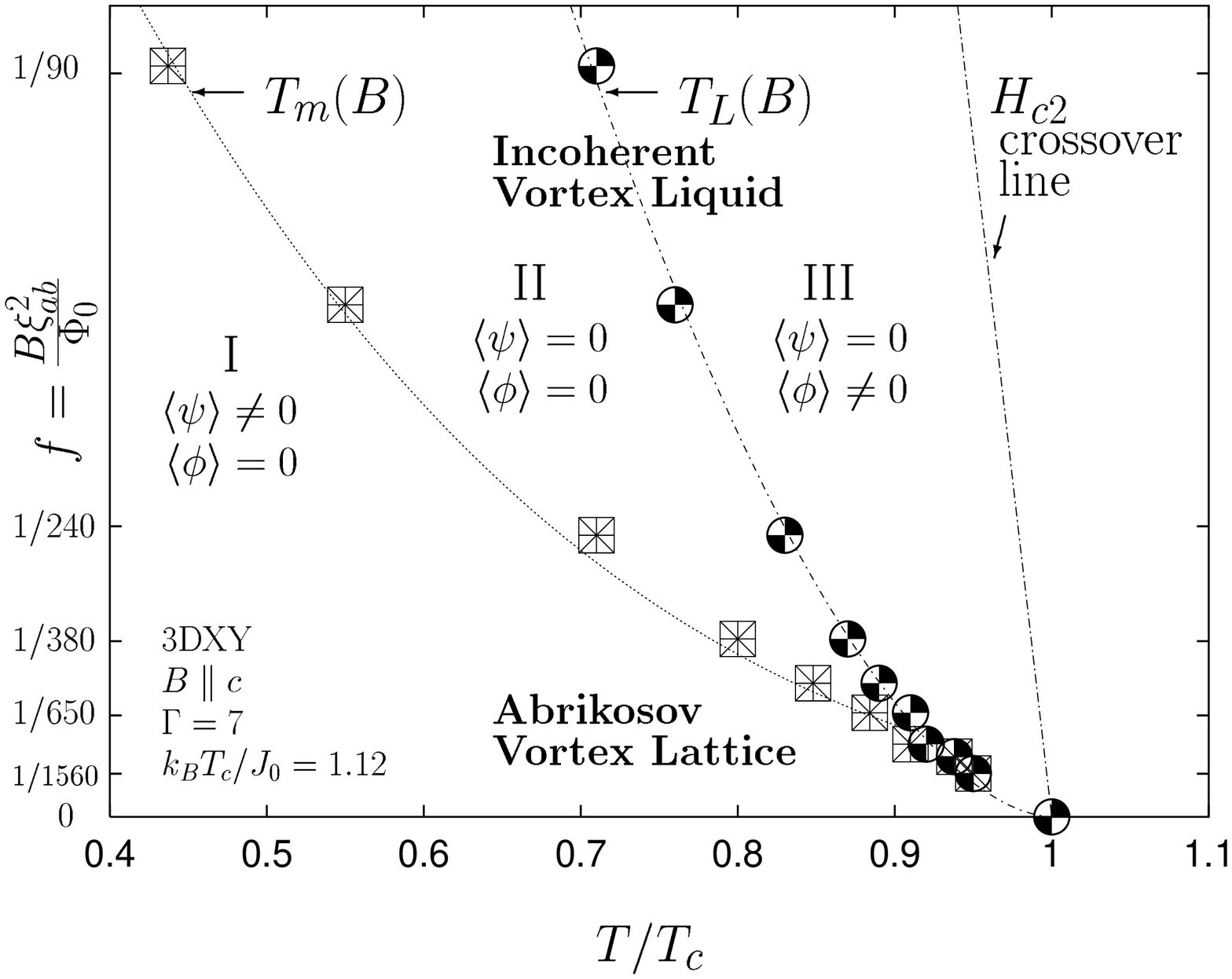}}
  \end{picture}
  {\small FIG. \ref{Phase.Diagram}.  $B-T$ phase-diagram for extreme
    type-II superconductors based on Monte-Carlo simulations of the
    $3DXY$-model with $B \parallel c$ and $\Gamma=7$. The phase-diagram
     splits into three differient regimes I, II, and III, characterized
     by the values of the Ginzburg-Landau and dual order parameters. }
  \refstepcounter{figure}
\label{Phase.Diagram}
\end{figure}

A summary of all of the above is contained in Fig.\ref{Phase.Diagram},
we have included results from filling fractions $1/f \in [90,..1560]$.
The results we have obtained pertain to an extreme type-II
superconductor in the absence of disorder, since we are primarily
interested in the intrinsic properties of this phase-diagram excluding
the severe complications due to disorder. There is a low-temperature
vortex-line lattice phase. When the vortex lattice melts, it melts
directly into an incoherent vortex-liquid with zero longitudinal
superfluid density. The transverse superfluid density has been
eliminated at temperature far below those where the VLL melts, by
choosing low enough filling fractions to eliminate an unwanted
commensuration effect due to the presence of the numerical lattice on
which the theory is defined.

At zero magnetic field, we have demonstrated that an alternative way
of describing the superconductor-normal metal transition, in addition
to the phase-disordering picture using the Ginzburg-Landau order
parameter, is in terms of an unbinding of vortex-loops. We emphasize
that although the quantity $O_L$ we have focused on is not an order
parameter, it may be {\it tied} to an order via the discussion in
Section IIF. By including amplitude fluctuations explicitly in the
Ginzburg-Landau theory, it is shown that this vortex-loop unbinding
does not lead to critical amplitude fluctuations. A generalized
``stiffness" characterizing the low-temperature phase which vanishes
at the transition, is the long-wavelength vortex-line {\it tension}
$\varepsilon(T)$, or equivalently the free energy per unit length of
the the thermally induced vortex loops of the system.

In a finite magnetic field, we find indications of a change in the
vortex-tangle connectivity across the system at a temperature
$T_L(B)$, whose zero-field endpoint is $T_c$. This has been done by
monitoring the quantity $O_L$ in the Ginzburg-Landau theory or the
$3DXY$-model in the same way as for the zero-field case. {\it $O_L$
  has precisely the same characteristica at finite fields and zero
  field.} In the regime $T_m(B)<T<T_L(B)$, the connectivity across the
system of the vortex-tangle of the molten phase is given entirely by
the field induced vortex lines. This appears to change across the line
$T_L(B)$. We have been able to tie $O_L$ to an order parameter even at
finite field, see Section IIF, involving a breaking of a
$U(1)$-symmetry across the line $T_L(B)$.

At low magnetic fields, to the accuracy of our calculations, we have
found that the VLL melting line and the line $T_L(B)$ {\it merge} at
low fields. Below these low magnetic fields, the picture of the molten
phase as a vortex-{\it line} liquid appears to be questionable. For
fields well above the point where $T_m(B)$ and $T_L(B)$ merge, we have
found that the position of the VLL melting line is well described by a
Lindemann criterion with Lindemann number $c \approx 0.25$, estimated
from the Debye-Waller factor.

Note that the rewriting of the theory Eq. \ref{Shma} to Eq.
\ref{Sphiha} is exact.  The onset of the expectation value $<\phi>$
takes place when vortex-loops unbind.  Moreover, the theory Eq.
\ref{Sphiha} exhibits an explicit $U(1)$-symmetry.  When this
connection is made, it seems very reasonable to tie the observed
change in the vortex-tangle connectivity to a vortex-loop unbinding
and hence an onset of $<\phi>$, {\it i.e. the order parameter and the
  symmetry being broken in the transition, have been identified.}

\section{Summary and discussion}

We have explored the $(B,T)$ phase diagram for extreme type II
superconductors using two simplified versions of the the
Ginzburg-Landau model: i) The frozen gauge (FG) approximation where
the gauge-field is fixed, while the phase and amplitude of the
superconducting order parameter are allowed to fluctuate, and ii) the
uniformly frustrated 3D XY model where only phase fluctuations are
allowed for. The former is obviously a more general model than the
latter, while the latter is a commonly accepted model in the studies
of fluctuation effects in extreme type-II superconductors. Our results
show that in the $\kappa \to \infty$-limit, where suppression of
gauge-fluctuations is an exact feature of a superconductor, amplitude
fluctuations are completely dominated by phase-fluctuations over a
{\it sizeable temperature regime}. The local order field
$<\psi'({\mathbf r})>$, as well as the helicity modulus (global phase
stiffness) $\Upsilon_\mu$ develop an expectation value for $T < T_c$,
and explicitly break the usual $U(1)$-symmetry present in the
Ginzburg-Landau theory. In contrast to this, the local Cooper-pair
density $<|\psi'|^2>$, is finite both above and below $T_c$.
Our precise calculations close to $T=T_c$ has brought out
clearly its singular temperature derivative at $T=T_c$. Below, we 
list the main results of this paper.

\underline{${\bf B = 0}$} \\
$\bullet$ In zero field, we have shown that the superconducting-normal 
metal phase-transition is described by a vortex-loop unbinding. This
is achieved by correlating a detailed study of qualitative changes in 
the vortex-loop distribution function $D(p)$ with calculations of 
superfluid density, condensate density, specific heat, amplitude 
fluctuations, and change in vortex-tangle connectivity, both including 
and excluding amplitude fluctuations of the Ginzburg-Landau  
order-parameter. The topological phase-fluctuations destroying the 
superconducting phase-coherence are thus unambiguously identified as 
thermally induced vortex loops. When amplitude fluctuations are included
explicitly, they are found to be far from critical. In other words, 
the vortex-loop unbinding may {\it not} be viewed as a reparametrization 
of critical amplitude fluctuations of the Ginzburg-Landau
order parameter, as is sometimes claimed.  

$\bullet$ The {\it vortex-content of the Ginzburg-Landau theory},
formulated in SectionIIE, is characterized by its own $U(1)$-symmetry
which becomes explicit on a further {\it exact} reformulation of the
vortex sector in terms of a new gauge-field, see Section IIF. The
low-temperature phase of the vortex-sector of the Ginzburg-Landau
theory, where all vortex-loops are confined, exhibits a
$U(1)$-symmetry. This symmetry of the vortex sector reflects the fact
that there is a number conservation of vortex loops extending across
the entire superconductor. In zero magnetic field, the conserved
number is zero, and the distribution function for closed vortex loops
of perimeter $p$ is an exponential function with length scale given by
$L_0(T)=k_B T/\varepsilon(T)$, where $\varepsilon(T)$ is the
vortex-line tension.  At the zero-field critical temperature, we find
vortex-loops with an algebraic distribution of perimeters, concomitant
with an abrupt change in the connectivity of the vortex tangle in
extreme type-II superconductors. The vortex line tension is found to
vanish as a power law as $T_c$ is approached from below,
$\varepsilon(T) \sim |T-T_c|^{\gamma}$, with $\gamma = 1.45 \pm 0.05$.

$\bullet$ {\it Both the change in the distribution function of
closed vortex loops, and the abrupt change in the connectivity of
the vortex tangle, shows that there is a diverging length in the
problem}, i.e. $L_0(T) \to \infty; ~ T \to T_c^{-}$.  At this point,
the number of closed vortex-loops extending through the system is no
longer a conserved number equal to zero, it becomes finite and
undergoes thermal fluctuations. Therefore, the $U(1)$-symmetry
characterizing the low-temperature vortex-phase is broken, the
vortex-system has suffered a vortex-loop blowout, or unbinding.

$\bullet$ The connection between the power-law behavior close to $T_c$
of the vortex-line tension and the anomalous dimension $\eta_{\phi}$
of the dual field $\phi$ was discussed in Section IV C.  Relating the
power-law for the vortex-line tension to the susceptibility exponent
$\gamma$ of the $\phi$-field, in conjunction with the Fisher scaling
law $\gamma=\nu(2-\eta_{\phi})$, allowed us to extract the value
$\eta_{\phi} = -0.18 \mp 0.07$. This result was compared to
renormalization group calculation performed directly on the dual
theory, for which the vortex-loop unbinding {\it is} the
phase-transition, and excellent agreement was found. Note the negative
sign of $\eta_{\phi}$ in the extreme type-II case.

\underline{${\bf B \neq 0}$} \\ $\bullet$ In finite field, we have
studied the phase-diagram over a wide range of filling fractions $f$,
corresponding to $1/f \in [20,...,1560]$.  Only a subset is shown
explicitly, but all results are summarized in Fig.
\ref{Phase.Diagram}. The VLL is found to melt in a first order phase
transition, for all filling fractions considered, into a {\it
  completely incoherent} vortex liquid characterized by zero global
phase coherence in all directions.  At intermediate fields, the VLL
melts into a liquid of vortex lines, whose position in the
$(B,T)$-phase diagram is well estimated by the Lindemann criterion
with a Lindemann number $\approx 0.25$.

$\bullet$ We have performed a scaling analysis for the melting line
for all filling fractions considered. We find a crossover from
mean-field type scaling at elevated fields to $3DXY$-scaling behavior
at small fields, showing that for the aniosotropies we have
considered, the melting line of the vortex-lattice at low fields is
significantly affected by zero-field critical fluctuations in a
sizeable region of the phase-diagram.

$\bullet$ Significantly, in addition to the VLL melting transition
line $T_m(B)$, we find indications of another transition line,
$T_L(B)$, inside the vortex liquid. This line is the finite-field
extension of the zero-field vortex-loop unbinding, and has an endpoint
which is the zero-field critical temperature $T_c$. Below $T_L$,
connectivity of the vortex-system is determined exclusively by the
field-induced vortex lines. All vortex-lines threading the entire
system are field-induced. Above $T_L$, this changes, as discussed in
Section IVB.  Above $T_L(B)$ there exist vortex lines that thread the
systen also perpendicular to the magnetic field, without using
periodic boundary conditions along the $z$-axis.

$\bullet$ We have performed a large-scale study of the finite-size
effects in $T_L$, {\it and found that the temperature where the
  vortex-tangle connectivity changes does not move up with system
  size, like it would have done in a vortex-line liquid}. In the $2D$
non-relativistic boson-analogy of vortex-liquid, such
vortex-configurations are never found. The symmetry broken at $T_L$ is
a global $U(1)$ symmetry, associated with the number-conservation of
vortex paths threading the entire system, the considerations are
similar to the zero-field case.  In a finite magnetic field, {\em this
  symmetry is hidden in the usual Ginzburg-Landau local order field
  representation}, but is brought out by a dual description of the
Ginzburg-Landau theory. In zero field, $T_L$ and $T_c$ are identical
and there is only one phase transition.

$\bullet$ We have found that the vortex-system in the clean limit
appears to be able to exhibit three distinct phases, I, II, and III
shown in Fig. \ref{Phase.Diagram}, characterized by the values of the
Ginzburg-Landau order parameter $<\psi>$ and its dual order-parameter
$<\phi>$. Here, we explicitly utilized the connection of Section IIE
between the vortex-tangle connectivity probe $O_L$ and the
$U(1)$-ordering in the dual field $\phi$. We found the three regimes
\begin{eqnarray} 
\rm{Region ~~  I    } &  : &  <\psi> \neq 0 ~~~~ <\phi>    = 0, \nonumber \\
\rm{Region ~~  II   } &  : &  <\psi>   =  0 ~~~~ <\phi>    = 0, \nonumber \\
\rm{Region ~~  III  } &  : &  <\psi>   =  0 ~~~~ <\phi> \neq 0. \nonumber 
\end{eqnarray}
At low fields, we have found that region II vanishes. Note that the
transition-line $T_L(B)$ separating the regions II and III inside the
vortex-liquid, was brought out solely through the dual description, it
could not have been detected by studying the Ginzburg-Landau order
parameter $<\psi>$, or any local function of it.

A few further comments are in order.
In the low-field regime, {\it within a lines-only picture of the
  molten phase}, one finds that the longitudinal correlation length of
field-induced vortex lines above melting increases, due to the
increased distance between field induced lines, being given by
\begin{eqnarray}
\xi_z = \frac{1}{\Gamma^2} ~ \sqrt{\frac{\Phi_0}{B}}.
\end{eqnarray}
It was therefore pointed out in Ref. \onlinecite{Nordborg:B98} that in
order to correctly predict the direct transition from the Abrikosov
vortex lattice to a phase-incoherent vortex liquid at low magnetic
fields, or equivalently predict the direct transition from the crystal
phase to the superfluid phase of $2D$ non-relativistic bosons at $T=0$
at low magnetic, sufficiently large systems in the $z$-direction must
be used. The use of too small systems could result in observing,
merely as a result of a finite-size effect, a {\it normal} $T=0$ $2D$
non-relativic bose-fluid, or equivalently a disentangled
vortex-liquid.  The former cannot exist in the thermodynamic limit in
the absence of disorder, on quite general grounds.

The above is a valid point of concern within the $2D$ boson-liquid
analogy of the vortex system when looking for entanglement. It is no
longer a point of concern if the lines-only approximation is abandoned
and the connectivity of the vortex tangle is probed rather than
entanglement.  (Precisely how to establish a criterion for when
field-induced vortex lines are entangled, also appears to be
problematic to say the least). For all fields we have considered, and
for all sample geometries we have used, it is clear from our results
that we have been able to correctly predict the direct transition from
the Abrikosov vortex lattice to the phase-incoherent vortex liquid.
The onset of $O_L$ and the change in the vortex-tangle connectivity is
a separate matter. The vortex-configurations dominating the
contribution to a change in $O_L$, are thermally induced unbound
vortex loops and not field-induced flux lines. Our results in the
low-field regime are therefore not artifacts of considering too small
systems in the $z$-direction. Quite the contrary, since we see the
change in $O_L$ also when making the system flatter, it supports the
proposition that there exists a regime in the $(B,T)$ phase-diagram,
beyond the line $T_L(B)$, where the notion of a vortex-{\it line}
liquid physics most probably should be revised.

The $U(1)$-transition line $T_L(B)$ has the zero-field superfluid
normal state transition $T_c$ as an endpoint. It is a feature of
extreme type-II superconductors, even homogenous, isotropic
three-dimensional ones, and should moreover occur in Helium$^4$ which
is a perfectly three-dimensional, homogenous, isotropic superfluid.
The proposed transition therefore is not in any obvious way connected
to various previously proposed quite intriguing scenarios leading to a
loss of {\it local line-tension} of field induced vortices, often
referred to as ``decoupling-transitions"
\cite{Glazman:B91,Bulaevskii:L92}.  These phenomena rely on the {\it
  layeredness} of the superconducting compounds, however they have no
symmetry-breaking or order parameter associated with them, but most
importantly do not have a zero-field counterpart. Moreover, probing
phase-coherence between {\it adjacent} layers in a layered
superconductor as was for instance done in Ref.  \cite{Glazman:B91}
(see their Eq. 28) probes {\it maximum} $q_z$-behavior, a part of
reciprocal space not usually associated with {\it critical phenomena},
which are infrared singularities.  Probing phase-coherence between
increasingly distant layers ultimately amounts to computing the
helicity modulus $\Upsilon_z$, which {\it cannot} vanish above the
melting-line of the vortex lattice in the thermodynamic limit, in the
absence of disorder.

The $T_L(B)$ line is potentially an important line in the $(B,T)$
phase diagram. It locates the position in the $(B,T)$-diagram where
the line-only approximation of the vortex-liquid breaks down. Pinning
of vortices by extended objects such as columnar pins may very well
turn out to be inefficient beyond the line $T_L(B)$. It also shows
that the line-only approximation can be used to describe the
vortex-liquid phase and the first order melting transition of the VLL
at $T_M(B)$ only for large and intermediate magnetic induction. In low
magnetic fields, on the other hand, $T_L(B)$ and $T_m(B)$ collapse
into a single line \cite{Nguyen:cm98}. Here, it would appear that a
line-only approximation does not describe the vortex liquid properly.
The fields where the line only approximation fails in the entire
liquid regime are expected to be of order $1T$ or less in YBCO
\cite{Nguyen:cm98}.

\section{Acknowledgments} Support from the Research Council of Norway 
(Norges Forskningsr{\aa}d) under Grants No. 110566/410, No. 110569/410, 
as well as a grant for computing time under the Program for
Super-computing, is gratefully acknowledged. We thank E. W. Carlson,
S. K. Chin, J. Hove, D. A. Huse, J. S. H{\o}ye, S. A. Kivelson, J. M.
Kosterlitz, F. Ravndal, A. M. J. Schakel, and N. C. Yeh for useful
communications. In particular, we thank Z.  Te{\v s}anovi{\'c} and P.
B. Weichman for discussions and critical readings of the manuscript.
Finally, we would like to express our sincere thanks to J{\o}rn
Amundsen for his invaluable and continuing assistance in optimizing
our computer codes for use on the CrayT3E.


\begin{thebibliography}{100}
\bibitem{Footnote:VLL} We use the abbreviation VLL rather than FLL
  (flux-line lattice) throughout this paper. The reason is that while
  flux lines, i.e. lines of confined magnetic flux, do not exist in
  the extreme type-II limit $\lambda \to \infty$, vortex lines do.
\bibitem{Emery:cm99} V. J. Emery and S. A. Kivelson, condmat-9902179,
  and references therein. In addition, the role of classical
  phase-fluctuations in high-$T_c$ cuprates have recently been
  considered using Monte-Carlo simulations of the 3DXY-model, i.e.
  phase-only approximation to the full Ginzburg-Landau theory, see
  Ref. \onlinecite{Carlson:cm99}.
\bibitem{Footnote:Houghton} Ref. \onlinecite{Houghton:B89} explicitly
  avoids a discussion of the low-field part of the melting curve. The
  calculation was performed from the low-temperature side using a
  Lindemann-criterion for melting, and the fluctuation propagator for
  the vortex system was obtained for the case where magnetic fields
  surrounding vortices strongly overlap, however with $\lambda$ being
  finite. At very low fields where the distance between field-induced
  vortex lines becomes large, the Lindemann criterion most likely
  becomes unphysical, and is therefore not useful for locating the VLL
  melting transition line as $B \to 0, T \to T_c$.
\bibitem{Footnote:dp} In Refs. \onlinecite{Li:B94,Nguyen:B98a} a
  slightly simpler form $D(p) \sim \exp(-\varepsilon(T) p/k_B T)$ was
  used. Adding an algebraic prefactor $p^{-\alpha}$ gives a better
  description of the numerical data over a wider temperature range.
\bibitem{Footnote:exponent} This value of the exponent is based on a
  mapping of the vortex-loop gas to a $3D$ non-interacting
  non-relativistic bose-system \cite{Hoye:JSP94}. It evidently gives a
  good fit to our numerical data. The fact that the bose-system is
  non-interacting reflects the fact that the vortices are well
  screened. This can only happen with infinite vortex-loops present,
  so the value $\alpha=5/2$ is expected to be a very precise value for
  $\alpha$ strictly speaking only immediately above $T_c$. Our
  simulation results are consistent with $\alpha = 2.4 \pm 0.10$, if
  one attempts a more precise determination of $\alpha$ at $T=T_c$.
  This is essentially consistent with the results of Antunes and
  Bettencourt, who find $\alpha=2.23 \pm 0.04$, see Ref.
  \onlinecite{Antunes:L98b}. It is possible that these authors have a
  more precise way of obtaining $\alpha$ than what Monte-Carlo methods
  offer. A very precise determination of $\alpha$ is however not our
  main concern in this paper, rather we wish to point out the fact
  that $D(p)$ clearly changes from exponential to power-law behavior.
  A simple scaling argument gives the too large value $\alpha=3$ which
  provides a moderately good fit to numerical data, see Ref.
  \onlinecite{Nguyen:B98a}. However as noted in Ref.
  \onlinecite{Williams:L98}, the value $\alpha=2.4$ fits the numerical
  data for $D(p)$ in Ref. \onlinecite{Nguyen:B98a} better. The reason
  for this is that the scaling argument assumes that the vortex-loop
  perimeter scales with its diameter. These perimeters are, however,
  ``crinkled" objects reducing the value of $\alpha$ from $3$ to
  $\approx 2.5$.
\bibitem{Footnote:Jagla} The authors of Ref. \onlinecite{Jagla:B96}
  concluded that percolation was necessary in order to destroy the
  longitudinal superfluid density regardless of magnetic field, not
  unlike our own previous claim of Ref.  \onlinecite{Nguyen:B98b}.
  However, as the present work and the results of Ref.
  \onlinecite{Nguyen:cm98} show, the loss of longitudinal superfluid
  density does not necessarily imply that the connectivity of the
  vortex tangle changes dramatically. The loss of superfluid density
  always occurs at the VLL melting transition, which at elevated
  magnetic fields is well below the temperature where we find a change
  in connectivity. We believe, however, that the remarks of Refs.
  \onlinecite{Jagla:B96,Nguyen:B98b} are essentially correct at low
  magnetic fields. Namely, at low fields the VLL melting transition,
  loss of longitudinal superfluid density, and change of vortex-tangle
  connectivity occur at the same temperature.
\end{thebibliography}


\end{multicols}
\end{document}